\documentclass[12pt,a4paper,final]{iopart} 
\usepackage{graphicx}
\pdfoutput=1
\usepackage{amsfonts, amsmath, amssymb, dsfont}

\usepackage{leftidx}
\usepackage{cite}
\usepackage{booktabs}
\usepackage[retainorgcmds]{IEEEtrantools}

\usepackage[breaklinks=true,colorlinks=true,linkcolor=blue,urlcolor=blue,citecolor=blue]{hyperref}
\numberwithin{equation}{section}
\usepackage{cleveref}

\def\limth{\lim\nolimits_\text{th}}

\def\epp{\: .}
\def\epc{\: ,}
\def\alp{\alpha}
\def\bet{\beta}

\def\lam{\lambda}

\def\cO{\mathcal{O}}
\def\boldx{{\boldsymbol{x}}}
\def\blam{{\boldsymbol{\lambda}}}
\def\brho{{\boldsymbol{\rho}}}
\def\be{\begin{equation}}
\def\ee{\end{equation}}
\def\barr{\begin{IEEEeqnarray}}
\def\earr{\end{IEEEeqnarray}}
\newcommand{\ket}[1]{{\left|#1\right\rangle}}
\newcommand{\bra}[1]{{\left\langle #1\right|}}

\newcommand{\skalarszorzat}[2]{{\langle #1 | #2 \rangle}}

\newcommand{\ftk}[2]{\text{FT}\big[#1\big](#2)}
\newcommand{\ft}[1]{\text{FT}\big[#1\big]}
\newcommand{\iftx}[2]{\text{FT}^{-1}\big[#1\big](#2)}

\begin{document}

\title{Quench action approach for releasing the N\'eel state into the spin-1/2 XXZ chain}

\author{M. Brockmann, B. Wouters, D. Fioretto, J. De Nardis, R.~Vlijm, and J.-S. Caux$^{1}$}
\address{$^1$Institute for Theoretical Physics, University of Amsterdam, Science Park 904,\\
Postbus 94485, 1090 GL Amsterdam, The Netherlands}
\ead{m.brockmann@uva.nl}

\begin{abstract}
The steady state after a quantum quench from the N\'eel state to the anisotropic Heisenberg model for spin chains is investigated. Two methods that aim to describe the postquench non-thermal equilibrium, the generalized Gibbs ensemble and the quench action approach, are discussed and contrasted. Using the recent implementation of the quench action approach for this N\'eel-to-XXZ quench, we obtain an exact description of the steady state in terms of Bethe root densities, for which we give explicit analytical expressions. 

Furthermore, by developing a systematic small-quench expansion around the antiferromagnetic Ising limit, we analytically investigate the differences between the predictions of the two methods in terms of densities and postquench equilibrium expectation values of local physical observables. Finally, we discuss the details of the quench action solution for the quench to the isotropic Heisenberg spin chain. For this case we validate the underlying assumptions of the quench action approach by studying the large-system-size behavior of the overlaps between Bethe states and the N\'eel state.
\end{abstract}

\section{Introduction}
The study of non-equilibrium quantum dynamics has been recently boosted by new experimental and theoretical advances~\cite{2008_Bloch, 2014_Langen_arxiv14086377, 2011_Polkovnikov_RMP_83}. From the experimental point of view it became possible to realize well-controlled isolated quantum systems using cold atoms and optical lattices~\cite{
2006_Kinoshita_NATURE_440, 
2012_Trotzky_NATPHYS_8, 
2012_Cheneau_NATURE_481, 
2012_Gring_SCIENCE_337}. 
In these systems, the quantum coherence of the time evolution is preserved on sufficiently long time scales, and as such it is possible to investigate the unitary dynamics of extended systems, neglecting the dissipation and decoherence due to the coupling with the external environment. In this context, the paradigm that has emerged is that of the so-called quantum quench~\cite{
2006_Calabrese_PRL_96, 
2007_Rigol_PRL_98, 
2006_Rigol_PRA_74, 
rigol_dunjko_08,
2008_Barthel_PRL_100, 
2010_Cramer_NJP_12, 
2011_Cassidy_PRL_106, 
2011_Calabrese_PRL_106, 
2012_Calabrese_JSTAT_P07016, 
2012_Calabrese_JSTAT_P07022, 
2014_Bucciantini_JPA_47, 
2014_Sotiriadis, 
2009_Barmettler_PRL_102, 
2010_Barmettler_NJP_12, 
2009_Rossini_PRL_102, 
2010_Rossini_PRB_82, 
2009_Faribault_JMP_50, 
2010_Fioretto_NJP_12, 
2010_Mossel_NJP_12, 
2011_Igloi_PRL_106, 
2011_Banuls_PRL_106,  
2011_Rigol_PRA_84, 
2012_Brandino_PRB_85, 
2012_Demler_PRB_86, 
2012_He_PRA_85, 
2013_He_PRA_87, 
2013_Caux_PRL_110,
2013_Mussardo_PRL_111, 
2013_Kormos_PRB_88,
2013_Pozsgay_JSTAT_P07003, 
2013_Fagotti_JSTAT_P07012, 
2013_Pozsgay_Jstat_10, 
2013_Liu,
2013_Marcuzzi_PRL_111,  
2014_DeNardis_PRA_89, 
2014_Essler_PRB_89, 
2014_Bertini_Sinegordon, 
2014_Fagotti_PRB_89,
2014_Sotiriadis_PLB_734,
2014_Wouters, 
2014_Pozsgay_Dimer, 
2014_Pozsgay_qbosons,
2006_Cazalilla_PRL_97,
2009_Uhrig_PRA_80,
2009_Iucci_PRA_80,
2010_Iucci_NJP_12,
2011_Dora_PRL_106,
2011_Mitra_PRL_107,
2012_Karrasch_PRL_109,
2012_Mitra_PRL_109,
2012_Mitra_PRB_85,
2012_Dora_PRB_86,
2012_Rentrop_NJP_14,
2013_Mitra_PRB_87,
2013_Bacsi_PRB_88,
2013_Ngo_Dinh_PRB_88,
2013_Adu_Smith_NJP_15,
2014_Bernier_PRL_112,
2013_Heyl_PRL_110, 
2013_Fagotti,
2013_Karrasch_PRB_87,
2014_Fugallo_PRB_89,
2014_Heyl, 
2014_Andraschko_PRB_89,
2014_PRB_Vajna_89,
2014_PRB_Kriel_90}. 
The system is prepared in a pure state with a finite energy density and then let evolve coherently. Particularly important is the issue of how to obtain a description of the steady state and of the mechanisms implementing relaxation.

The investigation of non-equilibrium dynamics of many-body quantum systems however represents a major theoretical challenge: the exponentially (in system size) large Hilbert space  severely limits brute-force approaches to small systems, while the simplifying techniques that enable us to understand equilibrium physics are generally not applicable. As such, an intriguing research direction is the study of integrable models, where the rich analytical structure available allows us to investigate quantum quenches directly in the thermodynamic limit. On the one hand, many integrable models can be realized in cold atom setups~\cite{2006_Kinoshita_NATURE_440, 2012_Gring_SCIENCE_337,2013_Fukuhara_NATURE_502}, so this line of research could  have direct experimental applications. On the other hand, integrable models are the first outpost to probe the effect of interactions on relaxation of thermodynamically large quantum systems, and their study is expected to lead to important insights into the generic underlying mechanism for equilibration.

A precise definition of integrability in quantum mechanics is not yet agreed upon~\cite{2011_Caux_JSTAT_P02023} although the general consensus agrees to classify as integrable all systems that have at least a set of order $N$ of local conserved charges, where $N$ is the number of constituents. These charges are expected to have much influence on local physical observables after the quench~\cite{ 2006_Kinoshita_NATURE_440} and, in particular, to characterize their steady state. In the same spirit of thermalization to a Gibbs ensemble (GE) where the Hamiltonian and the particle number are the only conserved charges, integrable models are expected to thermalize to a generalized Gibbs ensemble (GGE)~\cite{2007_Rigol_PRL_98,2006_Rigol_PRA_74} such that the entropy of the system is maximized under the constraint that the conserved charges are fixed by their expectation values in the initial state. This paradigm has been proven to be correct for free systems or systems mappable to free systems~\cite{
2008_Barthel_PRL_100, 
2010_Cramer_NJP_12, 
2011_Cassidy_PRL_106,
2011_Calabrese_PRL_106, 
2012_Calabrese_JSTAT_P07016, 
2012_Calabrese_JSTAT_P07022, 
2014_Bucciantini_JPA_47, 
2014_Sotiriadis}. 
Until recently~\cite{2014_Fagotti_PRB_89, 2014_Wouters, 2014_Pozsgay_Dimer, 2014_Pozsgay_qbosons} it was rarely tested for truly interacting systems \cite{2010_Fioretto_NJP_12}. 

A first-principles based approach, valid for generic quantum systems, has been introduced recently~\cite{2013_Caux_PRL_110,2014_DeNardis_PRA_89}. In the so-called quench action method the overlaps between the initial state and the eigenstates of the system, and in particular their scaling behavior in the thermodynamic limit, lead to an effective action whose saddle point characterizes the system at equilibrium. In Refs~\cite{2014_DeNardis_PRA_89,2014_Pozsgay_Dimer,2014_Wouters} this method was used to exactly predict the equilibrium expectation values of some local observables for some interaction quenches (where the system is prepared in the ground state of the Hamiltonian and the value of coupling constant is suddenly changed) in the Lieb-Liniger model of interacting bosons~\cite{2014_DeNardis_PRA_89} and in the anisotropic spin-1/2 Heisenberg chain~\cite{2014_Wouters,2014_Pozsgay_Dimer}. In the Lieb-Liniger case the GGE implementation was not feasible due to the divergence of expectation values of local conserved charges on the initial state \cite{2013_Kormos_PRB_88}, while in Ref.~\cite{2014_Wouters} the prediction of the GGE implemented with all known local conserved charges turned out to be incorrect. This was numerically verified by using linked-cluster expansions \cite{2014_Wouters, 2014_Rigol_PRL, 2014_Rigol_arxiv}. The same conclusion was obtained in Ref.~\cite{2014_Pozsgay_Dimer} where a different type of quench in the same model was also considered. 

In this paper we review and expand some of the results presented in Ref.~\cite{2014_Wouters}, providing a detailed implementation of the quench action method for the problem at hand. In Sec.~\ref{sec:XXZ} we introduce the spin-1/2 XXZ chain and in Sec.~\ref{sec:QuenchesinXXZ} we review the methods utilized to study quenches in integrable models. In Secs~\ref{sec:quench_overlaps}, \ref{sec:analytical_solution}, and \ref{sec:large_Delta} we focus on the implementation of the quench action approach to the N\'eel-to-XXZ quench. Finally, in Sec.~\ref{sec:XXX} we do the same for the N\'eel-to-XXX quench and provide for this specific quench in Sec.~\ref{sec:R:exotic} extra evidence for the validity of the quench action approach by analyzing the scaling properties of the overlaps between the N\'eel state and some classes of Bethe states.

\section{The spin-1/2 XXZ chain}\label{sec:XXZ}
The one-dimensional antiferromagnetic spin-1/2 XXZ chain is described by the Hamil\-tonian
\begin{equation}\label{eq:Hamiltonian_XXZ}
	H = \frac{J}{4}\sum_{j=1}^{N}\left[\sigma_{j}^{x}\sigma_{j+1}^{x}+\sigma_{j}^{y}\sigma_{j+1}^{y}+\Delta ( \sigma_{j}^{z}\sigma_{j+1}^{z}-1)\right]\epc
\end{equation}
where the Pauli matrices $\sigma_j^\alpha$ ($\alpha=x,y,z$) represent the spin-$1/2$ degrees of freedom at lattice sites $j=1,2,\ldots, N$. We assume periodic boundary conditions $\sigma_{N+1}^\alpha = \sigma_1^\alpha$. The exchange coupling $J>0$ sets the energy scale and $\Delta$ parametrizes the anisotropy of the nearest-neighbor spin-spin coupling.  Throughout the paper we focus on quenches to the gapped antiferromagnetic regime $\Delta>1$ and work in the zero-magnetization sector. Details about the quench to the isotropic point $\Delta=1$, where the theory is gapless, are given in Sec.~\ref{sec:XXX}.

\subsection{Bethe Ansatz solution} 
The XXZ Hamiltonian can be diagonalized by Bethe Ansatz~\cite{1931_Bethe_ZP_71,1958_Orbach_PR_112}. We choose the ferromagnetic state $\left|\uparrow \uparrow \ldots \uparrow \right\rangle =\left|\uparrow\right\rangle^{\otimes N} $ with all spins up as a reference state and construct interacting spin waves as excitations on this state. A state with $M$ down spins falls in the magnetization sector $\langle\sigma_\text{tot}^z\rangle/2 = N/2-M$ and is completely characterized by a set of complex quasimomenta $\blam=\{\lambda_j\}_{j=1}^M$, which are called rapidities. It is given by
\begin{subequations}\label{eq:BA_state}
\begin{equation}
	|\blam\rangle = \sum_{\boldx} \Psi_{M}\!\left(\boldx|\blam\right)\ \sigma_{x_1}^-\ldots\sigma_{x_M}^-\left|\uparrow\uparrow\ldots\uparrow\right\rangle \epc
\end{equation}
where the positions of the down spins are denoted by the coordinates $\boldx=\{x_j\}_{j=1}^M\subset \{1,\ldots,N\}$, and we assume $x_j<x_k$ for $j<k$. The explicit wave function in coordinate space takes a Bethe Ansatz form,
\begin{equation} 
	\Psi_{M}\!\left(\boldx|\blam\right)=\sum_{Q\in\mathcal{S}_M} (-1)^{[Q]} 
 \exp\left\{- i \sum_{j=1}^M x_j\, p(\lambda_{Q_j}) - \frac{i}{2} \sum_{\substack{ j,k=1\\ k>j}}^M \theta_2(\lambda_{Q_k} - \lambda_{Q_j}) \right\}\epp
\end{equation}
\end{subequations}
The sum runs over the set of all permutations of integers $1, \ldots, M$, denoted by $\mathcal{S}_M$, and $(-1)^{[Q]}$ is the parity of the permutation $Q\in\mathcal{S}_M$. The total momentum of the state~\eqref{eq:BA_state} is given by
\begin{equation}\label{eq:momentum}
P_\blam = \sum_{j=1}^M p(\lam_j) \epc \quad \text{where} \quad p(\lam) = - i \ln\left[\frac{\sin(\lambda + \frac{i\eta}{2})}{\sin(\lambda- \frac{i\eta}{2})}\right]
\end{equation}
is the momentum associated with a rapidity $\lam$. The parameter $\eta > 0$ is determined by the anisotropy $\Delta=\cosh(\eta) > 1$ (the limit $\eta\to 0$ is considered in Sec.~\ref{sec:XXX}). Throughout the paper we choose the branch $-\pi/2 \leq \text{Re}(\lam) <  \pi/2$. Furthermore, $\theta_2$ is the scattering phase shift defined by
\begin{equation} \label{eq:scateringphaseXXZ}
	\theta_2(\lambda) =2 \arctan \left( \frac{\tan(\lam)}{\tanh(\eta)} \right)\epp
\end{equation}
The state~\eqref{eq:BA_state} is called Bethe state if the rapidities $\blam$ satisfy the Bethe equations,
\begin{equation}\label{eq:BAE}
	\left[\frac{\sin(\lambda_j+\frac{i\eta}{2})}{\sin(\lambda_j-\frac{i\eta}{2})}\right]^N=-\prod_{k=1}^M\frac{\sin(\lambda_j-\lambda_k+i\eta)}{\sin(\lambda_j-\lambda_k-i\eta)} \epc
\end{equation}
for $j=1,\ldots,M$. Rapidities obeying these equations are called Bethe roots. A Bethe state is an eigenstate of the XXZ Hamiltonian~\eqref{eq:Hamiltonian_XXZ} with energy
\begin{equation}
	\omega_\blam = J \sum_{j=1}^M \left\{ \cos[p(\lam_j)] - \cosh(\eta) \right\} = -J\sum_{j=1}^M\frac{\sinh^2(\eta)}{\cosh(\eta)-\cos(2\lambda_j)}\epp
\end{equation}
Bethe states are orthogonal and their norm is given by $\|\left|\blam\right\rangle\| = \sqrt{\langle \blam | \blam \rangle}$ with~\cite{1981_Gaudin_PRD_23,1982_Korepin_CMP_86}
\begin{subequations}\label{eq:norm_Bethe_state}
\begin{align}
	\langle \blam| \blam \rangle &= \sinh^M(\eta) \prod_{\substack{j,k=1\\j\neq k}}^M \frac{\sin(\lambda_j - \lambda_k + i \eta)}{\sin(\lambda_j - \lambda_k)} \det{}_{\!M}(G) \epc\\
	G_{jk} &= \label{eq:Gaudin_matrix}  \delta_{jk}\left(NK_{\eta/2}(\lambda_j)-\sum_{l=1}^{M}K_\eta(\lambda_j-\lambda_l)\right) + K_\eta(\lambda_j-\lambda_k)\epc
\end{align}
\end{subequations}
where $K_\eta(\lambda)=\sinh(2\eta)/[\sin(\lambda+i\eta)\sin(\lambda-i\eta)]$ is the derivative of the scattering phase shift $\theta_2$.

\subsection{String hypothesis}
For large system size $N$, the question of how the rapidities organize themselves is addressed by the string hypothesis~\cite{1931_Bethe_ZP_71, 1971_Takahashi_PTP_46}. Rapidities of a Bethe state get grouped in strings,
\begin{equation}\label{eq:stringdef}
	\lam^{n,a}_\alpha=\lam^n_\alpha+\tfrac{i\eta}{2}(n+1-2a) + i \delta^{n,a}_\alpha
\end{equation} 
for $a=1,\ldots,n$, where $n$ is the length of the string and the deviations $\delta^{n,a}_\alpha$ vanish (typically) exponentially in system size. A more detailed discussion can be found in Sec.~\ref{sec:R:exotic}.

In the gapped regime ($\Delta>1$) the string centers $\lam^n_\alpha$ are real and lie in the interval $[-\pi/2,\pi/2)$. The physical interpretation of such an $n$-string is a bound state of $n$ magnons, which becomes in the Ising limit $\Delta\to\infty$ a block of $n$ adjacent down spins. Let $M_n$ be the total number of $n$-strings of a Bethe state, then $\alpha=1,2,\ldots,M_n$ labels the $n$-strings and $\sum_{n=1}^\infty n\,M_n=M$. In Ref.~\cite{1983_Tsvelik_AP_32} it is argued that the string hypothesis is valid if temperature and/or magnetization are nonzero.

Under the string hypothesis and for vanishing deviations a state is solely characte\-rized by its string centers $\lam_{\alpha}^n$. Neglecting the string deviations, the logarithmic form of the Bethe Eqs~\eqref{eq:BAE} can be recast into the Bethe-Gaudin-Takahashi (BGT) equations for string centers \cite{1971_Takahashi_PTP_46,1972_Takahashi_PTP_48,GaudinBOOK},
\begin{subequations}\label{eq:BGT}
\begin{equation}
	\theta_{n} \left( \lam_{\alp}^n \right) \ =\ \frac{2\pi}{N} I_{\alp}^n + \frac{1}{N} \sum_{\substack{(m,\beta)\,\neq \\ (n,\alp)}} \theta_{nm} \left( \lam_{\alp}^n - \lam_{\bet}^m \right)
\end{equation}
for $n\geq 1$ and $\alpha=1,2,\ldots,M_n$. Here,
\begin{equation}
	\theta_{nm}(\lam) =  (1-\delta_{nm})\theta_{|n-m|}(\lam) + 2 \theta_{|n-m| + 2}(\lam) + \ldots + 2 \theta_{n+m-2}(\lam) + \theta_{n+m}(\lam)
\end{equation}
and
\begin{equation}
	\theta_{n}(\lam)\ =\ 2 \arctan \left( \frac{\tan(\lam)}{\tanh(\frac{n\eta}{2})} \right) \epp
\end{equation}
\end{subequations}
Note that the function $\theta_2$ is the scattering phase shift~\eqref{eq:scateringphaseXXZ}. The quantum numbers $I_{\alpha}^n$ are integers (half-odd integers) if $N-M_{n}$ is odd (even).

\subsection{The thermodynamic limit}
By thermodynamic limit we mean the limit of infinite system size, $N\to\infty$, while keeping the fraction of down spins $M/N$ fixed. We will denote it by $\limth$. In this limit Bethe states are characterized by distributions of string centers. The density of $n$-strings is given by the function $\rho_n$, such that $N \rho_n(\lam) \, \mathrm{d}\lam$ is the number of $n$-strings in the interval $[\lam,\lam+\mathrm{d}\lam]$. 

In the thermodynamic limit, the BGT Eqs~\eqref{eq:BGT} become a set of integral equations for the density distributions~\cite{1971_Takahashi_PTP_46,1972_Takahashi_PTP_48,GaudinBOOK},
\begin{subequations}\label{eq:BTGthlim}
\begin{equation}
	\rho_{n,t}(\lam) \ =\ a_{n}(\lam) - \sum_{m=1}^{\infty}  (a_{nm} \ast \rho_{m}) (\lam)
\end{equation}
for $n\geq1$, where $\rho_{n,t}(\lambda) = \rho_{n}(\lambda) + \rho_{n,h}(\lambda)$ and $\rho_{n,h}$ is the hole density of $n$-strings. Further,
\begin{equation} \label{eq:kernelXXZsum}
	a_{nm}(\lam) =  (1-\delta_{nm}) a_{|n-m|}(\lam) + 2 a_{|n-m| + 2}(\lam) + \ldots + 2 a_{n+m-2}(\lam) + a_{n+m}(\lam)
\end{equation}
with
\begin{equation} \label{eq:kernelXXZ}
	a_{n}(\lam) = \frac{1}{2\pi} \frac{\mathrm{d}}{\mathrm{d}\lam} \theta_{n}(\lam) = \frac{1}{\pi} \frac{\sinh (n\eta)}{\cosh (n\eta) - \cos (2\lam)} \epp
\end{equation}
\end{subequations}
The convolution is defined by
\begin{equation} \label{eq:convolution}
	(f\ast g)\, (\lam) = \int_{-\pi/2}^{\pi/2} \mathrm{d}\mu \, f(\lam-\mu)\, g(\mu)\epp
\end{equation}
For both numerical and analytical evaluation of the integral equations, it is often convenient to get rid of the infinite sum over string types and to work with the ``partially decoupled'' set of equations. The partially decoupled form of the thermodynamic BGT equations can be derived~\cite{TakahashiBOOK},
\begin{subequations}\label{eq:BTGthlim_fact}
\begin{equation} \label{eq:BTGthlim_fact_a}
	\rho_{n}(1 + \eta_{n}) = s \ast (\eta_{n-1}\rho_{n-1} + \eta_{n+1}\rho_{n+1})
\end{equation}
for $n\geq 1$, where the $\lam$-dependence is left implicit and we use the conventions $\eta_0(\lam)=1$ and $\rho_{0}(\lam)=\delta(\lam)$. The kernel in Eqs~\eqref{eq:BTGthlim_fact_a} reads 
\begin{equation} \label{eq:defs}
	s(\lam) = \frac{1}{2\pi} \sum_{k\in\mathbb{Z}} \frac{e^{-2i k\lam}}{\cosh(k\eta)} \epp
\end{equation}
\end{subequations}

The set of positive, smooth functions $\brho=\{\rho_n\}_{n=1}^\infty$ represents an ensemble of states with Yang-Yang entropy
\begin{equation} \label{eq:YYentropyXXZ}
	S_{YY} \left[ \brho \right] = N \sum_{n=1}^{\infty}  \int_{-\pi/2}^{\pi/2} \mathrm{d}\lam \left[ \rho_{n,t}(\lambda) \ln\rho_{n,t}(\lambda) - \rho_{n}(\lambda) \ln \rho_{n}(\lambda) - \rho_{n,h}(\lambda)\ln\rho_{n,h}(\lambda) \right]\epp
\end{equation}

It is useful to introduce the notion of a representative state for a set of distributions $\brho$. It is defined as a Bethe state $| \blam \rangle$ for large finite system size $N$ such that we have for any smooth (local) observable $\cO$
\begin{equation}\label{eq:rep_state}
	\langle \blam| \cO | \blam \rangle = \langle \brho| \cO | \brho \rangle \big[ 1 + O(N^{-1}) \big] \epc
\end{equation}
where the quantity $\langle \brho|  \cO | \brho \rangle $ is a functional of the set of distributions. Given a set of densities $\brho$, there is an entropic number $e^{S_{YY}[\brho]}$ of possible choices for a representative state \cite{KorepinBOOK}. In Eq.~\eqref{eq:rep_state} and in the following we use the same symbol $\cO$ for operators both for finite system size and in the thermodynamic limit. It is clear from the context which one is meant.

\subsection{Conserved charges}
From the method of the algebraic Bethe Ansatz~\cite{KorepinBOOK} a set of conserved charges can be constructed~\cite{1994_Grabowski_MPLA_9}. Central in this construction is the transfer matrix $t(\lam)$, which commutes for any pair of spectral parameters $\lam$ and $\lam'$, $ [t(\lam),t(\lam')]=0 $. The transfer matrix is diagonal on the basis of Bethe states with eigenvalues
\begin{equation} \label{eq:transfermatrix}
	\tau(\lam) = \prod_{k=1}^M \frac{\sin(\lam - \lam_k - i \eta)}{\sin(\lam - \lam_k)} + \left[ \frac{\sin(\lam - \frac{i\eta}{2})}{\sin(\lam + \frac{i\eta}{2})} \right]^N \prod_{k=1}^M \frac{\sin(\lam - \lam_k + i \eta)}{\sin(\lam - \lam_k)} \epp
\end{equation}  
The conserved charges are defined via the coefficients of the operator expansion of the logarithm of the transfer matrix around the point $\lam=i\eta/2$,
\begin{equation} \label{eq:conservedcharges}
	Q_{m+1} = i \frac{\sinh^m(\eta)}{2^m}\left.\frac{\partial^m}{\partial \lam^m} \ln[t(\lam)]\right|_{\lam=i\eta/2} \epc\quad m\geq 0\epp
\end{equation}
They commute by construction. Note that $P = - Q_1$ and $H = JQ_2$. The range of the charge $Q_{m}$ is $m$ (where we assume $m < N$). This means that each element $Q_j^{(m)}$ in the decomposition $Q_{m}=\sum_{j=1}^{N}Q_j^{(m)}$ acts only nontrivially on a block of $m$ adjacent sites.

In the thermodynamic limit the charges $\{ Q_m \}_{m=1}^\infty$ form an infinite set of local conserved charges. Acting on a representative state $|\blam\rangle$, the eigenvalue of charge $Q_{m+1}$ is given by
\begin{subequations}
\begin{equation}
	\limth \langle \blam | \frac{Q_{m+1}}{N}| \blam \rangle = \sum_{n=1}^{\infty} \int_{-\pi/2}^{\pi/2}\mathrm{d}\lam \, \rho_n(\lam) \, c_{m+1}^{(n)}(\lam) \epc\quad m\geq 0\epc
\end{equation}
where
\begin{equation} \label{eq:defcn}
	c_{m+1}^{(n)}(\lam) = i(-1)^{m} \frac{\sinh^m(\eta)}{2^m} \frac{\partial^m}{\partial \lam^m} \ln \left[ \frac{\sin(\lam+\frac{i\eta}{2}n)}{\sin(\lam-\frac{i\eta}{2}n)}\right] \epp
\end{equation}
\end{subequations}
To see this, note that an $n$-string \eqref{eq:stringdef} with string center $\lam_\alpha^n$ and with neglected deviations $\delta_\alpha^{n,a}$ contributes a factor 
\begin{equation}
	\frac{\sin[\lam - \lam_\alpha^n - \frac{i\eta}{2}(n+1)]}{\sin[\lam - \lam_\alpha^n + \frac{i\eta}{2}(n-1)]}
\end{equation}
to the first term of the transfer-matrix eigenvalue~\eqref{eq:transfermatrix}. As long as $m<N$, the second term of Eq.~\eqref{eq:transfermatrix} does not contribute to the expectation values of charge $Q_{m+1}$. In the thermodynamic limit this is the case for any finite $m$.

\section{Methods for quenches in the XXZ model} \label{sec:QuenchesinXXZ}
For a general global quantum quench into the spin-1/2 XXZ chain of length $N$, one prepares an initial state $|\Psi_0\rangle$ and lets it evolve in time. We will also use $|\Psi_0\rangle$ as the symbol for the initial state in the thermodynamic limit. It will become clear from the context which state is meant. The unitary time evolution is governed by the Hamiltonian~\eqref{eq:Hamiltonian_XXZ}. At time $t$ after the quench, the state of the system can be expanded in the basis of Bethe states,
\begin{equation}
	\left| \Psi (t) \right\rangle = \sum_{\blam}\, e^{-i \omega_{\blam}t} \left\langle \blam | \Psi_{0} \right\rangle \left| \blam \right\rangle \epc
\end{equation}
where the sum runs over all Bethe states in the $2^N$-dimensional Hilbert space. The postquench time-dependent expectation value of a generic operator $\cO$ is exactly given by the double sum
\begin{equation} \label{eq:expect1}
	\left\langle \Psi(t) \right| \cO \left| \Psi (t) \right\rangle = \sum_{\blam,\blam'} e^{-S_{\blam}^*-S_{\blam'}}e^{i(\omega_{\blam} - \omega_{\blam'})t} \langle \blam| \cO |\blam'\rangle \epc 
	\end{equation}
where the quantities $S_{\blam} = - \ln \left\langle \blam | \Psi_{0} \right\rangle$ are called overlap coefficients. This double sum over the full Hilbert space is problematic, as the number of its terms grows exponentially with system size. 

In the thermodynamic limit a generic initial state is an infinite superposition of energy eigenstates. Due to dephasing in Eq.~\eqref{eq:expect1}, observables of such a closed, out-of-equilibrium, many-body quantum system are expected to relax to an equilibrium value. An important question is whether and how this system relaxes to a steady state, {\it i.e.}, whether and how equilibrium expectation values of these operators can effectively be computed on a specific thermodynamic Bethe state, called the steady state and denoted by $|\brho^{\Psi_0}\rangle$:
 \begin{equation} \label{eq:spev}
	\lim_{t \to \infty}\limth \left\langle \Psi (t) \right| \cO \left| \Psi (t) \right\rangle = \lim_{t\to\infty}\limth\langle \Psi_0 | e^{iHt} \mathcal{O} e^{-iHt}|\Psi_0\rangle = \left\langle\brho^{\Psi_0} \right| \cO \left| \brho^{\Psi_0}\right\rangle \epp
\end{equation}

\subsection{The generalized Gibbs ensemble}
For integrable systems, the presence of local conserved charges heavily constrains the time evolution after a quench. It is believed~\cite{2007_Rigol_PRL_98,2006_Rigol_PRA_74} that equilibrium expectation values of local observables are well-described by a generalized Gibbs ensemble (GGE) based on the local conserved charges present in the model. For the XXZ Hamiltionian, the infinite set $\left\{ Q_m \right\}_{m=1}^\infty$ defined in Eq.~\eqref{eq:conservedcharges} comprises all known local conserved charges. 
Given a local observable $\mathcal{O}$, the GGE predicts
\begin{equation}\label{eq:defGGE}
	\lim_{t \to \infty}\limth \left\langle \Psi (t) \right| \cO\, \left| \Psi (t) \right\rangle = \lim_{a\to\infty}\limth \frac{\text{Tr}\left( \cO  e^{- \sum_{m=1}^{a} \beta_{m} Q_{m}} \right) }{\text{Tr}\left( e^{- \sum_{m=1}^{a} \beta_{m} Q_{m}} \right) } \epc
\end{equation}
where the trace is over the full Hilbert space. The limit $a\to\infty$ after taking the thermodynamic limit $\limth$ indicates that we take infinitely many local conservation laws into account. The quantities $\left\{ \beta_m \right\}_{m=1}^{\infty}$ are the generalized chemical potentials associated with the charges. They are determined by the expectation values of the conserved charges on the initial state,
\begin{equation} \label{eq:GGEconstraints}
	\limth \frac{1}{N}\langle \Psi_0 | Q_n | \Psi_0 \rangle = \lim_{a\to\infty}\limth\frac{1}{N} \frac{\text{Tr}\left( Q_n e^{- \sum_{m=1}^{a}\beta_{m} Q_{m}} \right)}{\text{Tr}\left(e^{- \sum_{m=1}^{a}\beta_{m} Q_{m}} \right)} 
\end{equation}
for $n \geq 1$.
Recent years have seen numerous applications of the GGE formalism applied to lattice spin systems~\cite{
2010_Cramer_NJP_12,
2011_Cassidy_PRL_106,
2011_Calabrese_PRL_106, 
2012_Calabrese_JSTAT_P07016, 
2012_Calabrese_JSTAT_P07022, 
2013_Pozsgay_JSTAT_P07003, 
2014_Fagotti_PRB_89, 
2014_Bucciantini_JPA_47}. 
In general, obtaining the values of all chemical potentials is a highly nontrivial problem \cite{2012_Mossel_JPA_45, 2013_Fagotti_JSTAT_P07012} and one is often forced to work with a truncated subset of conserved charges~\cite{2013_Pozsgay_Jstat_10}.

At the level of root densities, the GGE is the set of distributions $\brho^{GGE}$ that maximizes the Yang-Yang entropy~\eqref{eq:YYentropyXXZ} under the constraint that the expectation values of all local conserved charges are fixed by the initial state. The resulting generalized thermodynamic Bethe Ansatz (GTBA) equations~\cite{2012_Mossel_JPA_45} are given by (for details see \ref{app:GTBAGGE})
\begin{subequations}\label{eq:GTBAGGE}
\begin{equation}\label{eq:GTBAGGE_a}
	\ln(\eta_n) = -\delta_{n,1} (s \ast d) + s\ast\left[\ln(1+\eta_{n-1}) + \ln (1+\eta_{n+1}) \right] 
\end{equation}
for $n\geq 1$, where $\eta_0(\lam)=0$ and $s$ is defined in Eq.~\eqref{eq:defs}. Note that the driving term is only present in the first integral equation and is specified by the chemical poten\-tials~$\beta_m$,~$m \geq 2$,
\begin{equation}\label{eq:GTBAGGE_b}
	d(\lam) = \sum_{k \in \mathbb{Z}} e^{-2ik\lam} \sum_{m=2}^\infty \beta_m \sinh^{m-1}(\eta) (ik)^{m-2} \epp
\end{equation}
\end{subequations}
Since the momentum of the initial state vanishes, we restrict ourselves to the zero-total-momentum sector and a term involving the Lagrange multiplier $\beta_1$ associated with the momentum charge $Q_1$ does not appear (see \ref{app:GTBAGGE}). Combined with the BGT Eqs~\eqref{eq:BTGthlim_fact}, the solution to these GTBA equations is a set of densities $\brho^{GGE} =\{\rho_n^{GGE}\}_{n=1}^\infty$. The claim of the GGE is that for any local operator $\cO$ this set of densities reproduces the steady state expectation value, {\it i.e.},
\begin{equation}\label{eq:GGE_assumption}
	\left\langle\brho^{\Psi_0} \right|\cO \left| \brho^{\Psi_0}\right\rangle = \left\langle\brho^{GGE} \right| \cO \left| \brho^{GGE}\right\rangle \epp
\end{equation}

\subsection{A one-to-one correspondence between local conserved charges and $\rho_{1,h}$}\label{sec:1to1}
In this section we show that for quenches in the spin-1/2 XXZ chain a GGE analysis based on an infinite number of local conserved charges is possible, despite the inaccessibility of the chemical potentials. As indicated in Ref.~\cite{2014_Wouters}, this is due to a one-to-one correspondence between the expectation values of the local conserved charges $\left\{ Q_m \right\}_{m=2}^\infty$ on the initial state and the density $\rho_{1,h}$ of 1-string holes. A detailed derivation of this correspondence is given here.

Since the postquench steady-state densities $\brho^{\Psi_0}$ should reproduce the (normalized) initial values of all local conserved charges, the steady-state distributions obey the constraints
\begin{equation}
	\limth \frac{\left\langle \Psi_0 \right| Q_{m+1} \left| \Psi_0 \right\rangle}{N} = \sum_{n=1}^{\infty} \int_{-\pi/2}^{\pi/2}\mathrm{d}\lam \, \rho_n^{\Psi_0}(\lam) \, c_{m+1}^{(n)}(\lam)
\end{equation} 
for $m\geq 0$ and $|\Psi_0\rangle$ the initial state. Obviously, this set of constraints is in general not very restrictive, there are infinitely many sets of densities $\brho$ that solve them, which was also observed in Refs~\cite{2014_Goldstein_failure_GGE,2014_Pozsgay_GGE}. However, it turns out that the set of initial expectation values of the local conserved charges $\{Q_{m}\}_{m=2}^\infty$ is in one-to-one correspondence with the density $\rho_{1,h}$ of $1$-string holes. 

The conventions that we use for the Fourier transform are
\begin{subequations}\label{eq:FourierTransform}
\begin{align}
	\hat{f}(k) &= \ftk{f}{k} = \int_{-\pi/2}^{\pi/2}\mathrm{d}\lambda e^{2ik\lam}f(\lam)\epc\quad k\in\mathbb{Z}\epc \\
	f(\lam) &= \iftx{\hat{f}}{\lam} = \frac{1}{\pi}\sum_{k\in\mathbb{Z}} e^{-2ik\lam}\hat{f}(k)\epc\quad \lambda\in[-\tfrac{\pi}{2}, \tfrac{\pi}{2})\epp \label{eq:inverse_FourierTransform}
\end{align}
\end{subequations}
For $m\geq1$, observe that partial integration ($m-1$ times) gives a simple expression for the Fourier transform of $c_{m+1}^{(n)}$,
\begin{align} \label{eq:cnFourier}
	\hat{c}_{m+1}^{(n)}(k) &= - 2\pi \frac{\sinh^m(\eta)}{2^m} (2ik)^{m-1} \int_{-\pi/2}^{\pi/2}\mathrm{d}\lam \, e^{2ik\lam} \, a_{n}(\lam) \nonumber \\
		&= -\pi \, \sinh^m (\eta)\, (ik)^{m-1}\, e^{-|k| n \eta} \epc
\end{align}
where we used that the Fourier transform of the XXZ kernel $a_n$ in Eq.~\eqref{eq:kernelXXZ} is $e^{-|k|n\eta}$. The eigenvalue of charge $Q_{m+1}$ can then be rewritten as
\begin{align} \label{eq:expecttemp}
	\sum_{n=1}^{\infty} \int_{-\pi/2}^{\pi/2}\mathrm{d}\lam \, \rho_n^{\Psi_0} (\lam) \, c_{m+1}^{(n)}(\lam) &= \frac{1}{\pi} \sum_{n=1}^{\infty} \sum_{k\in\mathbb{Z}} \, \hat{\rho}_n^{\Psi_0} (k)\, \hat{c}_{m+1}^{(n)}(k) \nonumber\\
		& = - \sinh^{m}(\eta)\, \sum_{k\in\mathbb{Z}} (ik)^{m-1} \sum_{n=1}^{\infty} \hat{\rho}_n^{\Psi_0} (k) \, e^{-|k|n\eta} \epp
\end{align}
We rewrite the sum over all string densities in terms of $\hat{\rho}_{1,h}^{\Psi_0}$ only,
\begin{equation}
	\sum_{n=1}^{\infty} \hat{\rho}_n^{\Psi_0} (k)\, e^{-|k|n\eta} = \frac{e^{-|k|\eta} - \hat{\rho}_{1,h}^{\Psi_0} (k) }{2 \cosh (k\eta)}\epp
\end{equation}
This identity~\cite{GaudinBOOK} can be derived from the Fourier transform of the partially decoupled form~\eqref{eq:BTGthlim_fact} of the BGT equations, which is (using the convolution theorem)
\begin{equation}
\hat{\rho}_{n,t}^{\Psi_0} (k) = \frac{1}{2 \cosh(k\eta)} \left[ \hat{\rho}_{n-1,h}^{\Psi_0} (k) + \hat{\rho}_{n+1,h}^{\Psi_0} (k) \right]
\end{equation}
for $n \geq 1$, where $\hat{\rho}_{0,h}^{\Psi_0}(k)=1$. The one-to-one correspondence between the expectation values of the charges $\{Q_m\}_{m=2}^\infty$ and $\hat{\rho}_{1,h}^{\Psi_0}$ is thus given by
\begin{equation}\label{eq:pinning_rho1h}
	\limth \left( \frac{\left\langle \Psi_0 \right| Q_{m+1} \left| \Psi_0 \right\rangle}{N \sinh^{m}(\eta)} \right)  = \sum_{k\in\mathbb{Z}}  \frac{ \hat{\rho}_{1,h}^{\Psi_0}(k) -  e^{-|k|\eta }}{2\cosh(k \eta)} (ik)^{m-1} \epc
\end{equation}
where it should be noted that this equation holds for all $m\geq 1$ and that the total-momentum charge is excluded.

We stress that the result~\eqref{eq:pinning_rho1h} is general, the 1-string hole density $\rho_{1,h}^{\Psi_0}$ of the steady state after any quench to the spin-1/2 XXZ chain is completely determined by the initial values of the local conserved charges $\{Q_{m}\}_{m=2}^\infty$. Note that the sum in Eq.~\eqref{eq:pinning_rho1h} is quickly converging due to the exponentially decaying factor for $\eta>0$, which ensures invertibility. 

To make this more explicit, following the method of Ref.~\cite{2013_Fagotti_JSTAT_P07012} one can define a generating function
\begin{equation} \label{eq:defgeneratingfunction}
	\Omega_{\Psi_0}(\lam)  = \limth \frac{i}{N} \bra{\Psi_0}  t^{-1} \! \left( \lam + \tfrac{i\eta}{2}\right) \partial_\lam t \! \left( \lam +\tfrac{i\eta}{2} \right)\ket{\Psi_0} \epc
\end{equation}
which has a Taylor series around $\lambda = 0$ whose coefficients are related to the expectation values of the local conserved charges on the initial state. Using Eq.~\eqref{eq:pinning_rho1h}, a direct relation between the generating function and the postquench steady-state density $\rho_{1,h}^{\Psi_0}$ can be established,
\begin{equation} \label{eq:generatingfunctiontorho1h}
	\rho_{1,h}^{\Psi_0} (\lam) = a_{1}(\lam) + \frac{1}{2\pi}\left[ \Omega_{\Psi_0} \left( \lam + \tfrac{i\eta}{2} \right) + \Omega_{\Psi_0} \left( \lam - \tfrac{i\eta}{2} \right) \right] \epp
\end{equation}
For initial states that are product states, {\it i.e.}, $| \Psi_0 \rangle = \otimes_{j=1}^{N/a} | \Psi_0^{(j)} \rangle$ where $ | \Psi_0^{(j)} \rangle$ comprises a finite number $a$ of spins, the generating function can easily be computed in the thermodynamic limit~\cite{2013_Fagotti_JSTAT_P07012}.

\subsection{Solution to the GGE}
As a consequence, a prediction for the GGE including all known local conserved charges can be obtained. Knowledge of $\rho_{1,h}^{\Psi_0}$ allows one to eliminate the first GTBA equation in Eqs~\eqref{eq:GTBAGGE} with the unknown driving term $d$. The GGE prediction for the steady-state densities $\brho^{GGE}$ can be found by solving the GTBA Eqs~\eqref{eq:GTBAGGE_a} for $n\geq 2$, combined with the Bethe Eqs~\eqref{eq:BTGthlim_fact} and the constraint $\rho_{1,h}^{GGE}=\rho_{1,h}^{\Psi_0}$. To implement this, one starts from an initial guess for the function $\rho_1$, denoted by $\rho_1^{(0)}$,  which determines the initial guess for $\eta^{(0)}_1 = \rho_{1,h}^{\Psi_0}/\rho_1^{(0)} $. Using this one solves the GTBA Eqs~\eqref{eq:GTBAGGE_a} for $n\geq 2$ and the BGT Eqs~\eqref{eq:BTGthlim_fact}. This computation can be performed by an application of the convolution theorem and a Fast Fourier Transform algorithm. One can truncate the infinite set of coupled equations by considering only the first $n_{\text{max}}$ equations of both the BGT and GTBA equations. This results in a new $\rho_1^{(1)}$ and a new $\eta^{(1)}_1$. The procedure can then be repeated until convergence is reached, $\lim_{l \to \infty}\eta_1^{(l)} = \rho_{1,h}^{\Psi_0}/\rho^{GGE}_1$, which automatically leads to the full solution of the GGE. With this procedure it is possible to obtain the GGE prediction for the steady state after any quench to the XXZ model starting from a product initial state. The functions $\eta_{n_{\text{max}}+1}$ and $\rho_{n_{\text{max}}+1}$ are needed as input for the last equations of the two truncated sets. It turns out that the functions become (approximately) constant with $\eta_n \thicksim n^2$ and $\rho_n \thicksim n^{-3}$. One can use this information to set the values of $\eta_{n_{\text{max}}+1}$ and $\rho_{n_{\text{max}}+1}$. For more details, see Refs~\cite{TakahashiBOOK, 1974_Takahashi_PTP_51, 2011_Klauser_PRA_84, 2014_Pozsgay_GGE}.

\subsection{The quench action approach}
There is an alternative approach that does not rely on the GGE assumption and that, besides predicting the steady state after a quantum quench, also gives access to the time evolution. This so-called quench action approach~\cite{2013_Caux_PRL_110} is based on first principles and in order to overcome the problem of the exponentially large sum in Eq.~\eqref{eq:expect1} it uses a saddle-point approximation. Here, the most important ingredients of the approach are briefly outlined. For details we refer to Refs~\cite{2013_Caux_PRL_110,2014_DeNardis_PRA_89, 2014_Bertini_Sinegordon,2014_De_Luca, 2014_De_Nardis, JS_QApaper}.

In the thermodynamic limit a single sum over the Hilbert space is replaced by a functional integral over the root distributions $\brho$.  For a generic quantity $\mathcal{A}_\blam$ that scales to a smooth function $\mathcal{A}[\brho]$ in the thermodynamic limit, the sum becomes
\begin{equation}
\limth \sum_{\blam\in\mathcal{H}} \mathcal{A}_\blam \sim \int \mathcal{D} \brho\,  e^{S_{YY}[\brho]} \, \mathcal{A}[\brho] \epp
\end{equation}
As explained in Ref.~\cite{2014_DeNardis_PRA_89}, for a large class of physical observables that have vanishing matrix elements between states that scale to different smooth root distributions, the double sum in Eq.~\eqref{eq:expect1} can be written in the thermodynamic limit as a functional integral,
\begin{subequations}
\begin{multline}\label{eq:QA_expectation}
	\limth\left\langle \Psi (t) \right| \cO \left| \Psi(t) \right\rangle = \frac{1}{\mathcal{Z}_{QA}} \int \!\mathcal{D}\boldsymbol{\rho} \:   e^{ -S_{QA}[\boldsymbol{\rho}] } \\
	\times\frac{1}{2}\sum_{ \mathbf{e} }
\Big(e^{ - \delta s_\mathbf{e} -  i \delta\omega_\mathbf{e} t } \langle \boldsymbol{\rho} | \cO | \boldsymbol{\rho} , \mathbf{e} \rangle +   e^{ - \delta s_\mathbf{e}^\ast +i\delta\omega_\mathbf{e} t } \langle \boldsymbol{\rho}, \mathbf{e}  | \cO | \boldsymbol{\rho} \rangle \Big) \epc 
\end{multline}
where $\sum_{\mathbf{e}}$ represents the sum over all discrete excitations on the state $|\boldsymbol{\rho}\rangle$. These excitations are obtained by displacing, creating, and annihilating a denumerable number of strings of the representative state for $|\brho\rangle$. The quantity $\mathcal{Z}_{QA}=\int\mathcal{D}\brho \, e^{-S_{QA}[\brho]}$ is the quench action partition function and $\delta s_\mathbf{e}$ is the non-extensive part of the overlap coefficient, while $\delta\omega_\mathbf{e}$ is the energy relative to $|\brho\rangle$,
\begin{align}
	\delta s_\mathbf{e} &=  -\ln \left[\frac{ \langle \boldsymbol{\rho} , \mathbf{e}  | \Psi_0 \rangle }{ \langle  \boldsymbol{\rho}   | \Psi_0 \rangle } \right] \epc \\
	\delta\omega_\mathbf{e} & = \omega[  \boldsymbol{\rho} , \mathbf{e} ] -  \omega[  \boldsymbol{\rho}] \epp
\end{align}
\end{subequations}
Defining $S[\boldsymbol{\rho}] = \limth \text{Re}\, S_\blam$ as the extensive real part of the overlap coefficient in the thermodynamic limit, the weight of the functional integral is given by the quench action $S_{QA}[ \boldsymbol{\rho} ] = 2 S[\boldsymbol{\rho}] - S_{YY}[\boldsymbol{\rho}]$. It should be noted that the overlap coefficients can vary wildly over the ensemble of states represented by the densities $\brho$ and therefore do not have a well-defined limit. However, the extensive part is universal and only depends on the smooth root distributions of these states. For a more detailed discussion see Sec.~\ref{sec:R:exotic}.

The quench action being extensive, real, and bounded from below, convergence of the functional integral is ensured and in the thermodynamic limit a saddle-point approximation of the functional integral becomes exact, leading to
\begin{align} \label{eq:expect2}
	\limth \left\langle \Psi (t) \right| \cO \left| \Psi (t) \right\rangle  = \frac12 \sum_\mathbf{e} \left( e^{-\delta s_\mathbf{e} - i \delta \omega_\mathbf{e} t } \left\langle \brho^{\text{sp}}  \right| \cO \left| \brho^{\text{sp}}, \mathbf{e} \right\rangle +  e^{-\delta s_\mathbf{e}^\ast + i \delta \omega_\mathbf{e} t } \left\langle \brho^{\text{sp}} , \mathbf{e}  \right| \cO \left| \brho^{\text{sp}}\right\rangle \right)\epp
\end{align}
Here, the saddle-point root distributions $\brho^{\text{sp}}$ are determined by the variational equations
\begin{equation}\label{eq:variation}
	0 = \left. \frac{\delta S_{QA}\left[ \brho \right] }{\delta \rho_n (\lam) } \right|_{\brho=\brho^\text{sp}}
\end{equation}
for $n\geq 1$, which form the set of GTBA equations. Equation~\eqref{eq:expect2} is valid for any time~$t$ after the quench. In particular, due to dephasing it predicts whereto time-dependent expectation values of the operator $\cO$ will relax at long times after the quench,
\begin{equation}
	\lim_{t \to \infty}\limth \left\langle \Psi (t) \right| \cO \left| \Psi (t) \right\rangle  = \left\langle \brho^{\text{sp}}  \right| \cO  \left| \brho^{\text{sp}}  \right\rangle \epp
\end{equation}
To summarize, the GTBA Eqs~\eqref{eq:variation}, whose driving terms are determined by the leading part of the overlap coefficient $S_{\blam} = - \ln \left\langle \blam | \Psi_{0} \right\rangle$ in the thermodynamic limit, give the quench action prediction for the steady state after a quantum quench with initial state $\ket{\Psi_0}$. 

In Ref.~\cite{2014_DeNardis_PRA_89} the saddle point state for an interaction quench in the Lieb-Liniger model was found analytically by means of the quench action approach. Both the density moments $g_2$ and $g_3$ and the static structure factor were computed on the steady state. For the quench to the Tonks-Girardeau gas, known exact results for the time-evolution of the density-density operator were reproduced using Eq.~\eqref{eq:expect2}.

\section{Quench action approach for the N\'eel-to-XXZ quench}\label{sec:quench_overlaps}

\subsection{Initial state}
Hitherto we have left the initial state unspecified, all the considerations above about the GGE and the quench action approach being completely generic. Now, we focus on quenches from the zero-momentum ground state in the antiferromagnetic Ising limit ($\Delta \to \infty$) to the gapped regime ($1<\Delta <\infty$) of the XXZ model. The quench to the isotropic point ($\Delta=1$) is discussed in Sec.~\ref{sec:XXX}. 

In the spin basis, the initial state is represented by
\begin{equation}\label{eq:Neel_state}
	|\Psi_0\rangle = \frac{1}{\sqrt{2}}\left(\left|\uparrow\downarrow\right\rangle^{\otimes N/2}+\left|\downarrow\uparrow\right\rangle^{\otimes N/2}\right) 
	\epp
\end{equation} 
Strictly speaking, this is the symmetric combination of the N\'eel and anti-N\'eel state, which is translationally invariant and has momentum zero. The quench action approach gives the same saddle-point prediction for any quench starting from an initial state that is a superposition of the N\'eel and anti-N\'eel state, since the extensive part of the overlap coefficient is always the same~\cite{2014_Brockmann_npi}. For convenience, we work with the zero-momentum N\'eel state~\eqref{eq:Neel_state} and simply call it the N\'eel state. 

Furthermore, in Refs~\cite{2014_Brockmann_Jstat_5,2013_Pozsgay_Neeloverlaps} it was shown that the overlaps of the N\'eel state are related to overlaps of other states of interest, namely the dimer state and the $q$-deformed dimer state. Recently, in Ref.~\cite{2014_Piroli_resursive_overlaps}, a recursive formula for overlaps of a larger class of initial states was derived. The quench action approach outlined here is therefore extendable to other initial states. For the dimer-to-XXZ quench, for example, see Ref.~\cite{2014_Pozsgay_Dimer}.

In the thermodynamic limit, the expectation values of the conserved charges on the N\'eel state are~\cite{2013_Fagotti_JSTAT_P07012}
\begin{equation} \label{eq:chargesonNeel}
	\limth\frac{\left\langle \Psi_0 \right| Q_{m+1} \left| \Psi_0 \right\rangle}{N} = - \frac{\Delta}{2} \left.{\frac{\partial^{m-1}}{\partial x^{m-1}} \left( \frac{1-\Delta^2}{\cosh \left( \sqrt{1-\Delta^2}x\right) - \Delta^2} \right) }\right|_{x=0} \epc
\end{equation}
which gives zero for odd $m+1$.

\subsection{Overlap formulas}
For convenience we take $N$ divisible by four, {\it i.e.}, the initial state is in the zero-magnetization sector $M=N/2$ with $M$ even. Since we are interested in the thermo\-dynamic limit, this choice is of no consequence and, using~\cite{2014_Brockmann_npi}, identical results can be obtained for chains with $N/2$ odd. 

The sums in Eq.~\eqref{eq:expect1} are taken over the complete set of Bethe states in the sector $M=N/2$. In Ref.~\cite{2014_Brockmann_npi} it was shown that the overlap between the zero-momentum N\'eel state and a Bethe state is zero if the Bethe state is not parity invariant. By parity invariant we mean that all rapidities come in pairs such that $\{\lambda_j\}_{j=1}^M= \{-\lambda_j\}_{j=1}^M$. Parity-invariant states with one pair of rapidities at $\{0,\,\tfrac\pi2\}$ are discarded since these Bethe states have total momentum $\pi$ [see Eq.~\eqref{eq:momentum}] and do not overlap with the zero-momentum N\'eel state.  We denote a parity-invariant state by
\begin{equation}
	|\tilde{\blam}\rangle= \ket{\{\pm\lambda_j\}_{j=1}^{M/2} } = \ket{ \{\lambda_j\}_{j=1}^{M/2}\cup \{-\lambda_j\}_{j=1}^{M/2} } \epp
\end{equation}
Besides having zero momentum, it turns out that also all other odd local conserved charges $Q_{2m+1}$ have zero eigenvalue on parity-invariant states,
\begin{subequations}
\begin{align}
	Q_{2m+1}|\tilde{\blam}\rangle &= \sum_{j=1}^M P_{2m+1}(\lambda_j) |\tilde{\blam}\rangle =0\epc\\
	P_{2m+1}(\lambda) &= i \frac{\sinh^{2m}(\eta)}{4^m}\frac{\partial^{2m}}{\partial \mu^{2m}} \ln\left[ \frac{\sin(\lambda - \mu + i \eta/2)}{\sin(\lambda - \mu - i \eta/2) }\right]_{\mu\to 0}\epc
\end{align}
\end{subequations}
since $P_{2m+1}$ is an odd function. This observation, combined with the fact that only parity-invariant Bethe states have nonzero overlap with $\left| \Psi_0 \right\rangle$, is in agreement with the vanishing of the expectation values of all odd conserved charges on the N\'eel state~\cite{2013_Fagotti_JSTAT_P07012}, see Eq.~\eqref{eq:chargesonNeel}.

Let us recall the nonzero overlaps for the quench we study, namely the overlaps of the zero-momentum N\'eel state $|\Psi_0\rangle$ with normalized parity-invariant Bethe states associated with the XXZ Hamiltonian \eqref{eq:Hamiltonian_XXZ}. In Refs~\cite{1998_Tsuchiya_JMathP,2012_Koslowski_JSTAT_P05021,2014_Brockmann_JPA} a formula for them was given. Interestingly, in Ref.~\cite{2014_Brockmann_JPA} a Gaudin-like form that is suitable in the thermodynamic limit was derived,
\begin{subequations}\label{eq:overlap}
\begin{equation}
	\frac{\langle \Psi_0 |\tilde{\blam}\rangle }{\sqrt{\langle \tilde{\blam}|\tilde{\blam}\rangle}}=\sqrt{2} \left[\prod_{j=1}^{N/4}\frac{\sqrt{\tan(\lambda_j+\frac{i\eta}{2}) \tan(\lambda_j-\frac{i\eta}{2})}}{2\sin(2\lambda_j)}\right]\sqrt{ \frac{\det_{N/4}(G^{+})}{\det_{N/4}(G^{-})}}
\end{equation}
where 
\begin{equation}\label{eq:overlap_b}
	G_{jk}^\pm = \delta_{jk}\left(NK_{\eta/2}(\lambda_j)-\sum_{l=1}^{N/4}K_\eta^+(\lambda_j,\lambda_l)\right) + K_\eta^\pm(\lambda_j,\lambda_k)\:, \quad j,k=1,\ldots,N/4\epc
\end{equation}
\end{subequations}
$K_\eta^\pm(\lambda,\mu)=K_\eta(\lambda-\mu) \pm K_\eta(\lambda+\mu)$, and $K_\eta(\lambda)$ as in norm formula \eqref{eq:norm_Bethe_state}. 
It should be noted that this overlap formula is completely general. In particular, it is valid for Bethe states with strings of rapidities. Furthermore, note that this overlap is connected to the Lieb-Liniger overlap formula for an initial state that describes a Bose-Einstein condensate of one-dimensional free Bosons~\cite{2014_DeNardis_PRA_89, 2014_Brockmann_Jstat_5}.

\subsection{GTBA equations}
The quench action approach uses a saddle-point approximation to overcome the double sum in Eq.~\eqref{eq:expect1}, where the overlaps in Eqs~\eqref{eq:overlap} serve as input. The resulting GTBA equations for the N\'eel-to-XXZ quench were derived in Ref.~\cite{2014_Wouters}. For the sake of completeness, this derivation is repeated in~\ref{app:GTBAQA}. The resulting quench action GTBA equations are given by
\begin{subequations}\label{eq:GTBA_QA}
\begin{equation}\label{eq:GTBAQAmain}
	\ln[\eta_{n}(\lam)]  = 2n \left[ \ln(4) - h \right] +  g_{n}(\lam)  + \sum_{m=1}^{\infty} \left[ a_{nm} \ast \ln \left( 1 + \eta_{m}^{-1} \right)\right] (\lam) \epc
\end{equation}
where $n \geq 1$, the parameter $h$ is a Lagrange multiplier fixing the total magnetization, and 
\begin{equation}
	g_n (\lambda) = \sum_{l=0}^{n-1} \ln\left[\frac{\sin^2(2\lambda)+\sinh^2[\eta(n-1-2l)]}{4\tan[\lambda+i\eta(\frac{n}{2}-l)]\tan[\lambda-i\eta(\frac{n}{2}-l)]}\right]\epp
\end{equation}
\end{subequations}
 They can be recast in simplified (partially decoupled) form~\cite{TakahashiBOOK} 
\begin{subequations}\label{eq:TBA_XXZ_fact}
\begin{equation}\label{eq:TBA_XXZ_fact_equation}
	\ln(\eta_n) = d_n + s \ast \big[\ln(1+\eta_{n-1})+\ln(1+\eta_{n+1})\big]\epc
\end{equation}
where $n\geq 1$ and $\eta_0(\lam)= 0$. The driving terms are given by 
\begin{align}\label{eq:TBA_XXZ_fact_driving}
	d_n(\lam) = \sum_{k\in\mathbb{Z}} e^{-2ik\lam}\frac{\tanh(k\eta)}{k}\left[(-1)^n-(-1)^k\right] = (-1)^n\ln\left[\frac{\vartheta_4^2(\lambda)}{\vartheta_1^2(\lambda)}\right] +\ln\left[\frac{\vartheta_2^2(\lambda)}{\vartheta_3^2(\lambda)}\right]\epc
\end{align}
\end{subequations}
where $\vartheta_j$, $j=1,\ldots,4$, are Jacobi's $\vartheta$-functions \cite{1989_Lawden} with nome $e^{-2\eta}$.

The GTBA Eqs~\eqref{eq:TBA_XXZ_fact} are an infinite set of coupled nonlinear integral equations and can, in principle, be solved recursively using a Fast Fourier Transform algorithm, as was the case for the GGE. Again, one truncates to only the first $n_{\text{max}}$ equations. By solving the system for different values of $n_{\text{max}}$, it can be observed that the solutions $\eta_n$ are converging for large $n$, where the solutions for odd and even $n$ must be treated separately,
\begin{subequations}
\begin{align}
	\lim_{n\to\infty} \eta_{2n}^{\text{sp}}(\lam) &= \eta^{\text{sp}}_{\text{even}}(\lam) \epc \\
	\lim_{n\to\infty} \eta_{2n+1}^{\text{sp}}(\lam) &= \eta^{\text{sp}}_{\text{odd}}(\lam) \epp
\end{align}
\end{subequations}
Here, $\eta^{\text{sp}}_{\text{even}}$ and $\eta^{\text{sp}}_{\text{odd}}$ are nonzero functions for any value of $\Delta > 1$. By setting $\eta_{n_\text{max}+1}(\lam)=\eta_{n_\text{max}-1}(\lam)$, this asymptotic behavior gets implemented into the numerical algorithm. 

As a consequence, the sum in Eq.~\eqref{eq:GTBAQAmain} evaluated on the saddle-point solution is infinite, corresponding to an infinite value of the Lagrange multiplier $h$. As opposed to what we find here, in Ref.~\cite{2014_Pozsgay_Dimer} it was stated that the integrals of $\eta^{\text{sp}}_n$ scale like $e^{\eta n^2}$ for large $n$. We note that this is an artifact of performing the numerical analysis at finite $h$ and with a truncated sum in the original form~\eqref{eq:GTBA_QA} of the GTBA equations. When the truncation level $n_{\text{max}}$ is increased, the observed asymptotic behavior sets in at longer string lengths and is therefore unphysical. Of course, by increasing the level of truncation the error can be pushed to longer and less significant strings and high-precision predictions for physical observables are still possible.

Substituting this solution of the GTBA equations into the BGT Eqs~\eqref{eq:BTGthlim_fact}, they can be solved numerically in a similar manner. One finds that the integrals of the functions $\rho_n^{\text{sp}}$ scale with $e^{-n\eta}$ for large $n$. Due to this exponential decay, the infinite set of Bethe equations can be safely truncated by setting $\rho_{n_{\text{max}}+1}(\lam)=0$.

\section{Analytical solution} \label{sec:analytical_solution}
As for the interacting quench in the Lieb-Liniger Bose gas, the GTBA equations derived from a quench action analysis can be solved analytically. Here, the solution can be found by mapping the GTBA Eqs~\eqref{eq:TBA_XXZ_fact} to well-known systems of functional equations, the Y- and T-system~\cite{1999_Suzuki, 1992_Kluemper}. Combining this with an analytic expression for $\rho_{1,h}$, which will be derived first using the results of Sec.~\ref{sec:QuenchesinXXZ} and is independent of any quench action analysis, also the BGT Eqs~\eqref{eq:BTGthlim_fact} can be solved analytically.

\subsection{Explicit expression for $\rho_{1,h}$} \label{sec:rho1h}
In Ref.~\cite{2013_Fagotti_JSTAT_P07012} the generating function~\eqref{eq:defgeneratingfunction} for the pure N\'eel state was computed in the thermodynamic limit. In this limit, matrix elements of local conserved charges between the N\'eel and anti-N\'eel states vanish and, therefore, the generating function for the zero-momentum N\'eel state is identical and reads
\begin{equation}\label{eq:generatingfunction_explicit}
	\Omega_{\text{N\'eel}}(\lam) = -\frac{ \sinh(2\eta)}{\cosh(2\eta)+1 - 2\cos(2\lam)} \epp
\end{equation}
Using Eq.~\eqref{eq:generatingfunctiontorho1h}, one arrives at an explicit expression for the density of $1$-holes,
\begin{equation}\label{eq:rho1h_exact_XXZ}
	\rho_{1,h}^{\text{N\'eel}}(\lambda) = a_1(\lambda)\left(1-\frac{\cosh^2(\eta)}{\pi^2 a_1^2(\lambda) \sin^2(2\lambda)+\cosh^2(\eta)}\right) \epc
\end{equation}
where $a_1$ is the usual XXZ kernel defined in Eq.~\eqref{eq:kernelXXZ}.

\subsection{Y-system}\label{sec:Y-system}
We consider a set of functional equations, the so-called Y-system~\cite{1999_Suzuki}, 
\begin{equation}\label{eq:y-system}
	y_n(x+\tfrac{i\eta}{2})y_n(x-\tfrac{i\eta}{2}) = Y_{n-1}(x)Y_{n+1}(x)\epc\quad n\geq 1\epc
\end{equation}
with $Y_n(x) = 1+y_n(x)$ for $n\geq 0$, where $y_0(x)=0$. In the following we denote arguments of functions by $x$ if these functions belong to a general structure (see Sec.~\ref{sec:T-system}), whereas we shall use $\lambda$ (as in Secs~\ref{sec:connection_to_GTBA} and \ref{sec:explicit_solution}) if the functions belong to the explicit solution of the special case~\eqref{eq:TBA_XXZ_fact} of GTBA equations. 

Fixing the analyticity properties of the $y$-functions in the physical strip (PS)
\begin{equation}
	PS = \{x\in\mathbb{C}|\ -\tfrac{\eta}{2} < \Im(x) < \tfrac{\eta}{2}\epc\  -\tfrac{\pi}{2} \leq \Re(x) < \tfrac{\pi}{2}\}
\end{equation}
and supposing $\pi$-periodicity in the real direction, the functional relations~\eqref{eq:y-system} can be written as nonlinear integral equations (NLIEs)
\begin{equation}\label{eq:NLIE_y-system}
	\ln[y_n(x)] = d_n(x) + s\ast [\ln(Y_{n-1})+\ln(Y_{n+1})](x)\epc\quad n\geq 1\epp
\end{equation}
The kernel function $s$ is given in Eq.~\eqref{eq:defs} and the driving terms $d_n$ are determined by the analytical behavior of the $y$-functions inside the PS. The NLIEs can be deduced by taking the Fourier transform~\eqref{eq:FourierTransform} of the logarithmic derivative of Eq.~\eqref{eq:y-system}, shifting the integration contours on the left hand side by $\pm i\eta/2$, collecting the explicit terms coming from the roots and poles of $y_n$ in the PS, dividing by $\cosh(k\eta)$, taking the inverse Fourier transform, and finally integrating over $x$. The integration constant can be usually fixed by analyzing the asymptotes of the functions.

\subsection{Connection to the GTBA equations of the N\'eel-to-XXZ quench}\label{sec:connection_to_GTBA}
The GTBA Eqs~\eqref{eq:TBA_XXZ_fact} are of the form~\eqref{eq:NLIE_y-system} and the driving terms in Eq.~$\eqref{eq:TBA_XXZ_fact_driving}$ can be considered as originating from the following analytical behavior:
\begin{subequations}\label{eq:analyticity_properties}
\begin{align}
	&\eta_n(\lambda) \sim \sin^2(2\lambda)\epc\quad\textnormal{for small $\lambda$ and $n$ odd}\epc\\
	&\eta_n(\lambda) \sim \cot^2(\lambda)\epc\ \quad\textnormal{for small $\lambda$ and $n$ even}\epc\\
	&\text{and no further roots or poles for all $\lambda\in PS\backslash \{0\}$}\epp
\end{align}
\end{subequations}
This can be shown by applying the steps described above. The Fourier transforms of the logarithmic derivatives are
\begin{subequations}
\begin{align}
	FT[\ln'(\sin^2(2\lambda))](k) &= -4\pi i \sinh(k\eta)[1+(-1)^k]\epc\\
	FT[\ln'(\cot^2(\lambda))](k) &= \phantom{-}4\pi i \sinh(k\eta)[1-(-1)^k]\epp
\end{align}
\end{subequations}
Dividing by $\cosh(k\eta)$, taking the inverse Fourier transform~\eqref{eq:inverse_FourierTransform} and integrating over $x$ yields exactly the driving terms~\eqref{eq:TBA_XXZ_fact_driving} of the GTBA Eqs~\eqref{eq:TBA_XXZ_fact_equation}. Therefore, a solution of the GTBA Eqs~\eqref{eq:TBA_XXZ_fact} is given by the solution of the Y-system~\eqref{eq:y-system} with analyticity properties~\eqref{eq:analyticity_properties}. 

The GTBA Eqs~\eqref{eq:GTBAGGE} for the GGE correspond to the same Y-system~\eqref{eq:y-system} but with different analyticity conditions, specified by the structure of the driving terms $d_{n\geq 1}$. It is reasonable to assume that the solution to the Y-system is unique as soon as the analytic behavior of all $y$-functions inside the physical strip is given.

\subsection{T-system}\label{sec:T-system}
Following the logic of~\cite{1999_Suzuki} and~\cite{1992_Kluemper} we write 
\begin{equation}\label{eq:def_yn}
	y_n(x) = \frac{T_{n-1}(x)T_{n+1}(x)}{f_n(x)}\epc\quad n\geq 1\epc
\end{equation}
where the functions $T_{n\geq 0}$, fulfill another system of functional equations, the so-called T-system,
\begin{equation}\label{eq:T-system}
	T_n(x-\tfrac{i\eta}{2})T_n(x+\tfrac{i\eta}{2}) = T_{n-1}(x)T_{n+1}(x)+f_n(x)\epc\quad n\geq 1\epc
\end{equation}
with $T_0(x)=1$. A general solution of the T-system is given by 
\begin{subequations}
\begin{align}
	T_0(x) &= 1\epc\\
	T_1(x) &= a_+(x)\frac{Q(x+i\eta)}{Q(x)} + a_-(x)\frac{Q(x-i\eta)}{Q(x)} =\lambda_1^{(1)}(x) + \lambda_2^{(1)}(x)\epc \label{eq:def_T1}\\
	T_{n+1}(x) &= T_{n}(x+\tfrac{i\eta}{2})T_1(x-\tfrac{i\eta n}{2}) - g_n(x+\tfrac{i\eta}{2})T_{n-1}(x+i\eta)\epc\quad n\geq 1\epc
\end{align}
\end{subequations}
with $g_n(x)=a_+(x-\frac{i\eta}{2}(n+1))a_-(x-\tfrac{i\eta}{2}(n-1))$. The functions $f_n$ then read 
\begin{equation}
	f_n(x)=\prod_{j=1}^n a_+(x+\tfrac{i\eta}{2}(n-2j))a_-(x-\tfrac{i\eta}{2}(n-2j))
\end{equation}
and fulfill the relations
\begin{equation}
	f_{n+1}(x)f_{n-1}(x) = f_n(x-\tfrac{i\eta}{2}) f_n(x+\tfrac{i\eta}{2})\epc\quad n\geq 1\epc
\end{equation}
which is necessary in order that the $y$-functions \eqref{eq:def_yn} are a solution of the Y-system~\eqref{eq:y-system} for a given solution of the T-system~\eqref{eq:T-system}.

Defining a new auxiliary function as the ratio of the two terms $\lambda_1^{(1)}$ and $\lambda_2^{(1)}$ in Eq.~\eqref{eq:def_T1}, 
\begin{equation}
	\mathfrak{a}(x) = \frac{\lambda_1^{(1)}(x)}{\lambda_2^{(1)}(x)} = \frac{a_+(x)Q(x+i\eta)}{a_-(x)Q(x-i\eta)}\epc
\end{equation}
it can be shown that $y_1$ is completely determined by this auxiliary function,
\begin{equation}\label{eq:a2y1}
	y_1(x) = \mathfrak{a}(x+\tfrac{i\eta}{2}) + \mathfrak{a}^{-1}(x-\tfrac{i\eta}{2}) + \mathfrak{a}(x+\tfrac{i\eta}{2})\mathfrak{a}^{-1}(x-\tfrac{i\eta}{2})\epp
\end{equation}
Together with $y_0(x)=0$ and the Y-system~\eqref{eq:y-system}, which can be interpreted as a recursion relation,
\begin{equation}\label{eq:y-system_recursion}
	y_{n+1}(x) = \frac{y_n(x+\tfrac{i\eta}{2})y_n(x-\tfrac{i\eta}{2})}{1+y_{n-1}(x)}-1\epc\quad n\geq 1\epc
\end{equation}
all higher $y$-functions $y_{n\geq 2}$ can be expressed in terms of the single function $\mathfrak{a}$.

\subsection{Explicit solution}\label{sec:explicit_solution}
One possible choice that gives the correct analytical behavior~\eqref{eq:analyticity_properties} of all $\eta$-functions is given by
\begin{equation}\label{eq:def_a}
	\mathfrak{a}(\lambda) = \frac{\sin(\lambda+i\eta)}{\sin(\lambda-i\eta)}\frac{\sin(2\lambda-i\eta)}{\sin(2\lambda+i\eta)}\epp
\end{equation} 
Using Eq.~\eqref{eq:a2y1} the function $\eta_1 \equiv y_1$ reads
\begin{equation}\label{eq:eta1_explicit}
	\eta_1(\lambda) = \frac{\sin^2(2\lambda)\left[\cosh(\eta) + 2\cosh(3\eta)-3\cos(2\lambda)\right]}{2\sin(\lambda-\tfrac{i\eta}{2})\sin(\lambda+\frac{i\eta}{2})\sin(2\lambda+2i\eta)\sin(2\lambda-2i\eta)}\epp
\end{equation}
Explicit expressions of all higher $\eta$-functions can be obtained using $\eta_0(\lambda)=0$ and the recursion relation~\eqref{eq:y-system_recursion} for $y_{n}\equiv \eta_n$, $n\geq 2$. They have the correct anayticity properties~\eqref{eq:analyticity_properties}. There are additional roots and poles at $\lambda = \pm\tfrac{\pi}{2}, \pm \tfrac{i\eta}{2}$, whose contributions cancel each other when taking the Fourier transform and shifting the contour as described in the paragraph right after Eqs~\eqref{eq:NLIE_y-system}. Therefore, the explicit function in Eq.~\eqref{eq:eta1_explicit} together with all higher functions $\eta_{n\geq 2}$ are a solution of the GTBA Eqs~\eqref{eq:TBA_XXZ_fact}. 

To get explicit expressions for the root distributions $\rho_{n}$ we use the explicit expressions of $\rho_{1,h}$ [Eq.~\eqref{eq:rho1h_exact_XXZ}] and of $\eta_n$ for $n\geq 1$ [Eqs~\eqref{eq:y-system_recursion} and~\eqref{eq:eta1_explicit}]. Together with the BGT Eqs~\eqref{eq:BTGthlim_fact}, which can be written as functional equations,
\begin{equation}
	\rho_{n+1,h}(\lambda) = \rho_{n,t}(\lambda+\tfrac{i\eta}{2}) + \rho_{n,t}(\lambda-\tfrac{i\eta}{2}) - \rho_{n-1,h}(\lambda)\epc\quad n\geq 1\epc
\end{equation} 
with $\rho_{0,h}(\lambda) \equiv 0$, $\rho_{n,t}(\lambda) = \rho_{n,h}(\lambda)\left[1+\eta_n^{-1}(\lambda)\right]$, they uniquely determine all $\rho_{n,h}$. Using the relations $\rho_{n}(\lambda) = \rho_{n,h}(\lambda)\eta_n^{-1}(\lambda)$ for $n\geq 1$ we finally obtain explicit expressions for all root distributions $\rho_n$. The first two functions, for example, read
\begin{subequations}
\begin{align}
	\rho_1(\lambda) &= \frac{\sinh^3(\eta)\sin(2\lam+2i\eta)\sin(2\lam-2i\eta)}{\pi f(\lam-\frac{i\eta}{2})f(\lam+\frac{i\eta}{2}) g(\lambda)}\epc\\
	\rho_2(\lambda) &= \frac{ 8\sin^2(\lambda)\sinh^3(\eta)\cosh(\eta) [3\sin^2(\lam) + \sinh^2(\eta)][\cosh(6\eta)-\cos(4\lam)]}{\pi f(\lam)g(\lam+\frac{i\eta}{2})g(\lam-\frac{i\eta}{2})h(\lambda)}\epc
\end{align}  
\end{subequations}
where $f(\lambda) = \cosh^2(\eta) - \cos(2\lam)$, $g(\lambda) = \cosh(\eta) + 2\cosh(3\eta) -3\cos(2\lam)$, and
\begin{equation*}
h(\lambda) = 2\cos(4\lam) - \cos(2\lam) [3 + 2\cosh(2\eta) + 3\cosh(4\eta)] + 2\cosh^2(2\eta)[2 + \cosh(2\eta)] \epp
\end{equation*}

The function $\mathfrak{a}$ can be interpreted as the auxiliary function corresponding to the quantum transfer matrix \cite{1992_Kluemper_b, 1993_Kluemper}. Using the standard contour $\mathcal{C}$, which encircles the only pole of $[1+\mathfrak{a}(\omega)]^{-1}$ at $\omega=-\pi/2$, one can compute the function $G$, defined for example in Refs~\cite{2004_Goehmann, 2005_Goehmann, 2009_Boos}, by explicitly performing the contour integral. This way we checked that the nontrivial relation (4.32) of Ref.~\cite{2013_Fagotti_JSTAT_P07012} that relates the auxiliary function $\mathfrak{a}$ to the generating function $\Omega_{\text{N\'eel}}$ [see Eq.~\eqref{eq:generatingfunction_explicit}] is fulfilled. Unfortunately, this explicit $G$ function does not give the correct values of short-range correlation functions as calculated in Ref.~\cite{2014_Wouters}, since the standard approach \cite{2009_Boos} fails due to the presence of higher nontrivial driving terms, $d_{n\geq 2}\neq 0$, in the GTBA equations. It remains an open problem to determine the correct correlation functions from this approach.

\section{The large-$\Delta$ expansion}\label{sec:large_Delta}
A natural analytical approach to the quench from the N\'eel state is a large-$\Delta$ expansion. In the (anti-ferromagnetic) Ising limit $\Delta\to \infty$ there is no quench, therefore  $\Delta^{-1}$ is expected to be  a good expansion parameter that governs the density of excitations in the postquench steady state.  The spirit of this expansion is close to the small-quench expansion in Refs~\cite{2011_Calabrese_PRL_106, 2012_Calabrese_JSTAT_P07016, 2014_Bertini_Sinegordon}.

The most convenient expansion parameter  is 
\begin{equation}
	z=e^{-\eta}=\Delta-\sqrt{\Delta^2-1}=\sum_{n=1}^{\infty} \frac{(2 \,n)!}{(2\,n-1)\, (n!)^2\,4^n} \left(\frac {1} {\Delta}\right)^{2n-1} =\frac 1 {2 \Delta}+O(\Delta^{-3})\epp
\end{equation}
For $\Delta > 1$, $z$ is in the interval $[0,1)$. The Ising limit corresponds to $z\to 0$, while the isotropic point ($\Delta=1$) is at $z=1$.  The aim of this section is to report our results for the large-$\Delta$ expansion of the quench action saddle-point state as well as for the GGE, and to show how the difference between these two ensembles can be approached analytically. In Sec.~\ref{sec:large_Delta_density} we present our results for the densities $\brho$, while in Sec.~\ref{sec:large_Delta_correlator} the expansions for the nearest-neighbor and next-to-nearest-neighbor correlators are reported. We illustrate some of the most significant details of these calculations in~\ref{app:SP_EXP}, \ref{app:GGE_EXP}, and \ref{app:CORR_EXP}. 

As a side remark we note that the expansions we found are mathematically not unique. However, we here present the only self-consistent and physically acceptable solution we found. In particular, our expansion for the solution of the GTBA equations leads to a consistent expansion for the solution of the BGT equations that also obeys the zero-magnetization condition (for details, see \ref{app:SP_EXP}).

\subsection{Large-$\Delta$ expansion for the densities}\label{sec:large_Delta_density}
For the saddle-point state, the large-$\Delta$ expansion of $\rho^\text{sp}_n$ can be derived by expanding systematically the GTBA Eqs~\eqref{eq:TBA_XXZ_fact} as well as the BGT Eqs~\eqref{eq:BTGthlim_fact}. The leading behavior of $\rho^\text{sp}_n$ is 
\begin{equation}\label{eq:LDELTA_rho}
	\rho^\text{sp}_{n} (\lam) = \left\{
\begin{array}{ll}
\frac{1}{2\pi} \left[ 1 + z\, \rho_{1}^{(1)}(\lam) + \ldots \right] \epc & \qquad \text{if } n=1 \epc \\
\\
\frac{1}{\pi}z^{n} \sin^{2}(\lam) \left[ 1 + z\, \rho_{n}^{(1)}(\lam) + \ldots \right] \epc & \qquad \text{if } n \text{ even} \epc \\
 \\
\frac{1}{4\pi}z^{n-1} \left[ 1 + z\, \rho_{n}^{(1)}(\lam) + \ldots \right]  \epc & \qquad \text{if } n\geq 3 \text{ odd} \epp
\end{array}\right. \phantom{\}}
\end{equation}
The $z^0$ order is a consequence of the fact that in the quenchless Ising limit the steady state coincides with the initial one. Since in this limit a string of length $n$ corresponds to a block of $n$ consecutive down spins, the (zero-momentum) N\'eel state is therefore a state with a constant density of 1-strings and no strings with length greater than one, {\it i.e.}, $\rho^\text{N\'eel}_{1}(\lam) = 1/(2\pi)$ and $\rho^\text{N\'eel}_{n>1}(\lam) = 0$. For a finite but large $\Delta$, we have a contribution also from strings with length $n> 1 $. However, their contributions are suppressed as $\Delta^{-n}$ for $n$ even or $\Delta^{-n+1}$ for $n$ odd, so longer strings have a negligible effect for large $\Delta$. For $\eta^\text{sp}_n$, the leading behavior is
\begin{equation}\label{eq:eta_lead}
	\eta^\text{sp}_n(\lam) = \left\{
\begin{array}{ll}
8 \,z^2 \sin^2(2\lam)  \left[ 1 + z\, \eta_{1}^{(1)}(\lam) + \ldots \right]\epc  & \qquad \text{for } n=1 \epc \\
\tan^{-2} (\lam)  \left[ 1 + z\, \eta_{n}^{(1)}(\lam) + \ldots \right]\epc & \qquad \text{for } n \text{ even} \epc\\
16 \,z^2\sin^2(2\lam)  \left[ 1 + z\, \eta_{n}^{(1)}(\lam) + \ldots \right]\epc & \qquad \text{for } n\geq 3 \text{ odd} \epp \\
\end{array}\right.\phantom{\}}
\end{equation}
Notice that Eq.~\eqref{eq:eta_lead} implies that the Lagrange parameter $h$ in Eq.~\eqref{eq:GTBA_QA} is actually divergent. Using {\it Mathematica}, we computed the expansion up to order $z^{16}$ for $\rho^\text{sp}_{n>1}$ and up to order $z^{19}$ for $\rho^\text{sp}_1$. For the hole densities  $\rho^\text{sp}_{n,h}$, we computed the expansion up to order  $z^{18}$ for $n>1$ and up to order $z^{21}$ for $n=1$. For all orders that were computed, the expansions agree with the exact formula for $\rho_{1, h}$ in Eq.~\eqref{eq:rho1h_exact_XXZ} as well as with the analytical solution presented in Sec.~\ref{sec:explicit_solution}. It is also consistent with all our numerical data. To give an idea of what the expansions look like, the saddle-point densities up to order $z^5$ are
\begin{subequations}\label{eq:sp_rho_exp}
\begin{align}
	\rho^\text{sp}_{1}(\lam) &=  \frac{1}{2\pi} \Big\{ 1 + 4z\cos(2\lam) + z^{2} \left[8\cos(4\lam) - \tfrac{7}{2}\right] + z^{3}\left[16 \cos(6\lam) - 15 \cos(2\lam)\right]\nonumber \\
&\qquad\quad + z^{4}\left[\tfrac{81}{4} - 48 \cos(4\lam) + 32 \cos(8\lam)\right] \nonumber \\
&\qquad\quad + z^5 \left[71 \cos (2\lam) - 126 \cos (6\lam) + 64 \cos(10\lam)\right]  \Big\} +  O(z^{6}) \epc  \\
\rho^\text{sp}_{2}(\lam)  &= \frac{z^{2}}{\pi} \sin^{2}(\lam) \Big\{1 + z^{2}\left[7\cos(2\lam) - 5\right]  \Big\} + O(z^{6}) \epc  \\
\rho^\text{sp}_{3}(\lam)  &=  \frac{z^{2}}{4\pi} \Big\{ 1 + 2z\cos(2\lam) + z^{2}\left[8\cos(4\lam) - \tfrac{13}{2}\right] \nonumber \\
& \qquad\quad +24z^{3}\left[\cos(6\lam) - \cos(2\lam)\right] \Big\} + O(z^{6}) \epc\\
\rho^\text{sp}_{4}(\lam) &=  \frac{z^{4}}{\pi} \sin^{2}(\lam) + O(z^{6}) \epc  \\
\rho^\text{sp}_{5}(\lam) &= \frac{z^{4}}{4\pi} \Big\{ 1 + 2z\cos(2\lam)\Big\}  + O(z^{6}) \epc
\end{align}
\end{subequations}
the other densities being at least $O(z^{6}) $. Similarly, for the hole densities we have
\begin{subequations}\label{eq:sp_rho_h_exp}
\begin{align}
	\rho^\text{sp}_{1,h}(\lam) &= \frac{4  z^2\sin^2 (2 \lam)}{\pi} \Big\{1+6 z \cos (2 \lambda )+z^2 \left[14 \cos (4 \lambda )+2\right] \nonumber \\
		& \qquad\qquad\qquad\ +z^3 \left[30 \cos (6 \lambda )-4 \cos (2 \lambda )\right]+z^4 \left[7-36 \cos (4 \lambda )+62 \cos (8 \lambda )\right] \nonumber \\
		&\qquad\qquad\qquad\ + 2 z^5 \left[25 \cos (2 \lambda )-66 \cos (6 \lambda )+63 \cos (10 \lambda )\right]\Big\}+O (z^8) \epc\\
	\rho^\text{sp}_{2,h}(\lam) &= \frac{z^2\cos ^2(\lambda )}{\pi }\Big\{1+ z^2 \left[1-\cos(2\lambda)\right]\nonumber \\
		& \qquad\qquad\qquad\ + z^4\left[\tfrac{49}{2} \cos (2 \lambda )-\tfrac{11}{2} \cos (4 \lambda )-18\right] \Big\}+O\left(z^8\right) \epc\\
	\rho^\text{sp}_{3,h}(\lam) &= \frac{4 z^4\sin^2 (2 \lam)}{\pi}\Big\{ 1+6 z \cos (2 \lambda )+z^2 \left[18 \cos (4 \lambda )+3\right]\nonumber \\
		& \qquad\qquad\qquad\ + z^3 \left[54 \cos (6 \lambda )-4 \cos (2 \lambda )\right] \Big\} +O\left(z^8\right) \epc\\
	\rho^\text{sp}_{4,h}(\lam) &= \frac{z^4\cos ^2(\lambda ) }{\pi }\Big\{ 1+z^2 \left[1-2 \cos (2 \lambda )\right]\Big\}+O\left(z^8\right) \epc\\
	\rho^\text{sp}_{5,h}(\lam) &= \frac{4 z^6\sin^2 (2 \lam) }{\pi}\Big\{ 1+6 z \cos (2 \lambda )\Big\}+O\left(z^8\right) \epc\\
\rho^\text{sp}_{6,h}(\lam) &= \frac{z^6\cos ^2(\lambda )}{\pi}+O\left(z^8\right) \epc
\end{align}
\end{subequations}
the other hole densities being at least $O(z^{8})$.

For the GGE, we can obtain a large-$\Delta$ expansion by expanding the GTBA Eqs~\eqref{eq:GTBAGGE_a} for $n\geq 2$ and the BGT Eqs~\eqref{eq:BTGthlim_fact} for $n \geq 1$, and by taking advantage of the explicit expression \eqref{eq:rho1h_exact_XXZ} for $\rho_{1,h}$. This way, we circumvent the problem of computing the chemical potentials that appear only in the driving term of the GTBA equation for $n=1$. The expansions for the densities are
\begin{subequations}\label{eq:GGE_exp}
\begin{align} 
\rho^{GGE}_1(\lam) &= \frac{1}{2\pi}\Big\{1 + 4z \cos(2\lam)+z^2\left[8\cos(4\lam) - 3\right]\\
&\qquad\quad + 16 z^3\left[\cos(6\lam)-\cos(2\lam)\right]+4z^4\left[4 - 12\cos(4\lam) + 7\cos(8\lam)\right] \Big\}+ O(z^5)\epc \nonumber\\
\rho^{GGE}_2(\lam) &= \frac{z^2}{3\pi} \Big\{1 + z^2 \left[\tfrac{9}{2}\cos(2\lam) - \tfrac{3}{2}\cos(4\lam) - \tfrac{20}{3}\right] \Big\}+O(z^5)\epc\\
\rho^{GGE}_n(\lam) &= \frac{2z^2}{\pi n (n^2-1)}\Big\{1 - 2z^2 \left[\tfrac{3}{2}+\tfrac{1}{n}+\tfrac{1}{n+1}+\tfrac{1}{n-1}\right]\Big\} +O(z^5)\epc\quad n\geq 3\epc
\end{align}
\end{subequations}
while for the hole densities we have 
\begin{subequations}
\begin{align}
	\rho^{GGE}_{2,h}(\lam) &= \frac{z^2}{\pi} \Big\{1 + z^2 \left[\tfrac{9}{2}\cos(2\lam) - \tfrac{3}{2}\cos(4\lam)-4\right] \Big\} + \mathcal{O}(z^5)\epc\\	
	\rho^{GGE}_{n,h}(\lam) &= \frac{2z^2}{\pi n} \Big\{1-2z^2\left[\tfrac{3}{2}+\tfrac{1}{n}\,\right]\Big\}+O(z^5)\epc\quad n\geq 3\epc
\end{align}
\end{subequations}
$\rho_{1,h}^{GGE}$ being given by Eq.~\eqref{eq:rho1h_exact_XXZ}.

The GGE densities differ qualitatively from the ones given by the quench action method. While for the saddle-point state the contributions of higher strings are suppressed by increasing powers of $ \Delta^{-1}$, the leading term of all $\rho_{n\geq 2}^{GGE}$ is of order $\Delta^{-2}$, and the higher-string contributions are suppressed only by the (algebraically decaying) prefactors. The difference between $\rho_{n}^{GGE}$ and $\rho_{n}^\text{sp}$  is of order $\Delta^{-2}$,
\begin{subequations}\label{eq:diffdist}
\begin{align} 
	&\rho_{1}^\text{GGE}(\lam) - \rho_{1}^\text{sp}(\lam) = \frac{1}{4\pi \Delta^2}+ O(\Delta^{-3}) \epc  \\
	&\rho_{2}^\text{GGE}(\lam) - \rho_{2}^\text{sp}(\lam) =  \frac{1-3 \sin^2 (\lam)}{3\pi \Delta^2}  + O(\Delta^{-3}) \epc \\
	&\rho_{3}^\text{GGE}(\lam) - \rho_{3}^\text{sp}(\lam) =- \frac{1}{24 \pi \Delta^2}+ O(\Delta^{-3}) \epc  \\
	&\rho_{n}^\text{GGE}(\lam) - \rho_{n}^\text{sp}(\lam) =\frac{1}{2 n (n^2-1) \pi \Delta^2}+O(\Delta^{-3})  \epc\qquad n\geq 4 \epp
\end{align}
\end{subequations}
Finally, in Ref.~\cite{2014_Pozsgay_Dimer} a nontrivial check for the quench action saddle point was suggested. If the saddle-point state is unique and if the saddle-point approximation of the functional integral is valid, then the quench action evaluated at the saddle-point must be zero,
\begin{equation} \label{eq:condition_value_quench_action}
	\limth \frac{S_{QA}[\brho^{\text{sp}}]}{N} = -\limth \frac{1}{N} \ln \skalarszorzat{\Psi_0}{\Psi_0}=0 \epp
\end{equation}
To derive this condition one writes the norm of the initial state $\skalarszorzat{\Psi_0}{\Psi_0}=1$ as a functional integral weighted by the quench action and subsequently performs a saddle-point approximation. Note that in the thermodynamic limit the ambiguity in the choice for the measure of the functional integral drops out of Eq.~\eqref{eq:condition_value_quench_action}. We evaluated the quench action on the large-$\Delta$ expansion of the saddle-point solution up to order $\Delta^{-16}$ and found perfect, nontrivial cancellation between the overlap coefficient and the Yang-Yang entropy,
\begin{align}\label{eq:sumrulecheck}
	\limth \frac{2 S[\brho^{\text{sp}}]}{N} &= \limth \frac{ S_{YY}[\brho^{\text{sp}}]}{2N} + o \left( \Delta^{-16} \right) \nonumber \\
		&= \frac{4\ln(2\Delta)-1}{8\Delta^{2}} - \frac{8\ln(2\Delta)-5}{32\Delta^4} + \ldots \nonumber \\
 		&\quad \ldots +\frac{3(6316800\ln(2\Delta)-6579767)}{18350080 \Delta^{16}} + o \left( \Delta^{-16} \right) \epp 
\end{align}
Note the extra factor $1/2$ in front of the Yang-Yang entropy due to parity invariance of the states with nonzero overlap with the N\'eel state (for details, see \ref{app:GTBAQA}). Also, notice that substituting the large-$\Delta$ expansion of the GGE solution into the quench action $S_{QA}[\brho]$ is not possible, since the quench action is not analytic in this point and therefore does not have a power-series expansion like Eq.~\eqref{eq:sumrulecheck}. Note that this finding is in agreement with the observed divergence of the quench action evaluated on the GGE solution in Ref.~\cite{2014_Pozsgay_GGE}.

\subsection{Large-$\Delta$ expansion for local correlators}\label{sec:large_Delta_correlator}
\begin{figure}[t]
\begin{center}
\includegraphics[width=0.73\columnwidth]{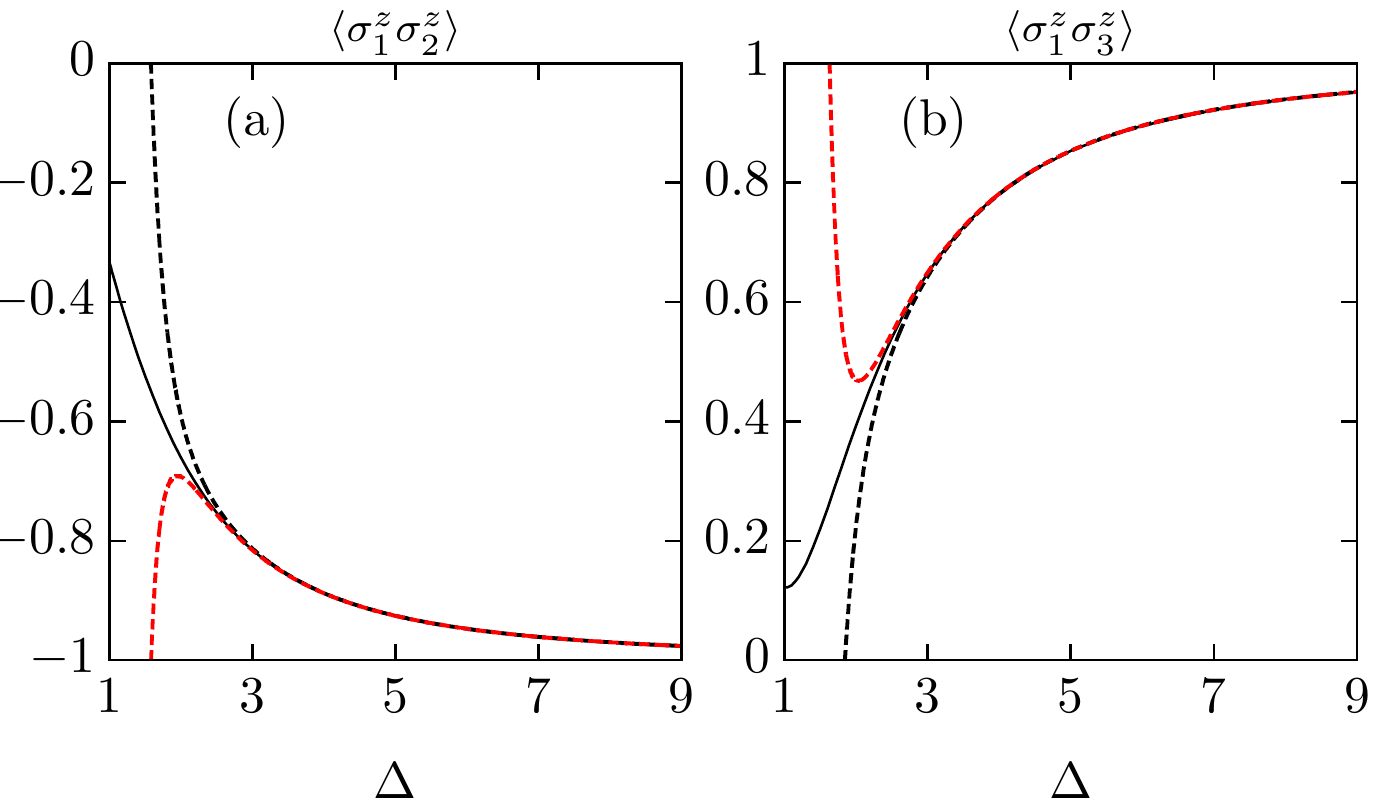}
\caption{\label{fig:correlators_exp} Numerical data for the saddle-point state correlators (solid line) \cite{2014_Wouters} compared with the large-$\Delta$ expansion up to the sixth (black dashed line) and the eight order (red dashed line). Increasing the order of the expansion, the agreement with the numerical data improves and extends to smaller $\Delta$.}
\end{center}
\end{figure}

In this subsection we report the large-$\Delta$ expansion for the local correlators $\langle \sigma_1^z \sigma_2^z \rangle$ and $\langle \sigma_1^z \sigma_3^z \rangle$. Given the root densities these correlators can be computed using the Hellman-Feynman theorem~\cite{2014_Wouters,2014_Mestyan} (for the nearest-neighbor correlators) or a recent conjecture presented in Ref.~\cite{2014_Mestyan} (for the next-to-nearest-neighbor correlators). More details on the expansion of the correlators are given in~\ref{app:CORR_EXP}. We find that
\begin{subequations}\label{eq:corr_exp}
\begin{align}
	&\langle \sigma_1^z \sigma_2^z \rangle_\text{sp}
		& \!\!\!\!\!\!\!\!\!\!\! &= -1+\frac{2}{\Delta^{2}} - \frac{7}{2\Delta^{4}}+\frac{77}{16 \Delta^{6}} - \frac{689}{128\Delta^{8}} + \frac{5769}{1024\Delta^{10}}+ \nonumber \\[0.5ex]
		& & \!\!\!\!\!\!\!\!\!\!\! &\qquad -\frac{50605}{8192\Delta^{12}}+\frac{462617}{65536\Delta^{14}} - \frac{4383949}{524288\Delta^{16}}+O\left(\Delta^{-17}\right)\epc \\[1.0ex]
	&\langle \sigma_1^z \sigma_2^z \rangle_\text{GGE} & \!\!\!\!\!\!\!\!\!\!\! &= -1+\frac{2}{\Delta^{2}} - \frac{7}{2\Delta^{4}} + \frac{43}{8\Delta^{6}} + O\left(\Delta^{-7}\right)\epc\\[1.0ex]
	&\langle \sigma_1^z \sigma_3^z \rangle_\text{sp}& \!\!\!\!\!\!\!\!\!\!\!  &=1-\frac{4}{\Delta^{2}} + \frac{35}{4\Delta^{4}}-\frac{195}{16\Delta^{6}} + \frac{773}{64\Delta^{8}}+O\left(\Delta^{-9}\right)\epc \label{eq:z1z3_sp} \\[1.0ex]
	&\langle \sigma_1^z \sigma_3^z \rangle_\text{GGE}& \!\!\!\!\!\!\!\!\!\!\! &= 1- \frac{4}{\Delta^{2}}+\frac{37}{4\Delta^{4}}+O\left(\Delta^{-5}\right)\epp\label{eq:z1z3_GGE}
\end{align}
\end{subequations} 
The expansions~\eqref{eq:corr_exp} agree nicely with our data for correlators~\cite{2014_Wouters}, obtained by solving the relevant integral equations numerically, as shown in Fig.~\ref{fig:correlators_exp}. By increasing the order of the expansion, the agreement with the correlators improves and the expansion becomes a better approximation for a larger range of $\Delta$. The fact that the large-$\Delta$ expansions blow up for small $\Delta > 1$ suggests that these series are not convergent in the whole complex plane. It is quite natural to assume that the radius of convergence in the $z$ plane is one, so that the series are not convergent in the gapless phase $\Delta<1$.
\begin{figure}[t]
\begin{center}
\includegraphics[width=0.73\textwidth]{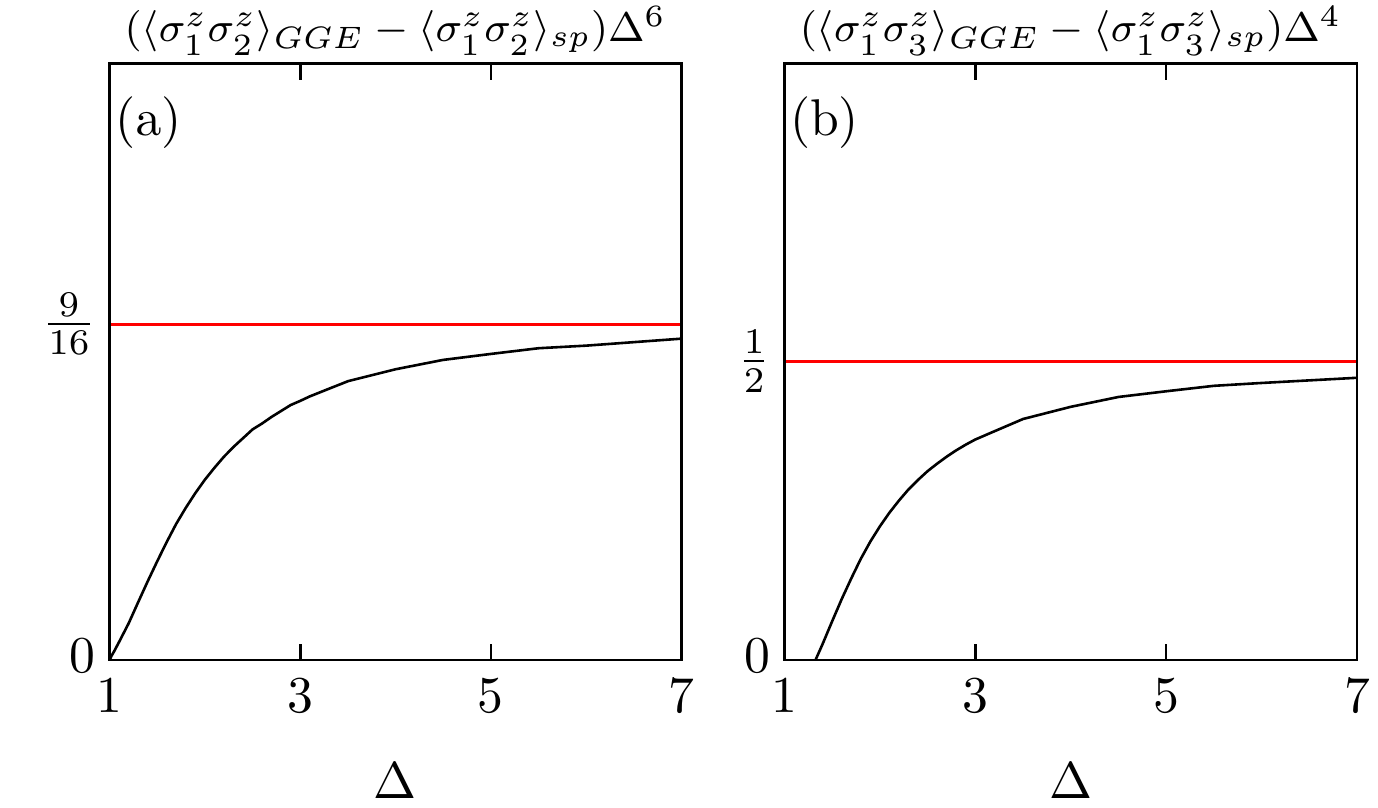}
\caption{\label{fig:correlators_diff} Rescaled difference between GGE and the saddle-point state for (a) $\langle \sigma^z_1 \sigma^z_{2} \rangle$  and (b) $\langle \sigma^z_1 \sigma^z_{3} \rangle$.  The numerical data (indicated by the black line, obtained in Ref.~\cite{2014_Wouters}) are consistent with the analytical prediction in Eq.~\eqref{eq:diffcorr}, which is indicated by the red line.}
\end{center}
\end{figure}

We noticed in Eqs~\eqref{eq:diffdist} that for the densities the difference between GGE and the saddle-point state is of order $O(\Delta^{-2})$. However, this is not necessarily the case for local correlators. Indeed, we have
\begin{subequations} \label{eq:diffcorr}
\begin{align}
	\langle \sigma^z_1 \sigma^z_{2} \rangle_\text{GGE} - \langle \sigma^z_1 \sigma^z_{2} \rangle_\text{sp}  &=  \frac{9}{16 \Delta^6} + O(\Delta^{-7}) \epc\\[0.8ex]
	\langle \sigma^z_1 \sigma^z_{3} \rangle_\text{GGE} - \langle \sigma^z_1 \sigma^z_{3} \rangle_\text{sp}  &=  \frac{1}{2 \Delta^4} + O(\Delta^{-5}) \epp
\end{align}
\end{subequations}
This behavior is consistent with our data from Ref.~\cite{2014_Wouters} as shown in Fig.~\ref{fig:correlators_diff}.

To summarize, for a (small) quench from the N\'eel state, the GGE is more effective in reproducing local correlators as $\langle \sigma^z_1 \sigma^z_{2} \rangle$ and $\langle \sigma^z_1 \sigma^z_{3} \rangle$ than the root densities $\brho$. This is especially true for the most local correlator $\langle \sigma^z_1 \sigma^z_{2} \rangle$, where the difference is of order $\Delta^{-6}$, while for $\langle \sigma^z_1 \sigma^z_{3} \rangle$ it is of order  $\Delta^{-4}$.

\section{The N\'eel-to-XXX quench} \label{sec:XXX}

\subsection{The scaling limit}
In this section the quench from the N\'eel state to the isotropic point $\Delta=1$ of the spin-1/2 XXZ model, where the theory is gapless, is studied. The Bethe Ansatz description of this XXX spin chain uses different conventions. They can be obtained from the gapped regime through a scaling limit. Rapidities of the gapped model go to zero with $\eta$, where $\Delta = \cosh(\eta)$. So, in order to have a description in terms of finite quantities, we scale all spectral parameters with a factor $\eta$, 
\begin{equation}
	\lam \to \eta \lam \epc
\end{equation}
where the rescaled rapidities and spectral parameters now lie in the interval $\big[\! -\!\tfrac{\pi}{2\eta}, \tfrac{\pi}{2\eta} \big)$. Subsequently, the XXX-limit $\eta \to 0$ is taken. After multiplication with the appropriate power of $\eta$ and taking this limit, XXZ quantities (indicated here by the tilde) scale to their XXX counterparts, for example,
\begin{subequations}
\begin{align}
	\theta_n (\lam) &= \lim_{\eta \to 0} \tilde{\theta}_n(\eta \lam) = 2 \arctan \left( \frac{2\lam}{n} \right) \epc \\
	a_n (\lam) &= \frac{1}{2\pi} \frac{\partial}{\partial \lam} \theta_n (\lam) = \lim_{\eta \to 0} \left[\eta\, \tilde{a}_n(\eta \lam)\right] = \frac{1}{2\pi} \frac{n}{\lam^2 + n^2/4} \epc \label{eq:defkernelXXX} \\
	\rho_n (\lam) &= \lim_{\eta \to 0} \left[\eta\, \tilde{\rho}_n(\eta \lam)\right]\epc\\
	K_\alpha(\lambda) &= \lim_{\eta\to 0} \left[\eta\, \tilde{K}_{\alpha\eta}(\eta\lambda)\right] = \frac{2\alpha}{\lambda^2+\alpha^2}\epp\label{eq:scaled_K}
\end{align}
\end{subequations}

The XXX Bethe equations and the eigenvalues of the transfer matrix are obtained from respectively Eq.~\eqref{eq:BAE} and Eq.~\eqref{eq:transfermatrix} through the scaling limit. The thermodynamic form of the Bethe equations is as in Eq.~\eqref{eq:BTGthlim}, with the appropriate kernels in Eq.~\eqref{eq:defkernelXXX} and convolution integrals over $\mathbb{R}$. The kernel in the partially decoupled form~\eqref{eq:BTGthlim_fact} becomes
\begin{equation} \label{eq:defsXXX}
	s(\lam) = \lim_{\eta\to 0}\left[\eta\tilde{s}(\eta\lambda)\right] = \frac{1}{2\cosh(\pi \lam)} \epp
\end{equation}
Note that for the XXX spin chain rapidities at infinity are allowed. They decouple from the Bethe equations and should be treated separately.

For the Fourier transform we use the conventions
\begin{subequations}
\begin{align}\label{eq:FourierTransformXXX}
	\hat{f}(k) &= \ftk{f}{k} = \int_{-\infty}^{\infty}\mathrm{d}\lam e^{ik\lam}f(\lam) \epc\quad k\in\mathbb{R}\epc \\
	f(\lam) &= \iftx{\hat{f}}{\lam} = \int_{-\infty}^{\infty}\frac{\mathrm{d}k}{2\pi} e^{-ik\lam}\hat{f}(k)\epc\quad \lambda\in\mathbb{R} \epp
\end{align}
\end{subequations}
If \begin{equation}
	f(\lam) = \lim_{\eta\to 0} \left[\eta^\alpha \tilde{f}(\eta \lam) \right]\epc
\end{equation}
then the Fourier-transformed relation between the XXZ and XXX quantity is
\begin{equation}
	\hat{f}(k) = \lim_{\eta\to 0} \left[\eta^{\alpha-1}\left.\hat{\tilde{f}}(k')\right|_{k=2k'\eta} \right] \epp
\end{equation}
In the XXX limit discrete sums in Fourier space become integrals,
\begin{equation}
	\lim_{\eta\to0} \frac{1}{\pi}\sum_{k'\in\mathbb{Z}} \eta \, f (2k'\eta) = \int_{-\infty}^\infty \frac{\mathrm{d}k}{2\pi} f(k) \epp
\end{equation}
Knowing this, our results for the N\'eel-to-XXZ quench are straightforwardly generalized to a quench to the spin-1/2 XXX chain. For the sake of completeness, we briefly outline the results for this quench. In the remainder of this section rapidities $\lam \in\mathbb{R}$ are always XXX quantities.

\subsection{Analytical solution of $\rho_{1,h}$}
The local conserved charges are defined by [see Eq.~\eqref{eq:conservedcharges}]
\begin{equation} \label{eq:conservedchargesXXX}
	Q_{m+1} = \frac{i}{2^m} \left.{\frac{\partial^m}{\partial \lam^m} \ln[t(\lam)]}\right|_{\lam=i/2} \epc
\end{equation}
and the relation with the generating function~\cite{2013_Fagotti_JSTAT_P07012} is [cf.~Eq.~\eqref{eq:defgeneratingfunction}]
\begin{equation}
	\frac{\bra{\Psi_0}Q_{m+1}\ket{\Psi_0}}{N} = \frac{1}{2^m} \left. \frac{\partial^{m-1}}{\partial \lam^{m-1}} \Omega_{\Psi_0}(\lam) \right|_{\lam=0} \epp
\end{equation}
This implies [cf.~Eqs~\eqref{eq:pinning_rho1h}, \eqref{eq:generatingfunctiontorho1h}] 
\begin{equation}
	\frac{1}{\pi} \hat{\Omega}_{\Psi_0} (k) = \frac{\hat{\rho}_{1,h}^{\Psi_0}(k) - e^{-|k|/2}}{\cosh(k/2)} \epc
\end{equation}
or in $\lambda$-space
\begin{equation} \label{eq:generatingfunctiontorho1hXXX}
	\rho_{1,h}^{\Psi_0} (\lam) = a_{1}(\lam) + \frac{1}{2\pi} \left[ \Omega_{\Psi_0} \left( \lam + \tfrac{i}{2} \right) + \Omega_{\Psi_0} \left( \lam - \tfrac{i}{2} \right) \right] \epp
\end{equation}
For the N\'eel-to-XXX quench the generating function in the thermodynamic limit is given by [$\tilde{\Omega}_\text{N\'eel}$ from Eq.~\eqref{eq:generatingfunction_explicit}]
\begin{equation}
	\Omega_\text{N\'eel}(\lam) = \lim_{\eta\to 0}\left[\eta\tilde{\Omega}_{\text{N\'eel}}(\eta\lambda)\right] = -\frac{1}{1+2\lam^2} \epc
\end{equation}
and the 1-string hole density of the steady state is
\begin{equation}\label{eq:rho1_XXX}
	\rho_{1,h}^\text{N\'eel} (\lam) = \frac{1}{2\pi} \frac{\lam^2}{(\lam^2+\frac{1}{4})(\lam^4+\frac{3}{2}\lam^2+\frac{1}{16})} \epp
\end{equation}

\subsection{The XXX overlaps}
For the specific quench to the isotropic point $\Delta=1$, the nonzero overlaps were also computed in Ref.~\cite{2014_Brockmann_JPA}. Bethe states can have an arbitary number of its rapidities at infinity, corresponding to zero-momentum spin excitations, which need to be treated separately. We denote a parity-invariant Bethe state with $N_\infty$ rapidities at infinity  by $|\{\pm\lambda_j\}_{j=1}^{m},\,n_\infty\rangle$, where the $m$ pairs of finite rapidities are denoted by $\{\pm\lambda_j\}_{j=1}^{m}$ and $M=N_\infty+2m =N/2$. Here, we assumed $N_\infty$ to be even, and we defined the fraction of rapidities at infinity by $n_\infty = N_\infty/M = 2 N_\infty/N$. 

The overlap between the zero-momentum N\'eel state and a normalized parity-invariant XXX Bethe state with $N_\infty$ rapidities at infinity is then given by
\begin{subequations}\label{eq:overlap_XXX_onshell}
\begin{align}\label{eq:overlap_XXX_onshell_a}
	&\frac{\langle \Psi_0 | \{\pm\lambda_j\}_{j=1}^{m},n_\infty \rangle }{\| |\{\pm\lambda_j\}_{j=1}^{m},n_\infty\rangle \|} = \frac{\sqrt{2}\,N_{\infty}!}{\sqrt{(2N_{\infty})!}}
	\left[\prod_{j=1}^m \frac{\sqrt{\lambda^2_j+1/4}}{4\lambda_j}\right]  
	\sqrt{ \frac{\det{}_{\!m}(\hat{G}^{+})}{\det{}_{\!m}(\hat{G}^{-})} }\epc\\
	&\hat{G}_{jk}^{\pm} = \delta_{jk}\left(NK_{1/2}(\lambda_j)-\sum_{l=1}^{m}K_1^{+}(\lambda_j,\lambda_l)\right) + K_1^{\pm}(\lambda_j,\lambda_k)\:, \quad j,k=1,\ldots,m
\end{align}
\end{subequations}
with $K_1^{\pm}(\lambda,\mu)=K_1(\lambda-\mu)\pm K_1(\lambda+\mu)$ and $K_\alpha(\lambda)$, $\alpha = \frac{1}{2},1$, as in Eq.~\eqref{eq:scaled_K}.

\subsection{The quench action GTBA equations}\label{sec:QAA_XXX}
In the thermodynamic limit a Bethe state of the spin-1/2 XXX chain is characterized by a set of root densities $\brho$, now defined as positive, smooth, bounded functions on $\mathbb{R}$, and the fraction of rapidities at infinity $n_\infty$. In order to determine the quench-action saddle point, one must also vary with respect to $n_\infty$.

As was the case for the XXZ quench, the ratio of determinants in Eqs~\eqref{eq:overlap_XXX_onshell_a} does not contribute to the extensive part of the overlap coefficient. Therefore, the thermodynamic overlap coefficient is given by
\begin{subequations}
\begin{align}
	S\left[ \brho , n_\infty \right] &= - \limth \ln\left( \frac{\langle \Psi_0 | \{\pm\lambda_j\}_{j=1}^{m},n_\infty \rangle }{\| |\{\pm\lambda_j\}_{j=1}^{m},n_\infty\rangle\|} \right)  \nonumber \\
	&= \frac{N}{2} \left( n_\infty \, \ln 2 + \sum_{n=1}^\infty \int_{0}^{\infty} \mathrm{d}\lam \, \rho_{n}(\lam) \big[ g_n(\lam) + 2n \ln (4) \big] \right) \epc
\end{align}
with
\begin{align}
g_n(\lam) &= \sum_{l=0}^{n-1} \Big[ f_{n-1-2l}(\lam) - f_{n-2l}(\lam) \Big] \epc \\ 
f_n (\lam) &= \ln \big(\lam^2 + n^2/4 \big) \epp
\end{align}
\end{subequations}
To fix the total magnetization, the Lagrange multiplier that needs to be added to the quench action $S_{QA}[\brho, n_\infty] = 2 S[ \brho,n_\infty] - \tfrac{1}{2} S_{YY}[ \brho]$ is
\begin{equation}
	- h\, N \left( 2 \sum_{m=1}^{\infty} \, m \int_{0}^{\infty} \mathrm{d}\lam\, \rho_{m}(\lam) + \frac{1}{2}n_{\infty} - \frac{1}{2} \right) \epp
\end{equation}
Unlike the XXZ case the Lagrange multiplier can be fixed immediately. Variation with respect to $n_\infty$ leads to the condition
\begin{equation}
	h = \ln(4) \epp
\end{equation} 
Variation with respect to $\rho_n$ gives the GTBA equations for the N\'eel-to-XXX quench,
\begin{equation} \label{eq:TBA_XXX}
	\ln[\eta_{n}(\lam)]  =  g_{n}(\lam)  + \sum_{m=1}^{\infty} \left[ a_{nm} \ast \ln \left( 1 + \eta_{m}^{-1} \right)\right] (\lam) 
\end{equation}
for $n\geq 1$. Since the Lagrange multiplier $h$ is already fixed, the saddle-point solution of the GTBA and Bethe equations will be independent of any free parameter. Instead, it will fix the fraction of rapidities at infinity of the steady state:
\begin{equation} \label{eq:infinity_rapidities}
	n_\infty = 1 - 2\sum_{m=1}^{\infty} \, m \,  \int_{-\infty}^{\infty} \mathrm{d}\lam \, \rho_{m}(\lam) \epp
\end{equation}
In analogy with Eq.~\eqref{eq:TBA_XXZ_fact0} one can factorize the GTBA equations into
\begin{equation}
	(a_0 + a_2) \ast \ln(\eta_n)  =  \tilde{d}_n  + a_1 \ast [\ln(1 + \eta_{n-1}) +\ln(1 + \eta_{n+1})  ]  \epc
\end{equation}
where $\tilde{d}_n(\lam) = (-1)^{n+1}[(a_0 - a_2)\ast f_0] (\lam)$, by convention $\eta_0(\lam)=0$, and we used that $a_m \ast f_n = f_{|n|+m}$. From this equation the asymptotic behavior of the function $\eta_n$ can be derived easily. Define $\eta_{n,\infty} = \lim_{\lam\to\infty} \eta_n(\lam)$, then $\eta_{n,\infty}^2 = (1+\eta_{n-1,\infty})(1+\eta_{n+1,\infty})$. The only physically meaningful solution is $\eta_{n,\infty}=n(n+2)$. Inverting the operation of $(a_0+a_2)\ast$ leads to
\begin{subequations}\label{eq:TBA_XXX_fact}
\begin{equation}\label{eq:TBA_XXX_fact_equation}
	\ln(\eta_n) = d_n + s \ast \big[\ln(1+\eta_{n-1})+\ln(1+\eta_{n+1})\big]\epc
\end{equation}
where $s$ was defined in Eq.~\eqref{eq:defsXXX} and the driving term is [cf.~Eq.~\eqref{eq:TBA_XXZ_fact_driving}]
\begin{align}\label{eq:TBA_XXX_fact_driving}
d_n(\lam) = (-1)^n \int_{-\infty}^\infty \mathrm{d}k \,  e^{-ik\lam} \frac{\tanh(k/2)}{k} = (-1)^{n+1}\ln\left[\tanh^2\left(\frac{\pi \lambda}{2}\right)\right]\epp
\end{align}
\end{subequations}

\subsection{Analytical solution} \label{sec:analytical_sol_XXX}
\begin{figure}[h]
\begin{center}
\includegraphics[width=0.9\columnwidth]{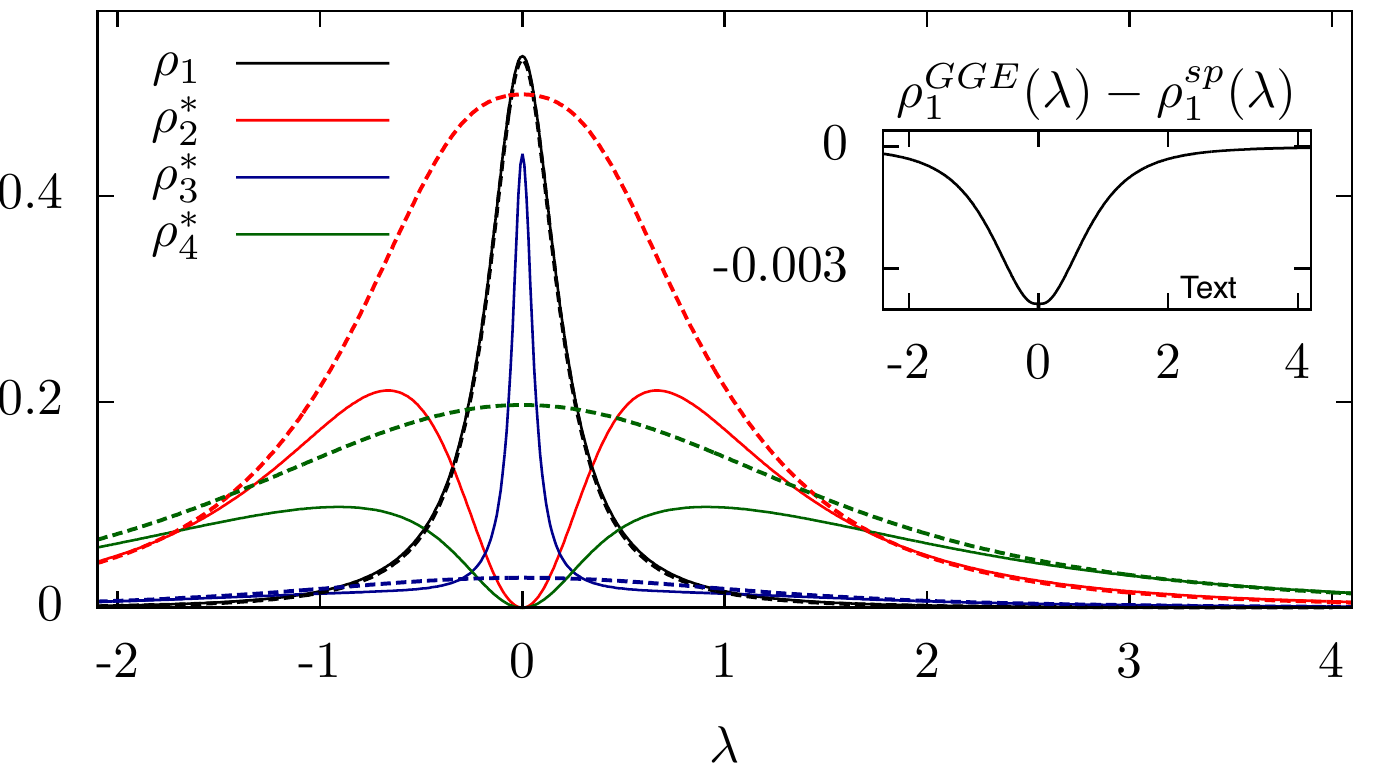}
\caption{\label{fig:rhos_XXX} Density functions $\rho_n$ with $n=1,2,3,4$ of the quench action saddle-point state (solid lines) and of the GGE equilibrium state (dashed lines) for the quench to the XXX model ($\Delta=1$). For $n>1$ the functions are rescaled as $\rho_n^*  = n^2 \rho_n$ for odd $n$ and $\rho_n^*  = 10n^2 \rho_n$ for even $n$.  Inset: Difference between the GGE prediction for the distribution $\rho_{1}$ of 1-strings and the quench action saddle-point result.}
\end{center}
\end{figure}

Explicit expressions for the solution of the GTBA Eqs~\eqref{eq:TBA_XXX_fact} are easily obtained from the explicit form~\eqref{eq:def_a} of the $\mathfrak{a}$ function. Replacing the spectral parameter $\lambda$ by $\eta \lambda$ and sending $\eta \to 0$ yields
\begin{equation}
	\mathfrak{a}(\lambda) = \frac{(\lambda+i)(2\lambda-i)}{(\lambda-i)(2\lambda+i)}\epp
\end{equation}
All functional relations of Sec.~\ref{sec:T-system} remain the same with the only difference that $i\eta$ in the arguments of the functions has to be replaced by $i$. This results in the explicit expressions
\begin{subequations}
\begin{align}
	\eta_1(\lambda) &= \frac{\lambda^2 (19 + 12 \lambda^2)}{(1 + \lambda^2) (1 + 4 \lambda^2)}\epc\\
	\eta_2(\lambda) &= \frac{8 (1 + 2 \lambda^2) (2 + 7 \lambda^2 + 2 \lambda^4)}{\lambda^2 (1 + \lambda^2) (9 + 4 \lambda^2)}\epc\\
	\eta_3(\lambda) &= \frac{\lambda^2 (19 + 12 \lambda^2) (509 + 520 \lambda^2 + 80 \lambda^4)}{(4 + \lambda^2) (1 + 
   4 \lambda^2)^2 (9 + 4 \lambda^2)}\epc\\
	\eta_4(\lambda) &= \frac{8 (2 + 7 \lambda^2 + 2 \lambda^4) (36 + 143 \lambda^2 + 65 \lambda^4 + 6 \lambda^6)}{\lambda^2 (1 + 
   \lambda^2)^2 (4 + \lambda^2) (25 + 4 \lambda^2)}\epc\\
   \vdots\notag
\end{align}
\end{subequations}
We obtain the root densities $\rho_n$ as described in Sec.~\ref{sec:explicit_solution} using the BGT Eqs~\eqref{eq:BTGthlim_fact} with the $s$-function calculated in Eq.~\eqref{eq:defsXXX} and using the explicit expression~\eqref{eq:rho1_XXX} of the 1-string hole density. The first four root densities read
\begin{subequations}
\begin{align}
	\rho_{1}(\lambda) &= \frac{32 (1 + \lambda^2)}{\pi (19 + 12 \lambda^2) (1 + 24 \lambda^2 + 16 \lambda^4)}\epc \\
	\rho_{2}(\lambda) &= \frac{\lambda^2 (1 + 3 \lambda^2) (9 + 4 \lambda^2)}{2 \pi (1 + 2 \lambda^2) (2 + 7 \lambda^2 + 2 \lambda^4) (16 + 33 \lambda^2 + 9 \lambda^4)}\epc \\
	\rho_{3}(\lambda) &= \frac{32 (\lambda^2+4) (4\lambda^2+1)^2 (5 + 4 \lambda^2) (21 + 
   20 \lambda^2)}{\pi (19 + 12 \lambda^2) (9 + 2496 \lambda^2 + 4192 \lambda^4 + 2048 \lambda^6 + 256 \lambda^8) (509 + 520 \lambda^2 + 
   80 \lambda^4)}\epc\\
	\rho_{4}(\lambda) &= \frac{\lambda^2 (\lambda^2+1)^2 (4\lambda^2+25) (12 + 5 \lambda^2) (4 + 
   15 \lambda^2 + 5 \lambda^4) [36 + 143 \lambda^2 + 65 \lambda^4 + 6 \lambda^6]^{-1}}{2 \pi (2 + 7 \lambda^2 + 2 \lambda^4) (576 + 2100 \lambda^2 + 1465 \lambda^4 + 350 \lambda^6 + 25 \lambda^8)}\epp
\end{align}
\end{subequations}
In Fig.~\ref{fig:rhos_XXX} the (scaled) densities of the first four string types are plotted. Apart from the infinite interval, they qualitatively exhibit the same features as the densities for the N\'eel-to-XXZ quench~\cite{2014_Wouters}. The $1$-strings are dominant and even-length-string densities have a zero at $\lam=0$. The predictions of the GGE, where no such zero is visible, are plotted as well. Since $\rho_{1,h}$ is fixed by the initial conditions (see Sec.~\ref{sec:1to1}), it is exactly the same for the quench action steady state and the GGE. Hence, the difference between the two predictions of $\rho_1$ is small (of order $\rho_{2,h}$, see Eqs~\eqref{eq:BTGthlim_fact_a} for $n=1$). Note that the curves for $\rho_2$ in Fig.~\ref{fig:rhos_XXX} are scaled by a factor 40. 

\subsection{String content of the saddle-point state}
Given the analytical solution of the GTBA equations in terms of the densities, the ``spin content'' of the saddle-point state can be studied. We define the quantity
\begin{equation} \label{eq:def_I_n}
	I_n = n \,  \int_{-\infty}^{\infty} \mathrm{d}\lam \, \rho_{n}(\lam) \epc
\end{equation}
which is the number of rapidities that form $n$-strings, normalized by the system size $N$. In Tab.~\ref{table:spin_content_XXX} they are given for $n=1,2,\ldots,9$. They are obtained via numerical integration of the root densities of Sec.~\ref{sec:analytical_sol_XXX}. The sum of these fractions converges to $1/2$. From Eq.~\eqref{eq:infinity_rapidities} it then follows that $n_\infty=0$ for the steady state, meaning that only a vanishing fraction of the rapidities is infinite. Supporting evidence of this finding can be found in \ref{app:spin_content} where the spin content of the N\'eel state is studied.
\begin{table} [h]
\begin{center}
\begin{tabular}{|c|c|c|c|c|c|c|c|c|c|}
\hline
n & 1 & 2 & 3 & 4 & 5 & 6 & 7 & 8 & 9 \\ \hline 
$I_{n}$ & 0.3097 & 0.0295 & 0.0458 & 0.0121 & 0.0203 & 0.0066 & 0.0115 & 0.0041 & 0.0074  \\ \hline
\end{tabular}
\centering
\caption{The spin content of the steady state after the N\'eel-to-XXX quench. $I_n$ is defined in Eq.~\eqref{eq:def_I_n} and represents the number of rapidities that form $n$-strings, normalized by the system size $N$. Data given up to $9$-strings.}\label{table:spin_content_XXX}
\end{center}
\end{table}

\section{Exotic states} \label{sec:R:exotic}
In the derivation of the GTBA Eqs~\eqref{eq:TBA_XXZ_fact}, see Ref.~\cite{2014_Wouters} or \ref{app:GTBAQA}, a representative state is chosen for the class of states that scale to the same macrostate $\brho$ in the thermodynamic limit. For the overlap of this specific state with the N\'eel state, the part exponential in system size is extracted. This procedure is valid under the assumption that the extensive part of the overlap coefficient is well-defined, regardless of the specific choice for a representative state. String deviations as mentioned in Eq.~\eqref{eq:stringdef} might, however, produce additional extensive contributions to the overlap coefficients. This possibility will be investigated in this section, restricted to the N\'eel-to-XXX quench, by examining in particular the behavior of the system-size scaling of the N\'eel overlap for various exotic string configurations.

\subsection{Possible deformations of the GTBA equations}
Unlike the reduced expressions for matrix elements of spin operators~\cite{2005_Caux_JSTAT_P09003} containing Bethe states consisting of strings, no reduced form for the N\'eel overlaps in terms of string centers is available. Explicit evaluation of the N\'eel overlap \eqref{eq:overlap_XXX_onshell} for a Bethe state at finite system size consequently requires the inclusion of string deviations.

As an example, the overlaps of all parity-invariant Bethe states for $N=12$ are computed and listed in~\ref{app:overlaps_N_12}. This was done by solving the Bethe equations numerically by an iterative procedure for all possible string configurations at this system size, parametrized in equations for the string centers and deviations separately~\cite{2007_Hagemans_JPA_40}. The resulting rapidities for each Bethe state are used directly in the evaluation of the overlap~\eqref{eq:overlap_XXX_onshell}.

Extraordinary string configurations arise when multiple odd (or even) strings have coinciding string quantum numbers at zero. Their central rapidities are pushed away from $\lam=0$, yielding perfectly regular Bethe states with deviated rapidities on the real axis. For $N=12$ this happens, for example, for the Bethe state containing one $3$-string and three $1$-strings (see \ref{app:overlaps_N_12}). If these deviations on the real axis vanish exponentially, the denominator in the overlap formula~\eqref{eq:overlap_XXX_onshell} produces an extra factor that is exponential in system size,
\begin{equation}\label{eq:R:dangeroustermoverlap}
	\frac{\sqrt{\lambda_j^2+1/4}}{4\lambda_j} \sim \frac{1}{8 \lambda} \sim \frac{1}{e^{-\alpha N}} \epc
\end{equation}
where $\lambda$ is the real rapidity pushed away from the coinciding string centers at zero and $\alpha$ is some positive constant. More details on the behavior of $\lambda$ will be given in Sec.~\ref{sec:R:closerlook}.

Furthermore, these exponentially vanishing rapidities could, in principle, produce another exponential factor coming from the ratio of the determinants. It is {\it a priori} unclear, however, whether this second exponential factor exists and whether the two factors have exactly cancelling exponential behavior or, when combined, will produce an extra extensive contribution to the overlap coefficient. This extra contribution would deform the driving terms of the GTBA Eqs~\eqref{eq:TBA_XXX_fact} and would require a modification of the quench action approach that is presented here and in Ref.~\cite{2014_Wouters}.

At present, it is not possible to rule out the appearance of deformations of the driving terms categorically, as this would require a survey of an exponentially growing number of states for large system size. However, we shall look at some very simple examples of states where deformations might show up. Here, we consider states with one $1$-string and one $3$-string centered at zero and assume this is a prototypical example of coinciding strings at zero. The other rapidities are put in a Fermi-like sea of $1$-strings. Subsequently, the exponential behavior of the overlaps of this state is compared with the state without the $1$- and $3$-string centered at zero. 

The same types of states but with $\sqrt{N}/2$ rapidities at infinity (denoted by **) were also studied, as well as states where the sea of remaining $1$-strings is symmetrically divided in two and separated as far as possible (these states are denoted by ``extr''). The choice for $\sqrt{N}/2$ rapidities at infinity is motivated by the fact that the expectation value of the number of rapidities at infinity for the N\'eel state is of the same order, see~\ref{app:spin_content}.

Maximally dividing the Fermi sea of $1$-strings is unnatural and unlike the steady state, where the $1$-strings are clustered around zero. However, the assumptions of the quench action approach ought to be valid for all states and therefore examining their validity for this extremal type of state is useful. 
\begin{figure}[t]
\centering
\includegraphics[width=\textwidth]{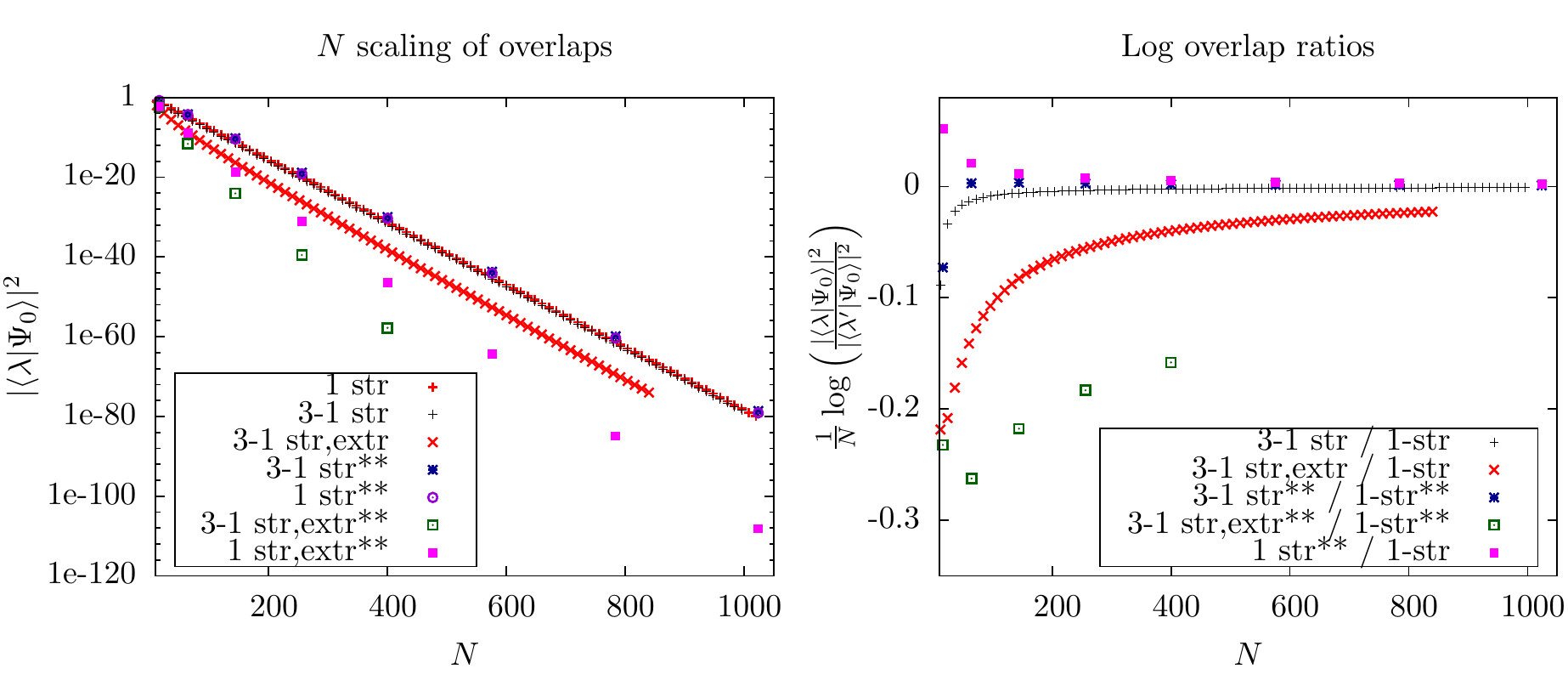}
\caption{Left: scaling of the N\'eel overlap squared with Bethe states of various string content, with both $N_\infty=0$ and $N_\infty=\sqrt{N}/2$ (denoted by **). The extremal case (denoted by ``extr'') refers to the configuration of one string quantum numbers put at the edges of the allowed range. Right: logarithm of the ratios between overlaps squared of a state with coinciding 1- and 3-strings with a state containing of only 1-strings.}
\label{fig:R:overlapplot}
\end{figure}

In Fig.~\ref{fig:R:overlapplot} the squared overlaps for the states described above are plotted as a function of system size. The overlaps were computed up to system size $N\sim 1000$ and the evaluation was done using arbitrary precision numerics due to divergencies in the determinants when encountering exponentially small string deviations. The scaling of the overlaps is indeed exponential in system size. Since all the considered states converge to the same macroscopic description in terms of densities $\brho$, ({\it i.e.}, they are representative states of the same $|\brho\rangle$), the extensive parts of the overlap coefficients are expected to be the same. To test this more thoroughly, we took two states $\ket{\blam}$ and $\ket{\blam'}$ of different type and plotted the difference between the extensive parts of their respective overlaps, up to finite size corrections, {\it i.e},
\begin{equation}
	\frac{1}{N} \ln \left( \frac{\left| \left\langle \blam | \Psi_0 \right\rangle \right|^2}{\left| \left\langle \blam' | \Psi_0 \right\rangle \right|^2} \right) \epp
\end{equation}
In the right panel of Fig.~\ref{fig:R:overlapplot} it can be observed that this quantity scales to zero for all different combinations of states considered here, indicating that the extensive part of the overlap coefficient is indeed universal. Note that for the maximally split Fermi seas the convergence is significantly slower and the range of data points is limited.

\subsection{A closer look at string deviations}\label{sec:R:closerlook}
In this section, the coinciding 1- and 3-string at the origin will be considered as a prototypical example of a coinciding string configuration, while for this case the behavior of the string deviations and important parts of the N\'eel overlap formula will be examined in more detail. Further parity-invariant Bethe states with exotic string configurations can be constructed by placing an even number of odd-strings or even-strings respectively at coinciding string quantum numbers at zero. The first example of two even-strings at the origin contains a 2- and a 4-string, whose overlap for $N=12$ can be found in \ref{app:overlaps_N_12}. This configuration with an even number of even-strings at the origin however contains no rapidities on the real axis and will be left outside of consideration in the further analysis.

A coinciding 1- and 3-string at the origin, obtained by placing their respective string quantum numbers at zero, can be parameterized as
\begin{subequations}
\begin{align}
	\lambda^{(3)}&=-\lambda^{(1)}=\lambda \epc \\
	\lambda^{(3,\pm)}&=\pm i(1+\delta^{(3)}) \epp
\end{align}
\end{subequations}
The real rapidities of the 1- and 3-string are pushed away from each other, described by the parameter $\lambda>0$. The 3-string deviations of the outermost rapdities are parametrized by $\delta^{(3)}$. A converging iterative procedure to obtain the roots of the Bethe equations \eqref{eq:BAE} for this case is obtained in Ref.~\cite{2007_Hagemans_JPA_40} by adding up the logarithmic form of the Bethe equations for $\lambda$ and $\delta^{(3)}$ and will be used here. Furthermore, we quote its result for the system-size scaling of real deviation $\lambda$ by approximating the Bethe equations for $\lambda \ll 1$ and $\delta \ll 1$,
\begin{equation}\label{eq:R:lambdascaling}
	\lambda=\sqrt{\frac{12}{F} }3^{-N/2}, \quad \text{where} \quad F=\prod_{\substack{ \lambda_\beta \not\in \{\pm\lambda,\lambda^{(3,\pm)}\}}} \frac{|\lambda_\beta|}{\sqrt{\lambda_\beta^2+4}},
\end{equation}
yielding intrinsically exponential behavior of $\lambda$ in Eq.~\eqref{eq:R:dangeroustermoverlap}. 
\begin{figure}[t]
\centering
\includegraphics[width=\textwidth]{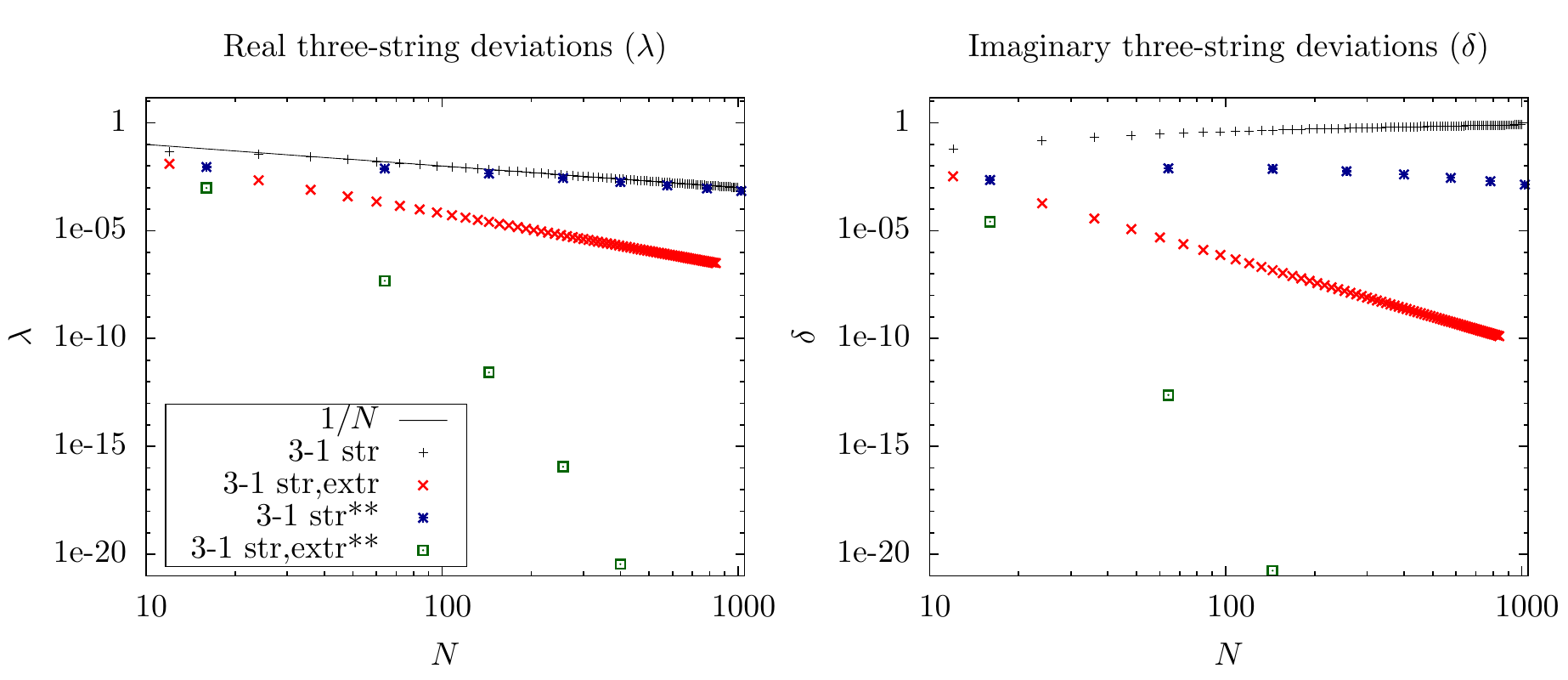}
\caption{Scaling of the coinciding 1- and 3-string deviations on the real axis $\lambda$ (left) and imaginary axis $\delta^{(3)}$ (right), with both $N_\infty=0$ and $N_\infty=\sqrt{N}/2$ (denoted by~**). The extremal case (denoted by ``extr'') refers to the configuration of 1-string quantum numbers put at the edges of the allowed range. }
\label{fig:R:deviationsplot}
\end{figure}
However, a macroscopic number of 1-strings contained in the scattering term $F$ can push the innermost rapidities further apart. Precisely this case is what we want to analyse. Therefore, we will obtain the Bethe roots by an iterative procedure for increasing system size. Figure~\ref{fig:R:deviationsplot} shows the results for the behavior of $\lambda$ and $\delta^{(3)}$ with respect to system size $N$ for distinguishing situations of no rapidities at infinity and $\sqrt{N}/2$ infinite rapidities.  For a macroscopic number of remaining 1-strings, the real string deviations scale algebraically with system size, in particular as $1/N$ when there are no infinite rapidities present in the Bethe state. For states containing a macroscopic number of 1-strings, the deviations $\delta^{(3)}$ turn out to be of~$\mathcal{O}(1)$, rendering the approximation in Eq.~\eqref{eq:R:lambdascaling} invalid. 

The configuration of the 1-strings is taken to be the Fermi sea in the former case, but putting the 1-strings further outwards to the edge of the sea results in a different effect on the scaling of the deviations. The number of free quantum numbers for holes is $2+N_\infty$, therefore the (positive, symmetric) quantum numbers for this case are $I^+_j=I^{+,\text{Fermi}}_j+2+N_\infty$. The deviations in this extremal case tend to scale much faster to zero.
the Gaudin-like determinants.
\begin{figure}[t]
\centering
\includegraphics[width=\textwidth]{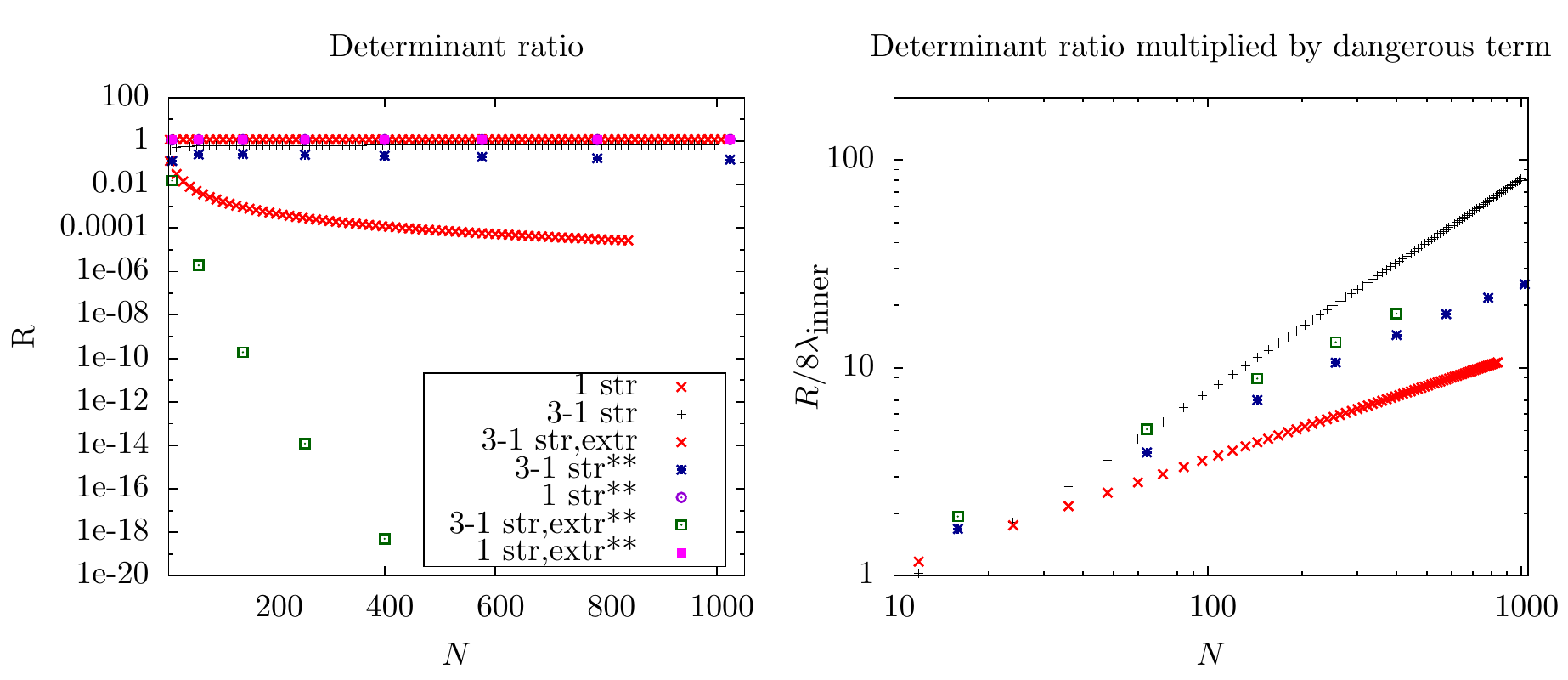}
\caption{Left: plot of the ratio of determinants in the N\'eel-overlap formula for different Bethe states at both $N_\infty=0$ and $N_\infty=\sqrt{N}/2$ (denoted by **). The extremal case (denoted by ``extr'') refers to the configuration of 1-string quantum numbers put at the edges of the allowed range. Right: Multiplication of the ratio $R$ of determinants with the possibly exponentially large term coming from a single factor of the prefactor of the overlap formula.}
\label{fig:R:ratioplot}
\end{figure}

Finally, we proceed with analyzing the system size scaling for separate parts of the N\'eel-overlap formula for a Bethe state. Figure~\ref{fig:R:ratioplot} (left panel) plots the square root of the ratio of the Gaudin-like determinants,
\begin{equation}
	R=\sqrt{\frac{\det_m(\hat G^+)}{\det_m(\hat G^-)}}.
\end{equation}
\begin{figure}[t]
\centering
\includegraphics[width=\textwidth]{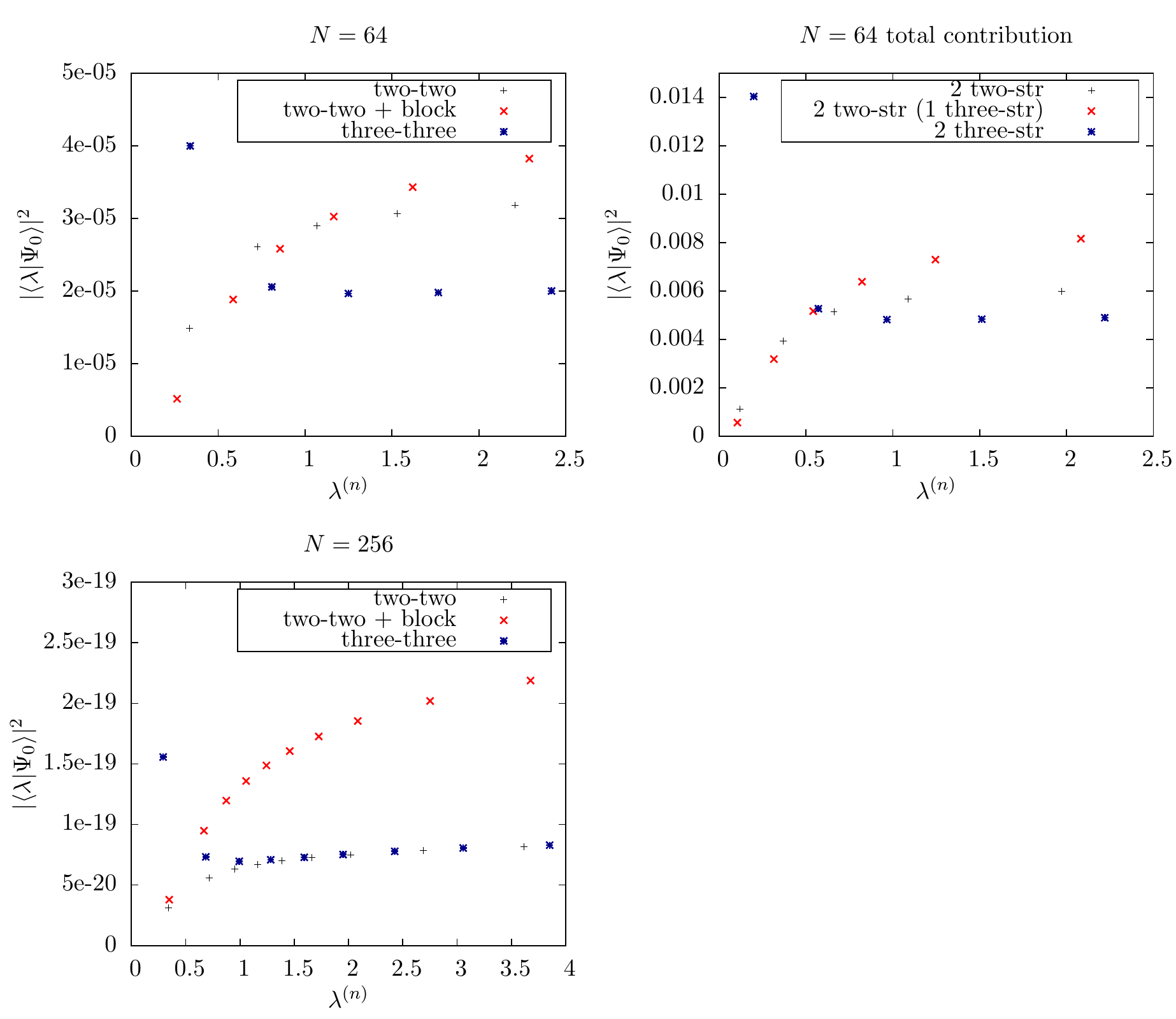}
\caption{Left (upper and lower): the overlaps for a state with a Fermi sea of 1-strings and, respectively, two 2-strings (black crosses), two 2-strings and a 1- and 3-string centered at zero (red crosses), and two 3-strings (blue crosses). The position of the symmetrically lying pair of 2- or 3-strings can vary and depends on the choice of quantum numbers for their string centers. The horizontal axis gives the position $\lambda^{(n)}$ of these (positive) string centers. Each data point represents the overlap of one Bethe state with the N\'eel state. For all states we have $N_{\infty}=\sqrt{N}/2$. Upper right: same as in the upper left panel, but now summed over all possible configurations of 1-strings.}
\label{fig:numerical_overlaps}
\end{figure}
For several cases the ratio $R$ can become exponentially small, in particular for the cases with (exponentially) small real deviations from a coinciding string configuration. The right panel of Fig.~\ref{fig:R:ratioplot} therefore multiplies the ratio $R$ with the possibly dangerous term from Eq.~\eqref{eq:R:dangeroustermoverlap}, $R/8\lambda_\text{inner}$, showing explicitly that the effect of exponentially small coinciding string deviations can be (at least algebraically) cancelled against the ratio of 

To summarize, from the analysis of this typical state there is no implication that the quench action approach presented in Sec.~\ref{sec:QAA_XXX} has to be modified, as the product of $R$ and $1/\lambda$ is always subleading in the thermodynamic limit. The leading part coming from the rest of the prefactor $\gamma$ remains universal and leads via the GTBA equations to the same saddle point state presented in this paper and in Ref.~\cite{2014_Wouters}. However, further numerical studies are needed to exclude the possibility that towers of strings and higher accumulations of rapidities around the origin lead to extra exponential contributions to the prefactor. That said, in view of the structure of the initial N\'eel state, in which downturned spins are never found in neighboring blocks, it is not expected that such degenerate string states develop a sufficiently large overlap to overhaul the contributions from regular strings.

An additional confirmation of the correctness of the quench action saddle-point state is presented in Fig.~\ref{fig:numerical_overlaps}. Here, we show the dependence of the overlap as function of the position $\lambda^{(n)}$ of one specific pair of string centers (either 2-strings or 3-strings). One can observe that the overlap vanishes if the center $\lambda^{(2n)}$ of an even-length string approaches zero. The behavior of the curves qualitatively agrees with the saddle-point distributions shown in Fig.~\ref{fig:rhos_XXX}.

\section{Conclusions}\label{sec:conclusions}
In this paper we reviewed and extended some of the results of Ref.~\cite{2014_Wouters}, where a quantum quench into the gapped regime $\Delta>1$ and to the isotropic point $\Delta=1$ of the integrable spin-1/2 XXZ chain was studied. Starting from the zero-momentum ground state of the anti-ferromagnetic Ising model, the steady state for long times after the quench was computed using the recently developed quench action method~\cite{2013_Caux_PRL_110,2014_DeNardis_PRA_89}, as well as physical spin-spin correlators on this steady state. It was shown that the GGE based on all known local conserved charges fails to give a correct description of the steady state for this particular quench.

Here, we gave a detailed account of how to compute the densities of roots predicted by the GGE based on all known local conserved charges, as was done in Ref.~\cite{2014_Wouters}. Note that in the meantime this method was also applied to the quench from the dimer state~\cite{2014_Pozsgay_GGE}. We showed that this method can easily be applied to any initial state that is of product form. Regarding the quench action approach, we investigated in more detail the derivation of the driving terms for the GTBA equations. By looking at specific examples of states with multiple strings centered at zero, we argued that the choice for a representative state is indeed valid.

One of the main results of this paper is the analytical solution of the quench action GTBA equations, which are found by solving related systems of functional equations, the Y- and T-system~\cite{1999_Suzuki, 1992_Kluemper}. Using this we derived explicit expressions for the Bethe root densities, which describe the quench action steady state. An interesting open question is how this approach can be extended to calculate spin-spin correlators and other physical observables.

Furthermore, we elaborated in great detail on solving the GTBA equations of both the quench action approach and the GGE, and on computing spin-spin correlation functions in terms of a large-$\Delta$ expansion. All evaluated orders of the expansion for the root densities of the quench action steady state are in perfect agreement with the analytical solution. The expansions for GGE distributions and for correlators prove very useful as a check for numerical computations. The large-$\Delta$ expansion also confirms the correct prediction of the conserved charges by the quench action method and the vanishing of the quench action on its steady state solution. In addition, it gives analytical evidence and an order-of-magnitude estimation of the differences between the quench action and GGE predictions, in particular for local spin-spin correlation functions.

Finally, we also presented the analysis of the N\'eel-to-XXX quench, which shows the same qualitative features as the quenches to the gapped regime.

These results, in combination with~\cite{2013_Caux_PRL_110,2014_DeNardis_PRA_89,2014_Wouters,2014_Bertini_Sinegordon,2014_Pozsgay_Dimer}, establish the broad applicability of the quench action approach to integrable quantum systems. This method, which is based on first principles, turns out to be a powerful way to predict the postquench steady state. It would be interesting to extend its range further, for example to the gapless regime $-1<\Delta
<1$, to different initial states~\cite{2014_Piroli_resursive_overlaps}, or to non-translationally invariant initial states whose steady state is believed to exhibit currents. Furthermore, in order to improve our understanding of the dynamics of integrable quantum systems, studying the postquench time evolution by means of the quench action approach could reveal some similarly unexpected physical behavior.

At a more fundamental level, the research conducted here and in Refs~\cite{2014_Wouters,2014_Pozsgay_Dimer} has raised the question of the validity and the general applicability of the GGE for interacting integrable quantum systems. We stress that in these studies the GGE was based on all known local conserved charges, but little is known about the exhaustiveness of this list of charges and whether and how quasi- and nonlocal charges could affect the steady state. The report~\cite{2013_Prosen_PRL_111,2014_Prosen_quasilocal_charges,2014_Pereira_quasilocal_charges} of so-called quasilocal exactly conserved charges for the spin-1/2 XXZ chain could be an interesting first step in this direction.

In Refs~\cite{2014_Goldstein_failure_GGE,2014_Pozsgay_GGE} the failure of the GGE was tied to the existence of bound states, since due to the appearance of strings the local conserved charges alone do not fully determine the root densities of the steady state. This is of course a necessary condition for failure of the GGE, but we do not believe it to be a sufficient one. In its essence, the GGE is a statistical ensemble that is determined by maximization of the (Yang-Yang) entropy, while the conserved charges only constrain this maximization procedure. In principle, including other (non)local charges could shift the extremum and lead to a correct steady-state prediction.

Answers to these pressing open problems are likely to yield new fundamental insights into the physics of integrable quantum systems and, in particular, their out-of-equilibrium phenomena.

\section*{Acknowledgements}
We would like to thank 
P. Calabrese,
F.~Essler, 
M.~Fagotti,
F.~G{\"o}hmann, 
V.~Gritsev, 
A.~Kl{\"u}mper, 
R.~Konik, 
M.~Kormos, 
B.~Pozsgay, 
J.~Suzuki, and
A.~Tsvelik 
for useful discussions. We acknowledge support from the Foundation for Fundamental Research on Matter (FOM), the Netherlands Organisation for Scientific Research (NWO). For their support and hospitality, MB and J-SC thank the Perimeter Institute, and BW, JDN and J-SC thank CUNY (where the main results of this work were first made public). This work forms part of the activities of the Delta Institute for Theoretical Physics (D-ITP).

\appendix

\section{Derivation of GTBA equations for GGE} \label{app:GTBAGGE}
To derive the GTBA equations for the GGE, which was done in Ref.~\cite{2012_Mossel_JPA_45} for the Lieb-Liniger model, we start from its definition in Eq.~\eqref{eq:defGGE} and assume that for a given initial state $\ket{\Psi_0}$ the chemical potentials are determined such that Eqs~\eqref{eq:GGEconstraints} holds. In the thermodynamic limit the trace over the full Hilbert space can be replaced by a functional integral over the root densities,
\begin{subequations}
\begin{equation}\label{eqnGGEfuncintegral}
	\left\langle\brho^{\Psi_0} \right| \cO \left| \brho^{\Psi_0}\right\rangle = \frac{1}{Z_{GGE}} \text{Tr}\left( \cO\, e^{- \sum_{m=1}^\infty \beta_{m} Q_{m}} \right)= \int \mathcal{D}\boldsymbol{\rho} \:\: \cO [\brho]\,  e^{- N d_{GGE} [\boldsymbol{\rho}] + S_{YY}[\boldsymbol{\rho}]} \epc
\end{equation}
where the term $d_{GGE}$ in the exponent is given by
\begin{equation}
	d_{GGE}[\boldsymbol{\rho}] = \frac{1}{N}\sum_{m=1}^\infty \beta_m Q_m[\brho] \epp
\end{equation}
\end{subequations}
This functional integral can be approximated by its saddle point. So, the GGE for integrable models is given by a set of GTBA equations whose solution is the set $\boldsymbol{\rho}$ of root densities that maximizes the entropy under the constraint that expectation values of the local conserved charges are fixed by the initial conditions~\cite{2012_Mossel_JPA_45}. The solution can be found by minimizing the effective generalized free energy per lattice site
\begin{equation}
	f^{GGE}  [\boldsymbol{\rho}] =    d_{GGE} [\boldsymbol{\rho}]  - \sum_{n=1}^\infty \int_{-\pi/2}^{\pi/2} \mathrm{d}\lambda\left[ \rho_n(\lambda) \ln( 1+ \eta_n(\lambda)) + \rho_{n,h}(\lambda) \ln( 1+ \eta_n^{-1}(\lambda)) \right]\epp
\end{equation}

For the XXZ model, $d_{GGE}$ can be rewritten as a functional of ${\rho}_{1,h}(\lambda)$ only,
\begin{align}\label{eq:dGGE_calc}
 	d_{GGE} [\boldsymbol{\rho}] & = \sum_{n=1}^\infty \int_{-\pi/2}^{\pi/2} \mathrm{d}\lambda \: \rho_n(\lambda) \sum_{m=1}^\infty \beta_{m} c_{m}^{(n)}(\lam)  \nonumber \\
 		&= \frac{1}{\pi} \sum_{n=1}^\infty  \sum_{k \in \mathbb{Z} }\widehat{\rho}_n(k)\sum_{m=2}^\infty \beta_{m} \widehat{c}_{m}^{(n)}(k) \nonumber \\
		&= \sum_{k \in \mathbb{Z} } \frac{\widehat{\rho}_{1,h}(k) - e^{- |k| \eta}}{2 \cosh (k \eta)}  \sum_{m=2}^\infty  \beta_{m} \sinh^{m-1}(\eta) (ik)^{m-2}  \epc
\end{align}
where the $c_m^{(n)}$ are defined in Eq.~\eqref{eq:defcn} and we used their Fourier transforms~\eqref{eq:cnFourier}. Note that a term involving $\beta_1$ does not appear as we restrict our analysis to the zero-total-momentum sector, {\it i.e.}, $0=\limth\langle\blam |Q_1/N|\blam\rangle =\sum_{n=1}^\infty \int_{-\pi/2}^{\pi/2} \mathrm{d}\lambda \: \rho_n(\lambda) c_{1}^{(n)}(\lam) $ in the first step of Eq.~\eqref{eq:dGGE_calc}. We conclude that the full GGE solution, obtained by including all known local conserved charges, corresponds to the set $\brho$ that maximizes the entropy under the constraint of fixing the density of holes for the 1-strings, ${\rho}_{1,h}(\lambda) =\rho_{1,h}^{\Psi_0}(\lambda)$. 

To minimize the generalized free energy it is convenient to work in Fourier space. We vary with respect to the $\brho_h$ and constrain the $\brho$ in terms of the hole densities using the Bethe Eqs~\eqref{eq:BTGthlim_fact}, {\it i.e.},
\begin{subequations}
\begin{align}
	\delta \hat{\rho}_1(k) &= \frac{1}{2 \cosh (k \eta)} ( 1+ \delta \hat{\rho}_{2,h}  )  - \delta \hat{\rho}_{1,h} \epc\\
	\delta \hat{\rho}_n(k) &= \frac{1}{2 \cosh (k \eta)} ( \delta \hat{\rho}_{n-1,h} + \delta \hat{\rho}_{n+1,h} ) - \delta \hat{\rho}_{n,h} \epc \qquad \text{for }n\geq 2 \epp
\end{align}
\end{subequations}
Variation of the generalized free energy gives the condition
\begin{subequations}
\begin{align}
	\delta f^{GGE} = & \sum_{k \in \mathbb{Z} } \frac{\widehat{d}(k) }{2 \cosh (k \eta)}  \delta \hat{\rho}_{1,h} (k) \nonumber \\
		& - \sum_{n=1}^\infty \sum_{k \in \mathbb{Z} } \left[   \delta\hat{\rho}_n(k) \ftk{\ln( 1+ \eta_n)}{k} + \delta \hat{\rho}_{n,h}(k) \ftk{\ln( 1+ \eta_n^{-1})}{k} \right]= 0 \epc
\end{align}
where we defined 
\begin{equation}
	\widehat{d}(k) = \sum_{m=2}^\infty  \beta_{m} \sinh^{m-1}(\eta) (ik)^{m-2}  \epp
\end{equation}
\end{subequations}
After some manipulations we arrive at the GTBA equations in Fourier space
\begin{subequations}
\begin{align}
	\ftk{\ln \eta_1}{k} &  = -\frac{  \widehat{d}(k) }{2 \cosh(k\eta)}   + \frac{1}{2 \cosh (k \eta)} \ftk{\ln( 1+ \eta_2)}{k} \epc \\
	\ftk{\ln \eta_n}{k} & =  \frac{1}{2 \cosh (k \eta)} \big\{ \ftk{\ln( 1+ \eta_{n-1})}{k}  + \ftk{\ln( 1+ \eta_{n+1})}{k}   \big\} \epc
\end{align}
\end{subequations}
which can be rewritten in $\lambda$-space as
\begin{subequations} 
\begin{align}\label{eqn:GGE_eqn}
	[(a_0 + a_2)\ast \ln(\eta_1)](\lambda) &= - (a_1 \ast d)(\lambda)   + [a_1 \ast \ln( 1+ \eta_2)](\lambda) \epc \\
	[(a_0 + a_2)\ast \ln(\eta_n)](\lambda) &=  [a_1 \ast\ln( 1+ \eta_{n-1})](\lambda)  + [a_1\ast\ln( 1+ \eta_{n+1})](\lambda) \epp
\end{align}
\end{subequations}
Together with the Bethe equations, they uniquely determine the full GGE solution for the quench problem, provided the values of the chemical potentials are known. Another formulation of the GTBA equations for the GGE is given in Eqs~\eqref{eq:GTBAGGE}.

\section{GTBA equations for the N\'eel-to-XXZ quench} \label{app:GTBAQA}
In this section we derive the GTBA equations for the N\'eel-to-XXZ quench, as prescribed by the quench action method. Furthermore, we put the GTBA equations in the more convenient partially decoupled form. This derivation was presented earlier in Ref.~\cite{2014_Wouters}. Since elements of this calculation are needed in Sec.~\ref{sec:R:exotic} and for the sake of completeness, we repeat this derivation here.

\subsection{Thermodynamic limit of the overlaps}
For the implementation of the quench action approach the leading extensive parts of the overlap coefficients in the thermodynamic limit are needed, 
\begin{equation}
	S[\boldsymbol{\rho}]= \limth S_{\boldsymbol{\lambda}} = - \limth \ln \frac{\langle \Psi_0 | \{\pm\lambda_j\}_{j=1}^{M/2} \rangle }{\| |\{\pm\lambda_j\}_{j=1}^{M/2}\rangle \|} \epp
\end{equation}
One needs to consider the overlap coefficient for a generic finite size Bethe state $| \{ \lambda_j\}_{j=1}^M \rangle$ that in the thermodynamic limit, $N \to \infty$ with $ M/N=1/2$ fixed, flows to a set of densities $| \{ \lambda_j\}_{j=1}^M \rangle \to | \boldsymbol{\rho} \rangle $. This means that in the thermodynamic limit the eigenvalue of a smooth diagonal observable $\mathcal{A}$ is determined by a sum of integrals weighted by the distributions $\boldsymbol{\rho}  = \{ \rho_n\}_{n=1}^\infty$:
\begin{equation}\label{eqn:therm_observable}
	\mathcal{A} | \{ \lambda_j\}_{j=1}^M \rangle   =\Big[\sum_{j=1}^M A_j \Big]| \{ \lambda_j\}_{j=1}^M \rangle   \to \Big[N \sum_{n=1}^\infty \int_{- \pi/2}^{\pi/2} \mathrm{d}\lambda \:\: \rho_n(\lambda) \tilde{A}_n(\lambda) \Big]  | \boldsymbol{\rho} \rangle \epp
\end{equation}
It is assumed that the extensive part of the overlap coefficients $S[\boldsymbol{\rho}]$ is smooth and Bethe states that scale to the same densities $\brho$ have the same extensive part, regardless of finite-size differences. Each set of distributions $\boldsymbol{\rho}$ represents a number of Bethe states that is given by the extensive Yang-Yang entropy~\eqref{eq:YYentropyXXZ}: $e^{S_{YY}[\boldsymbol{\rho}]}$. To determine $S[\brho]$, we are then free to select a representative finite size Bethe state from the set of states that scale to the same $\brho$. Let us choose as a representative state $| \{ \lambda_j\}_{j=1}^M \rangle $ one consisting of $2n_s$ strings such that $2n_s = \sum_{n=1}^\infty M_n$, where $M_n$ is the number of $n$-strings and we choose all $M_n$ to be even. Note that different choices for the fillings $\{M_n \}_{n=1}^\infty$ lead to different expressions for the exact overlap formula~\eqref{eqn:Overlap_Exact}, but are believed~\cite{2013_Caux_PRL_110} to have the same extensive smooth part $S[\boldsymbol{\rho}]$. In Sec.~\ref{sec:R:exotic} additional evidence in the case of some very simple Bethe states was given.

For any finite size $N$, the string hypothesis tells us that Bethe states are organized in deviated strings. We label the rapidities of such states as follows,
\begin{equation}
	\lambda_j \to \lam^{n,a}_\alpha =   \lambda^{n}_\alpha + \tfrac{i\eta}{2}(n+1 -2 a) + i\delta^{n, a}_\alpha \epc
\end{equation}
where $a= 1, \ldots n$ and $\alpha= 1, \ldots , M_n$. In the thermodynamic limit the string deviations $\delta^{n, a}_\alpha$ vanish. Although the string hypothesis is not systematically verified around the ground state of the zero-magnetized spin chain~\cite{1982_Woynarovich_JPA_15,1983_Babelon_NPB_220_1}, it has been effectively verified away from the ground state, for example at finite temperatures~\cite{1983_Tsvelik_AP_32}. Since the non-thermal steady state we obtain is far away from the ground state, by extension the string hypothesis is valid here as well. 

The finite size overlap formula between the N\'eel state and our class of representative states can be written as~\cite{2014_Brockmann_JPA},
\begin{equation}\label{eqn:Overlap_Exact}
	\frac{\langle \Psi_0 | \{\pm\lambda_j\}_{j=1}^{M/2} \rangle }{\| |\{\pm\lambda_j\}_{j=1}^{M/2}\rangle\|}=  \gamma\ \sqrt{ \frac{\det_{M/2}(G^{+})}{\det_{M/2}(G^{-})}} \quad \text{with}\ \gamma = \sqrt{2} \prod_{j=1}^{M/2}\frac{\sqrt{\tan(\lambda_j+\tfrac{i\eta}{2}) \tan(\lambda_j - \tfrac{i\eta}{2})}}{2\sin(2\lambda_j)} \epp
\end{equation}
For our representative state the prefactor $\gamma$ has to leading order no explicit system size dependence from the string deviations $\delta \to 0$, but is exponentially vanishing when the particle number $M$ is sent to infinity due to the product over all rapidities.

For the moment, let us focus on the ratio of the two determinants, where the matrices are given by
\begin{equation}
	G^\pm_{(n,\alpha,a),(m,\beta,b)} = \delta_{(n,\alpha,a),(m,\beta,b)}\Big[NK_{\eta/2}(\lambda^{n,a}_\alpha)-\sum_{(\ell,\gamma,c)}K_\eta^+(\lambda^{n,a}_\alpha, \lambda^{\ell,c}_\gamma)\Big]+ K_\eta^\pm(\lambda^{n,a}_\alpha ,\lambda^{m,b}_\beta) \epp
\end{equation}
Here, $K_\eta^\pm(\lambda,\mu)=K_\eta(\lambda-\mu) \pm K_\eta(\lambda+\mu)$ and $K_\eta(\lambda)=\sinh(2\eta)/[\sin(\lambda+i\eta)\sin(\lambda-i\eta)]$.
One finds divergencies in system size going like $1/\delta$ in each string block $(n=m,\alpha= \beta)$ when $b= a+1$ in the term $K_\eta(\lambda^{n,a}_\alpha - \lambda^{n,a+1}_\alpha) \sim i/(\delta^{n, a+1}_\alpha -  \delta^{n, a}_\alpha )$. On the other hand, for our representative state with all $M_n$ even the terms $\pm K_\eta(\lambda+\mu)$ in $G^\pm$ are never divergent, since all string centers in the matrices $G^\pm_{jk}$ are strictly positive. The divergencies in $1/\delta$ in $\det_{M/2}(G^+)$ will therefore cancel exactly the divergencies in $\det_{M/2}(G^-)$, as they occur in precisely the same form. A similar cancellation occurs for divergencies appearing in $K_\eta(\lambda- \mu)$, when two rapidities from different strings get close in the thermodynamic limit $\mu \to \lambda \pm i \eta + g(N)$ with $\limth g(N) = 0$. The thermodynamic limit $\limth$ for the overlap coefficients can thus be performed analogously to Ref.~\cite{2014_DeNardis_PRA_89}. 

Since non-exponential in system size, the contribution from the ratio of the two determinants is non-extensive and therefore negligible. The thermodynamic overlap coefficients are then given by
\begin{equation}
	S[\brho] = \limth S_{\boldsymbol{\lambda}}  =   \frac{N}{2} \sum_{n=1}^\infty \int_{0}^{\pi/2} \mathrm{d}\lam \, \rho_n(\lam) \big[ g_n(\lam) + 2n \ln (4) \big]  \epc
\end{equation}
where
\begin{subequations}\label{eq:gTBA}
\begin{align}\label{eq:gTBAa}
	g_n &= \sum_{l=0}^{n-1} \ln\left[\frac{s_{n-1-2l}c_{n-1-2l}s_{-n+1+2l}c_{-n+1+2l}}{t_{n-2l}t_{-n+2l}}\right]\epc \\
	t_n &= \frac{s_n}{c_n}\epc\quad s_n(\lam) = \sin\left(\lam+\tfrac{i\eta  n}{2}\right)\epc\quad c_n(\lam) = \cos\left(\lam+\tfrac{i\eta n}{2}\right)\epp
\end{align}
\end{subequations}

\subsection{Derivation of GTBA equations}
In this section we focus on the derivation of the saddle point state, specified by the set of distribution $\brho^\text{sp}$ obtained by varying the quench action $S_{QA}\left[ \brho \right] = 2 S[\brho] -\tfrac{1}{2} S_{YY}\! \left[ \brho \right]$ with respect all root densities. Since only states in the magnetization sector $\langle\sigma_\text{tot}^z\rangle/2 = N/2-M=0$ have nonzero overlap with the initial N\'eel state, we need to add a Lagrange-multiplier term to the quench action in order to vary with respect to all $\rho_n(\lam)$ independently,
\begin{equation}
	-h\,N \left( \sum_{m=1}^{\infty} m \,  \int_{-\pi/2}^{\pi/2} \mathrm{d}\lam \, \rho_{m}(\lam) - \frac{1}{2}  \right) \epc
\end{equation}
where $h$ is the Langrange multiplier. For the variation of the Yang-Yang entropy the BGT Eqs~\eqref{eq:BTGthlim_fact} can be used~\cite{KorepinBOOK}. In front of the Yang-Yang entropy there is an unusual factor $1/2$. Since only parity-invariant Bethe states contribute, the number of microstates in the ensemble $\brho$ is the square root of the usual number. The saddle-point conditions are then obtained through variation with respect to $\rho_n(\lam)$,
\begin{equation} \label{eq:TBA_XXZ}
	\ln[\eta_{n}(\lam)]  = 2n \left[ \ln(4) - h \right] +  g_{n}(\lam)  + \sum_{m=1}^{\infty}  a_{nm} \ast \ln \left( 1 + \eta_{m}^{-1} \right) (\lam) \epc
\end{equation}
where $n\geq 1$. The parts $2n[\ln(4)-h]+g_n$ are called driving terms. For each fixed value of $h$ this set of GTBA equations has a solution in terms of the functions $\eta_n$. Substituting these into thermodynamic Bethe Eqs~\eqref{eq:BTGthlim_fact} leads to the saddle point distribution $\brho^\text{sp}$. Subsequently, the parameter $h$ is fixed by the zero-magnetization condition of the initial state,
\begin{equation}
	\sum_{m=1}^{\infty} m \,  \int_{-\pi/2}^{\pi/2} \mathrm{d}\lam \, \rho_{m}^\text{sp}(\lam) = \frac{1}{2} \epp
\end{equation}

\subsection{Partially decoupled GTBA equations}
It is often convenient to work with a form of the GTBA equations where there is no infinite sum over string types. We will derive this partially decoupled form, as was already done for the TBA equations at finite temperature~\cite{TakahashiBOOK}. The Fourier transform [Eqs~\eqref{eq:FourierTransform}] of the kernels in Eq.~\eqref{eq:kernelXXZ} is $\hat{a}_{n,k} =e^{-|k|n\eta}$ and, using the convolution theorem, this implies $a_m \ast a_n = a_{m+n}$. From this a set of identities for the kernels follows easily~\cite{TakahashiBOOK}
\begin{subequations} \label{eq:kernelidentities}
\begin{equation}
	(a_0+a_2)\ast a_{nm} = a_1\ast (a_{n-1,m}+a_{n+1,m}) + (\delta_{n-1,m}+\delta_{n+1,m})\,a_1\epc \quad n>1,\, m \geq 1 \epc
\end{equation}
and
\begin{equation} \label{eq:2ndkernelidentity}
	(a_0+a_2)\ast a_{1,m} = a_1 \ast a_{2,m} + a_1 \, \delta_{2,m}  \epc \quad m\geq 1 \epc
\end{equation}
\end{subequations}
where we used the convention $a_0(\lam) = \delta(\lam)$. The infinite sum over string types can be removed by convolving the GTBA Eqs~\eqref{eq:TBA_XXZ} with $(a_0 + a_2)$,
\begin{equation}\label{eq:gTBA_fact_prev}
	(a_0+a_2)\ast\ln(\eta_n) = (a_0+a_2)\ast g_n - a_1\ast(g_{n-1}+g_{n+1}) + a_1 \ast \big[\ln(1+\eta_{n-1})+\ln(1+\eta_{n+1})\big] \epp
\end{equation}
Defining $g_0 (\lam) = 0$ and $\eta_0 (\lam) = 0$, Eq.~\eqref{eq:gTBA_fact_prev} holds for $n\geq 1$. In order to rewrite the new driving terms $\tilde{d}_n =  (a_0+a_2)\ast g_n - a_1\ast(g_{n-1}+g_{n+1})$, we first rewrite $g_n$ such that only positive indices are present:
\begin{align}
	g_n = 2\delta_{n\,\text{mod}\,2,1}\ln\left[s_0^{(2)}\right] + 4 \sum_{l=1}^{\lfloor n/2 \rfloor}\ln\left[s_{n+1-2l}^{(2)}\right] + 2 \sum_{l=1}^{n-1}\ln\left[\frac{c_{l}^{(2)}}{s_{l}^{(2)}}\right] + \ln\left[\frac{c_{0}^{(2)}}{s_{0}^{(2)}}\right] + \ln\left[\frac{c_{n}^{(2)}}{s_{n}^{(2)}}\right] \epc
\end{align}
where $s_l^{(2)} = s_ls_{-l}$, $c_l^{(2)} = c_lc_{-l}$ or, explicitly,
\begin{subequations}
\begin{align}
	s_l^{(2)}(\lam) &= \sin\left(\lam+\tfrac{i\eta}{2}l\right)\sin\left(\lam-\tfrac{i\eta}{2}l\right) = \sin^2\left(\lam\right)+\sinh^2\left(\tfrac{\eta l}{2}\right)\epc\\
	c_l^{(2)}(\lam) &= \cos\left(\lam+\tfrac{i\eta}{2}l\right)\cos\left(\lam-\tfrac{i\eta}{2}l\right) = \cos^2\left(\lam\right)+\sinh^2\left(\tfrac{\eta l}{2}\right)\epp
\end{align}
\end{subequations}
Now we use that for $\tilde{a}_\alpha(\lam) = (2\pi)^{-1}\sinh(2\alpha)/[\sin^2(\lam)+\sinh^2(\alpha)]$ and $f_\beta(\lam) = \ln\left[\sin^2(\lam)+\sinh^2(\beta)\right]$ the following relation holds ($\alpha,\beta >0$):
\begin{equation}
	\tilde{a}_{\alpha}\ast f_\beta = f_{\alpha+\beta}-2\alpha\epp
\end{equation}
Similarly, for $g_\beta(\lam) = \ln\left[\cos^2(\lam)+\sinh^2(\beta)\right]$ we find $\tilde{a}_{\alpha}\ast g_\beta = g_{\alpha+\beta}-2\alpha$. From this we can calculate $\tilde{d}_{2n}$ and $\tilde{d}_{2n-1}$ for all $n\geq 1$:
\begin{equation}
	\tilde{d}_{2n} =  \ln\left[\frac{c_0^{(2)}}{c_2^{(2)}}\right] - \ln\left[\frac{s_0^{(2)}}{s_2^{(2)}}\right] \epc\qquad
	\tilde{d}_{2n-1} =\ln\left[\frac{c_0^{(2)}}{c_2^{(2)}}\right] + \ln\left[\frac{s_0^{(2)}}{s_2^{(2)}}\right] \epc
\end{equation}
where we used the identities
\begin{equation}
	a_m\ast \ln\left[\frac{c_l^{(2)}}{s_l^{(2)}}\right] = \ln\left[\frac{c_{l+m}^{(2)}}{s_{l+m}^{(2)}}\right] \epc\quad a_l\ast \ln\left[\frac{s_{0}^{(2)}}{s_{2}^{(2)}}\right] = \ln\left[\frac{s_{l}^{(2)}}{s_{l+2}^{(2)}}\right]\epc\quad a_l\ast \ln\left[\frac{c_{0}^{(2)}}{c_{2}^{(2)}}\right] = \ln\left[\frac{c_{l}^{(2)}}{c_{l+2}^{(2)}}\right] \epp
\end{equation}
\begin{subequations}
More explicitly, the driving terms are given by
\begin{equation}
	\tilde{d}_{n}(\lam) = \ln\left[\frac{\cos^2(\lam)}{\cos^2(\lam)+\sinh^2(\eta)}\right] - (-1)^n \ln\left[\frac{\sin^2(\lam)}{\sin^2(\lam)+\sinh^2(\eta)}\right]\epp
\end{equation}
and the GTBA equations can be written compactly as
\begin{equation}\label{eq:TBA_XXZ_fact0}
	(a_0+a_2)\ast\ln(\eta_n) = \tilde{d}_n + a_1 \ast \big[\ln(1+\eta_{n-1})+\ln(1+\eta_{n+1})\big]\epc
\end{equation}
where $n\geq 1$, the $\lam$-dependence is left implicit and by convention $\eta_0(\lam)= 0$ and $a_0(\lam)=\delta(\lam)$. 
\end{subequations}
The operation of $(a_0+a_2)\ast$ can be inverted and brougth to the right hand side of Eq.~\eqref{eq:TBA_XXZ_fact0} by another application of the convolution theorem. The Fourier transformed driving terms are
\begin{equation}
	\hat{\tilde{d}}_{n,k} = 2\pi \frac{(1-e^{-2|k|\eta})}{|k|}\left[\frac{(-1)^n-(-1)^k}{2}\right] \epp
\end{equation}
Defining 
\begin{align}
	\hat{d}_{n,k} &= \frac{\hat{\tilde{d}}_{n,k} }{\hat{a}_{0,k}+\hat{a}_{2,k}} = 2\pi \frac{\tanh(k\eta)}{k}\left[\frac{(-1)^n-(-1)^k}{2}\right]\epc \nonumber \\
	\hat{s}_k & = \frac{\hat{a}_{1,k}}{\hat{a}_{0,k}+\hat{a}_{2,k}} = \frac{1}{2\cosh(k\eta)} \epc
\end{align}
the GTBA equations in Fourier space are
\begin{equation}
	\ft{\ln(\eta_n)}(k) = \hat{d}_{n,k} + \hat{s}_k\Big( \ft{\ln(1+\eta_{n-1})}(k) + \ft{\ln(1+\eta_{n+1})}(k) \Big) \epp
\end{equation}
After applying the inverse Fourier transform, this eventually leads to Eqs~\eqref{eq:TBA_XXZ_fact}.

\section{Large-$\Delta$ expansion of the saddle-point state.} \label{app:SP_EXP}
In this appendix we would like to discuss briefly the derivation of the large-$\Delta$ expansion for the saddle-point state. In particular, we would like to discuss the derivation of the leading term of the expansion of $\eta_n$, which is the non-straightforward point of this calculation. As stated in Sec.~\ref{sec:large_Delta},  we need to expand the GTBA Eqs~\eqref{eq:TBA_XXZ_fact} and the BGT Eqs~\eqref{eq:BTGthlim_fact}. We assume the following analytical ansatz for $\{\eta_{n}(\lam)\}$ 
\begin{equation}
	\eta_{n}(\lam)\ =\ z^{\alp_{n}} \eta_{n}^{(0)}(\lam)\, \exp \left[ \Phi_{n}(\lam) \right] \epc \qquad \Phi_{n}(\lam)\ \equiv\ \sum_{j=1}^{\infty} z^{j}\, \eta_{n}^{(j)}(\lam) \epc \quad n\geq 1,
\end{equation}
where $z=e^{-\eta}$, $\Delta = \cosh{\eta}$, and $\alpha_n$ are integer numbers. The functions $\eta_n^{(j)}(\lam)$ with $j=0,1,2,\ldots$  characterize the solution at order $z^j$ in the expansion. From the leading behaviors of $\rho_1$ and of the exact solution \eqref{eq:rho1h_exact_XXZ} for $\rho_{1,h}$, we know that $\alpha_1=2$. This is the only information about $\rho_{1, h}$ we use in our expansion. 
The driving terms $\tilde{d}_n(\lam)$ in Eqs~\eqref{eq:TBA_XXZ_fact} have a very simple expansion in $z$,
\begin{equation}\label{eq:driving_fact}
	\tilde{d}_{n}(\lam) = \left\{
\begin{array}{ll}
4 \ln z + \ln \left( 4 \, \sin^{2}(2\lam) \right) + 2 \sum_{k=1}^{\infty}\frac{1}{k} \cos(4 k \lam) z^{4k} \epc & \qquad  n \text{ odd} \epc \\
\\
- \ln \tan^{2}(\lam)  - 4 \sum_{k=1}^{\infty}\frac{1}{2k-1} \cos[2 (2k-1) \lam] z^{2(2k-1)} \epc & \qquad n \text{ even} \epp
\end{array}\right.\phantom{\}}
\end{equation}
The leading order of the small-$z$ expansion of Eqs~\eqref{eq:TBA_XXZ_fact} is a $\ln(z)$-divergence. Since $\rho_{1,h}(\lambda)$ in Eq.~\eqref{eq:rho1h_exact_XXZ} does not exhibit exponential behavior in $\lambda$, we expect (possible) divergencies in $\eta_n(\lam)$ to be power law. This means that for the convolutions of the right-hand side of Eqs~\eqref{eq:TBA_XXZ_fact}
\begin{equation}
	s \ast \ln (1 + \eta_n) = s \ast \ln\left(1 + z^{\alpha_n} \eta_n^{(0)}\right) + O(z) = \Theta(-\alpha_n) \alpha_n + O(z^0) \epc
\end{equation}
where $\Theta(x)$ is the Heaviside step function. This leads to a set of conditions on the parameters $\alpha_n$,
\begin{align}
	2 \alpha_1 & = 4 + \Theta(-\alpha_2)\,\alpha_2 \epc \nonumber \\
	2 \alpha_n & = \Theta(-\alpha_{n-1})\,\alpha_{n-1} + \Theta(-\alpha_{n+1})\,\alpha_{n+1}  \epc \qquad\ \, n\geq 2 \text{ even} \epc \nonumber \\
	2 \alpha_n & = 4 +  \Theta(-\alpha_{n-1})\,\alpha_{n-1} + \Theta(-\alpha_{n+1})\,\alpha_{n+1}  \epc \quad n\geq 3 \text{ odd} \epp
\end{align}
Notice that $\alpha_n\le 0$ for $n$ even, and so from $\alpha_1=2$ we have $\alpha_2=0$. However, this set of equations does not have a unique solution. The general form of the solution for integers $\alpha_n$ is the following,
\begin{equation}
	\{ \alpha_1, \alpha_2, \alpha_3, \ldots \} = \{2,0,2,0,\ldots,2,0,\alpha_{2k+1}<2,\alpha_{2k+2}(\alpha_{2k+1}),\alpha_{2k+3}(\alpha_{2k+1}),\ldots \}  ,
\end{equation}
where $k$ is a positive integer (or infinite), $\alpha_{2k+1}=1,0$ and  $\alpha_{n>2k+1}<0$ and they are unequivocally determined by $\alpha_{2k+1}$. Our intuition is that this freedom in our ansatz is apparent and it disappears when we take into account the BGT Eqs~\eqref{eq:BTGthlim_fact}. Indeed, we checked explicitly that the two $k=1$ solutions are not consistent with Eqs~\eqref{eq:BTGthlim_fact}. Therefore, the most natural choice is 
\begin{equation}
	\alpha_n = \left\{
\begin{array}{ll}
2 & \qquad \text{for } n \text{ odd} \epc \\
0& \qquad \text{for } n \text{ even} \epp
\end{array}\right.\phantom{\}}
\end{equation}
This means that the leading scaling exponent of $\eta_n(\lam)$ is only due to the $\ln(z)$ part of the driving term~\eqref{eq:driving_fact}.
At order $z^0$, the convolutions on the right-hand side of Eqs~\eqref{eq:TBA_XXZ_fact} are independent of $\lam$, and therefore the functional behavior of $\eta_n^{(0)}$ is determined by the driving terms only, {\it i.e.},
\begin{equation}
	\eta_n^{(0)}(\lam) = \left\{
\begin{array}{ll}
c_n \sin^2(2\lam) \epc & \qquad \text{for } n \text{ odd} \epc \\
c_n\tan^{-2} (\lam) \epc & \qquad \text{for } n \text{ even} \epc
\end{array}\right.\phantom{\}}
\end{equation}
where $c_n\geq 0$ on physical grounds (densities cannot be negative). The convolutions $s\ast \ln (1 + \eta_n)$ at order $z^0$ are zero if $n$ is odd and $2 \ln \left( 1 + \sqrt{c_n} \right)  + O(z)$ if $n$ is even. Substituting this into Eqs~\eqref{eq:TBA_XXZ_fact}, we have
\begin{equation}
	c_n = \left\{
\begin{array}{ll}
4 \left( 1 + \sqrt{a_{n-1}} \right) \left( 1 + \sqrt{a_{n+1}} \right)  \epc & \qquad \text{for } n \text{ odd} \epc \\
1 \epc & \qquad \text{for } n \text{ even} \epc
\end{array}\right.\phantom{\}}
\end{equation}
where by convention $a_0=0$. We find that 
\begin{equation}
	\eta_n^{(0)}(\lam) = \left\{
\begin{array}{ll}
8 \sin^2(2\lam) \epc & \qquad \text{for } n=1 \epc \\
16 \sin^2(2\lam) \epc & \qquad \text{for } n\geq 3 \text{ odd} \epc \\
\tan^{-2} (\lam) \epc & \qquad \text{for } n \text{ even} \epc
\end{array}\right.\phantom{\}}
\end{equation}
The functions $\eta_n^{(j)}$ for $j>0$ can  then be computed. Up to $j=3$ we have
\begin{align}
	\Phi_{1}(\lam) &= 2z \cos (2\lam) + z^{2}\left[ \cos(4\lam) + \tfrac{1}{2}\right] + z^{3} \left[ \tfrac{2}{3}\cos(6\lam) - 3 \cos(2\lam)\right] + O(z^{4}) \epc \nonumber \\[0.4ex]
	\Phi_{2}(\lam) &=  z^{2}\left[-8 \cos(2\lam) + 6\right] + \cO(z^{4}) \epc \nonumber \\[0.4ex]
	\Phi_{3}(\lam) &= 4z \cos (2\lam) + z^{2}\left[ 2 \cos(4\lam) + \tfrac{3}{2}\right] + z^{3} \left[ \tfrac{4}{3}\cos(6\lam) - 5 \cos(2\lam)\right]+ O(z^{4}) \epc \\[0.4ex]
	\Phi_{n}(\lam) &= z^{2}\left[-8 \cos(2\lam) + 8\right] + \cO(z^{4}) \epc \hspace{33.3ex} n\geq 4 \text{ even} \epc \nonumber \\[0.4ex]
	\Phi_{n}(\lam) &= 4z \cos (2\lam) + z^{2}\left[2 \cos(4\lam) + 2 \right] + z^{3} \left[ \tfrac{4}{3}\cos(6\lam) - 4 \cos(2\lam)\right]+ O(z^{4}) \epc
	\notag\\ & \hspace{64.9ex} n\geq 3 \text{ odd} \epp \nonumber
\end{align}
Using this expansion and the BGT Eqs~\eqref{eq:BTGthlim_fact}, the expansion for the densities [Eqs~\eqref{eq:sp_rho_exp} and~\eqref{eq:sp_rho_h_exp}] can then be computed as well.

\section{Large-$\Delta$ expansion of the GGE state} \label{app:GGE_EXP}
In this appendix we would like to discuss briefly the derivation of the large-$\Delta$ expansion for the GGE. In particular, we derive the leading terms of the expansion, making the computation of the next-leading terms straightforward.

As stated in Sec.~\ref{sec:large_Delta},  we need to expand the GTBA Eqs~\eqref{eq:GTBAGGE} for $n\geq 2$ and the BGT Eqs~\eqref{eq:BTGthlim_fact} for $n \geq 1$, and use the exact formula \eqref{eq:rho1h_exact_XXZ} for $\rho_{1, h}$. All information about the expectation values of the local charges is thus encoded in $\rho_{1,h}$, and we do not need to to compute the chemical potentials that appear only in the driving term of the $n=1$ GTBA Eq.~\eqref{eq:GTBAGGE}. Two useful sum rules to check the correctness of our assumptions are
\begin{subequations} \label{eq:norm_cond}
\begin{align}
	2\sum_{m=1}^{\infty}\int_{-\pi/2}^{\pi/2}\mathrm{d}\lam\: \rho_m(\lam) &= 1-  \int_{-\pi/2}^{\pi/2}\mathrm{d}\lam\: \rho_{1,h}(\lam)\epc \\
	2\sum_{m=1}^{\infty}m\int_{-\pi/2}^{\pi/2}\mathrm{d}\lam\: \rho_m(\lam) &= 1\epp
\end{align}
\end{subequations}
The first one is a consequence of the BGT Eqs~\eqref{eq:BTGthlim_fact}, while the second one expresses the conservation of the total magnetization. Our analytical ansatz is
\begin{subequations}
\begin{align}
	\eta_n(\lambda) &= z^{\alpha_n}\eta_{n}^{(0)}(\lambda) e^{\Phi_n(\lambda)}\epc  \qquad \Phi_n(\lambda) = \sum_{l=1}^\infty z^l\eta_n^{(l)}(\lambda)\epc \label{eq:GGE_ans_eta}\\
	\rho_{n,h}(\lambda) &= z^{\gamma_n}\rho_{n,h}^{(\gamma_n)}(\lambda) \left[1+\sum\nolimits_{l=1}^\infty z^l\rho_{n,h}^{(l+\gamma_n)}(\lambda)\right]\epc
\end{align}
\end{subequations}
where $\gamma_n \in \mathbb{N}$. Since $z=0$ corresponds to the quenchless point, we have $\rho_1(\lambda) =1/(2 \pi)+O(z)$.
Since $\rho_{1,h}(\lambda) =4 z^2 \sin^2 (2\lam)/\pi+O(z^3)$ [Eq.~\eqref{eq:rho1h_exact_XXZ}], we have $\gamma_1=\alpha_1=2$. Inserting the ansatz \eqref{eq:GGE_ans_eta} into the GTBA Eqs~\eqref{eq:GTBAGGE} for $n\geq 2$ and isolating the terms proportional to $\ln (z)$, we obtain
\be
	2 \alpha_n=\theta (-\alpha_{n-1}) \alpha_{n-1}+\theta (-\alpha_{n+1}) \alpha_{n+1}\epc \qquad n\geq 2 \epp
\ee
From here it follows that, for $n\ge 2$, $\alpha_n\le0$ and hence $\alpha_n=(n-1)\alpha_2$. Let us now expand the BGT Eqs~\eqref{eq:BTGthlim_fact} for $n\geq 2$. The leading term of the l.h.s.~is proportional to $z^{\gamma_n}+z^{\gamma_n-\alpha_n}\sim  z^{\gamma_n}$, while the r.h.s is proportional to $z^{\gamma_{n-1}}+ z^{\gamma_{n+1}}$. Notice that the term proportional to $z^{\gamma_n}$ in $s\ast \rho_{n,h}$ is always strictly positive as $\rho_{n,h}$ is always positive while $s(\lambda) = 1/(2 \pi)+ O(z)$. Therefore, we can conclude that $\gamma_n=\gamma_2 \leq 2$ for $n\geq 2$. Because of our analyticity hypothesis $\gamma_n\in \mathbb{N}$, there are three possible values for $\gamma_2$: 0, 1 and 2. Let us now expand the $n=1$ BGT Eq.~\eqref{eq:BTGthlim_fact} up to the second order. The case $\gamma_2=0$ can be excluded because $\rho_1(\lambda)=1/(2 \pi)+{\cal O} (z)$. Similarly, $\gamma_2\neq 1$ because if $\gamma_2=1$ we would have that $\int_{-\pi/2}^{\pi/2} \mathrm{d}\lam\, \rho^{(1)}(\lam)>0$, in contradiction with the sum rules \eqref{eq:norm_cond}. Therefore, we conclude that $\gamma_n = \gamma_2=2$. Moreover, we can conclude that $\alpha_{n\geq 2} = \alpha_2 = 0$, because otherwise $\rho_n\to +\infty$ for $ z\to 0$ and $n$ sufficiently large.

We are now in the position to compute all $\eta^{(0)}_n$. As we can see by expanding Eq.~\eqref{eq:GTBAGGE}, they are actually constant and obey the recursive relations
\begin{subequations}	
\begin{align}
	\ln(\eta_2^{(0)}) &= \frac{1}{2}\ln(1+\eta_3^{(0)})\epc\\
	\ln(\eta_n^{(0)}) &= \frac{1}{2}\left[\ln(1+\eta_{n-1}^{(0)}) + \ln(1+\eta_{n+1}^{(0)})\right]\epp
\end{align}
\end{subequations}	
The solution
\be
	\eta_{n\geq 2}^{(0)}=n^2-1
\ee
is the only one consistent with the sum rules \eqref{eq:norm_cond}. Expanding now the BGT Eqs~\eqref{eq:BTGthlim_fact} for $n\geq 2$ up to the second order, we have
\begin{subequations}	
\begin{align}
	\rho_{2,h}^{(0)}\left(1+(\eta_2^{(0)})^{-1}\right) &= \frac{1}{\pi} + \frac{1}{2}\rho_{3,h}^{(0)}\epc\\
	\rho_{n,h}^{(0)}\left(1+(\eta_n^{(0)})^{-1}\right) &= \frac{1}{2}\left[\rho_{n-1,h}^{(0)} +\rho_{n+1,h}^{(0)}\right]\epp
\end{align}
\end{subequations}	
The only solution to this system of recursion relations is $\rho_{n\geq 2,h}=2/(\pi n) + c \left(n^2-1 \right)$, where $c$ is an arbitrary constant. The only value of $c$ consistent with the sum rules~\eqref{eq:norm_cond} is $c=0$. Summarizing, we have
\begin{subequations}
\begin{align}
	&\eta_n=\left( n^2-1\right)+ O(z) \epc \qquad n\geq 2 \epc\\
	&\rho_{n,h}=\frac{2 z^2}{\pi n}+ O(z^3) \epc \qquad n\geq 2 \epc
\end{align}
\end{subequations}
Therefore,
\begin{equation}
	\rho_{n}=\frac{2 z^2}{\pi n \left( n^2-1\right)}+ O(z^3) \epc \qquad n\geq 2 \epc
\end{equation}
while $\rho_1$ can be computed using the $n=1$  BGT Eq.~\eqref{eq:BTGthlim_fact}
\begin{equation}
	\rho_1(\lam)=s(\lam)+ (s \ast \rho_{2,h})(\lam)-\rho_{1,h}(\lam)= \frac{1}{2\pi}\Big\{1 + 4z \cos(2\lam)+z^2[8\cos(4\lam) - 3] \Big\}+ \mathcal{O}(z^3) \epp
\end{equation}
Similarly, we can compute subleading orders of the expansion. The next-leading order vanishes for $n\geq 2$, while the next-next-leading order terms are reported in Eqs~\eqref{eq:GGE_exp}. As for the leading term, computing the GGE expansion involves the solutions of a set of recursion relations (one for $\eta_n$, another for $\rho_{n,h}$). Hence, the large-$\Delta$ expansion is technically more involved than the one for the quench action saddle-point state.

\section{Large-$\Delta$ expansion for local correlators} \label{app:CORR_EXP}
In this appendix, we would like to summarize the basic formulas for computing the local correlators $\langle \sigma^z_1 \sigma_2^z \rangle$ and  $\langle \sigma^z_1 \sigma_3^z \rangle$ as well as some intermediate results of their large-$\Delta$ expansion.

\subsection{The nearest-neighbors correlator $\langle \sigma^z_1 \sigma_2^z \rangle$ } 
The correlator $\langle \sigma^z_1 \sigma_2^z \rangle$ can be computed thanks to the Hellman-Feynman theorem~\cite{2014_Wouters, 2014_Mestyan}. We have
\begin{align} 
	\langle \sigma^z_1 \sigma^z_{2} \rangle  &=  1+ 4\left\{ \frac{\cosh(\eta)}{\sinh^2(\eta)} \frac{E}{J}  + \sum_{k \in \mathbb{Z}} |k | \left[\frac{ e^{- |k| \eta} }{2 \cosh(k\eta)} +\tanh(|k| \eta)  \left(   \frac{ e^{- |k|\eta } - \hat{\rho}_{1,h}(k) }{2  \cosh(k \eta)} \right) \right] \right.\notag \\
	&\qquad\qquad\qquad\qquad\left. - \pi \int_{-\pi/2}^{\pi/2} \mathrm{d}\lambda  \: \rho_1^h(\lambda) \sigma_1(\lambda)     \frac{\partial}{\partial \lambda}
s(\lambda) \right\} \epc
\end{align}
where $E$ is the energy of the state, $\hat{\rho}_{1,h}$ is the Fourier transform of $\rho_{1,h}$, while $s$ is defined in Eq.~\eqref{eq:defs}. The auxiliary function $\sigma_1$ satisfies the following set of equations
\begin{subequations}
\begin{equation} \label{eq:h_fact}
	(\rho_n +\rho_{n,h})\, \sigma_n = \left[  d_n  - s \ast (d_{n-1} + d_{n+1})\right] + s \ast (\sigma_{n-1} \,{\rho_{n-1,h}} + \sigma_{n+1} \,{\rho_{n+1,h}})  \epc
\end{equation}
with $\sigma_0 = d_0 = 0$. Here, $d_n$ is defined as
\begin{equation}
	d_n(\lambda) = \tilde{a}_n(\lambda) - \sum_{m=1}^\infty  \tilde{a}_{nm}\ast \rho_m \epc
\end{equation}
where
\begin{align}
	\tilde{a}_n(\lam) &= - \frac{n}{\pi}\sum_{k=1}^\infty \sin(2k\lam)z^{nk} \epc \\
	\tilde{a}_{nm}(\lambda) &= (1-\delta_{nm})\tilde{a}_{|n-m|}(\lambda) + 2 \tilde{a}_{|n-m| + 2}(\lambda) + \ldots + 2 \tilde{a}_{n+m-2}(\lambda) + \tilde{a}_{n+m}(\lambda)  \epp
\end{align}
\end{subequations}
The large-$\Delta$ expansion of the auxiliary functions $\sigma_n$ does not present any difficulty. The first difference between the saddle-point state and the GGE manifests itself at the $z^3$ order in $\sigma_1$, as it can be seen by the expansions
\begin{subequations}
\begin{align}
	\sigma_1^\text{sp}(\lam)  &= -2 \sin (2\lam) z + 2 \sin (4\lam) z^2 - 2 \sin(6\lam)z^3  +\frac{3}{2} \sin(2\lam) z^3  + O (z^4)  \epc\\
	\sigma_1^\text{GGE}(\lam)  &= -2 \sin (2\lam) z + 2 \sin (4\lam) z^2 - 2 \sin (6\lam) z^3 + O (z^4) \epp
\end{align}
\end{subequations}
This leads to a difference in the correlators only at the $z^6$ order, as stated in Eq.~\eqref{eq:diffcorr}.

\subsection{The next-to-nearest-neighbors correlator $\langle \sigma^z_1 \sigma_3^z \rangle$ } 
The correlator $\langle \sigma^z_1 \sigma_3^z \rangle$ can be computed thanks to a conjecture proposed in Ref.~\cite{2014_Mestyan}. However, it is necessary to compute two sets of auxiliary functions, and not only one as for $\langle \sigma^z_1 \sigma_2^z \rangle$. Given $\eta_n=\rho_{n,h}/\rho_n$, let us define the functions $\rho_{n, h}^{(a)}$ and $\rho_n^{(a)}=\rho_{n,h}^{(a)}/\eta_n$ ($a=0,1,2,\ldots$), determined by the set of equations
\begin{equation}
	\rho_{n, h}^{(a)}(\lambda) \left[1+\eta_n^{-1}(\lambda)\right]=\delta_{n ,1} \frac{\mathrm{d}^{a}}{\mathrm{d}\lambda^{a}}s(\lambda) + \left[s \ast \left(\rho_{n-1, h}^{(a)}+\rho_{n+1, h}^{(a)} \right)\right](\lambda) \epc
\end{equation}
where $\rho_{0, h}^{(a)}(\lambda)=0$. Notice that $ \rho_{n, h}^{(0)}=\rho_{n, h}$ and $ \rho_{n}^{(0)}=\rho_{n}$.  Now, we are ready to introduce the functions $\sigma_n^{(a)}$ satisfying
\begin{equation}
	(\rho_n +\rho_{n,h})\,\sigma_n^{(a)} =\left[d_n^{(a)}-s\ast \left(d_{n-1}^{(a)}+d_{n+1}^{(a)}\right) \right] + s \ast  \left[\sigma_{n-1}^{(a)} \rho_{n-1, h}+\sigma_{n+1}^{(a)}\rho_{n+1, h} \right],
\end{equation}
where $\sigma^{(a)}_0(\lambda) = d^{(a)}_0(\lambda) = 0$ and $d_n^{(a)}(\lambda) =\partial_\lambda^{a} \tilde{a}_n(\lambda) -\sum_{m=1}^{\infty} (\tilde{a}_{n m} \ast \rho_m^{(a)})(\lambda)$. For $a=0$, $\sigma^{(a)}_n$ reduces to the function $\sigma_n$ defined in Eq.~\eqref{eq:h_fact}. Given these sets of auxiliary functions, $\langle \sigma^z_1 \sigma_3^z \rangle$ can be expressed as
\begin{equation}\label{eq:z1z3_poly}
	\langle \sigma^z_1 \sigma^z_3 \rangle = \langle \sigma^z_1 \sigma^z_2 \rangle - \tanh(\eta) \frac{4 \Omega_{0,0} - \Omega_{0,2} + 2 \Omega_{1,1} }{4} + \frac{\sinh^2(\eta)}{4} \Gamma_{1,2} \epp
\end{equation}
The quantities $\Omega_{a b}$ and $\Gamma_{a b}$ are defined as
\begin{subequations}\label{eq:Omega_Gamma}
\begin{align}
	\Omega_{a b} &=4 \pi \int_{-\frac \pi 2}^{\frac \pi 2} \mathrm{d}\mu\, s^{(b)}(-\mu) \left[(-1)^a a_1(\mu) + (-1)^{b+1} \rho_{1, h}^{(a)} \right] \epc\\
	\Gamma_{a b} &= (-)^b 4 \pi  \int_{-\frac \pi 2}^{\frac \pi 2} \mathrm{d}\mu\,\Big[ s^{(a+b)}(-\mu) \,\tilde{a}_1(\mu)+ g^{(a+b)}(-\mu) \tilde{a}_1 (\mu)\nonumber \\
		& \qquad\qquad\qquad \qquad\qquad+ \tilde{g}^{(b)}(-\mu) \rho_{1,h}^{(a)} (\mu) -s^{(b)}(-\mu) \rho_{1, h}(\mu) \sigma^{(1)}_1 (\mu) \Big] \epc
\end{align}
\end{subequations}
where the superscript $\!^{(a)}$ stands for the $a$-th derivative with respect to $\lambda$, and
\begin{subequations}
\begin{align}
	g(\lambda) &= \frac 2 \pi \sum_{k=1}^{\infty} \frac{\tanh(k \eta)} {2 \cosh(k \eta)} \cos(2  k \lambda) \epc\\
	\tilde{g}(\lambda) &= \frac 1 \pi \sum_{k=1}^{\infty} \frac{\tanh(k \eta)} {2 \cosh(k \eta)} \sin(2  k \lambda) \epp
\end{align}
\end{subequations}
In order to compute $\langle \sigma_1^z \sigma_3^z \rangle$ we need $\rho_n^{(1)}$ (to compute $d^{(a)}_n$) and $\rho_{1,h}^{(1)}$ and $\sigma_1^{(1)}$. The leading behavior of $\rho_{n,h}^{(1)}$ is
\begin{subequations}
\begin{align}
	&  \rho^{(1)\,\text{sp}}_{n,h}(\lambda)\ \, \,\,\sim  -  32\, 6^{\frac {n -1}2} \, z^{2n+1} \,\sin^3(2 \lam)+O(z^{2n+2})\epc & &\textrm{$n$ odd} \epc \\
	&  \rho^{(1)\, \text{sp}}_{n,h}(\lambda)\ \,\,\, \sim  -  48\, 6^{\frac n 2 -1} \, z^{2n} \,\cos^3(\lam)\, \sin(\lam)+O(z^{2n+1})\epc & & \textrm{$n$ even} \epc\\
	&  \rho^{(1)\,\text{GGE}}_{1,h}(\lambda) \sim  -  \frac {32}{\pi}\,  z^{3} \,\sin^3(2 \lam)+O(z^{4})\epc \\
	&  \rho^{(1)\,\text{GGE}}_{n,h}(\lambda) \sim  -  \frac{12}{\pi}\,\frac{n+1}{n}  z^{n+2} \,\sin(2 \lam)+O(z^{n+3})\epc & &n\geq 2 \epp
\end{align}
\end{subequations}
and the resulting expansion for $\sigma_1^{(1)}$ is thus
\begin{subequations}
\begin{align}
	\sigma_1^{(1),\text{sp}}(\lambda) &= -4 z \cos (2 \lambda ) + 8 z^2 - 4z^3 [\tfrac{5}{2}\cos (2 \lambda )+\cos (6 \lambda )] - z^4 [2\cos (4 \lambda ) -7]+O(z^5) \\
	\sigma_1^{(1),\text{GGE}}(\lambda) &= -4 z \cos (2 \lambda ) + 8 z^2 - 4z^3 [2\cos (2 \lambda )+\cos (6 \lambda )] - z^4 [8\cos (4 \lambda )+2]+O(z^5)\epp
\end{align}
\end{subequations}
Knowing the small-$z$ expansions of the functions $\rho_{1,h}^{(a)}$, $a=0,1$, and $\sigma_{1}^{(1)}$, plugging them into Eqs~\eqref{eq:Omega_Gamma}, and afterwards the results into Eq.~\eqref{eq:z1z3_poly}, gives finally the large-$\Delta$ expansions \eqref{eq:z1z3_sp} and \eqref{eq:z1z3_GGE} of the next-to-nearest neighbor correlator.

\section{Spin content of the N\'eel state} \label{app:spin_content}

\subsection{Global spin operators}\label{sec:total_spin_operators}
It is well-known that the spin-1/2 XXX Hamiltonian ($\Delta=1$) exhibits a global $SU(2)$ symmetry. Let us consider the global $SU(2)$ operators (here and in the following we choose $N$ even, such that zero magnetization states are always possible)
\begin{equation}\label{eq:S_def}
	S^\alpha = \sum_{j=1}^N s_j^\alpha\epc \qquad\text{for}\quad \alpha = x,y,z,+,- \epp
\end{equation}
The operators $s_j^\alpha = \sigma_j^\alpha/2$ represent the local spin degrees of freedom and act locally as $SU(2)$ operators. They have the usual commutation relations
\begin{equation}
	[s_j^\alpha, s_k^\beta] = i\delta_{jk}\epsilon_{\alpha\beta\gamma}s_k^\gamma \qquad\text{for}\quad \alpha, \beta, \gamma \in \{x,y,z\}
\end{equation}
where $\epsilon_{\alpha\beta\gamma}$ is the total anti-symmetric epsilon tensor. Using the definitions $s_j^{\pm}=s_j^x \pm i s_j^y$ these commutation relations transform into $[s_j^z,s_k^\pm] = \pm \delta_{jk} s_k^\pm$ and $[s_j^+,s_k^-] = 2\delta_{jk}s_k^z$. Similar relations hold for the global operators,
\begin{equation}\label{eq:SU2_relations}
	[S^z,S^\pm] = \pm S^\pm \qquad\text{and}\qquad [S^+,S^-] = 2S^z\epp
\end{equation}    
The total spin operator 
\begin{equation}\label{eq:S_squared}
	S^2 \equiv \vec{S}^2 = \sum_{\alpha=x,y,z} S^\alpha S^\alpha = \frac{1}{2}\left(S^+S^- + S^-S^+\right) + \left(S^z\right)^2 = S^+S^- - S^z + \left(S^z\right)^2
\end{equation}
is a central element of $SU(2)$, {\it i.e.}, $[S^2,S^\alpha]=0$ for all $\alpha = x,y,z,+,-$.

The Hilbert space of the XXX chain is given by an $N$-fold tensor product of local spin-1/2 $SU(2)$ representation spaces. Due to the global $SU(2)$ symmetry, we can choose simultaneous eigenstates of $S^z$ and $S^2$ with eigenvalues $s^z$ and $s(s+1)$, respectively, as an orthonormal basis of the Hilbert space. The eigenstates are denoted by $|s,s^z, a\rangle$, where the integer values $s$, $s^z$, and $a$ are restricted by $0\leq s \leq N/2$, $-s\leq s^z \leq s$, and $1\leq a \leq A_N(s)$. Here, $A_N(s)$ is the number of $(2s+1)$-multiplets in the $N$-fold tensor product of $SU(2)$ spin-1/2 representations, 
\begin{equation}
	A_N(s) = \begin{pmatrix}
		N \\ \frac{N}{2}-s
	\end{pmatrix} - \begin{pmatrix}
		N \\ \frac{N}{2}-s-1
	\end{pmatrix}\epp
\end{equation}

The Bethe states, which are constructed as eigenstates of the operator $S^z$, form multiplets of the global $SU(2)$ symmetry. A highest-weight state $|s,s, a\rangle$ is a Bethe state with $N/2-s$ finite rapidities. Other states of the multiplet, with $s^z < s$, are constructed by repeatedly applying ($s-s^z$ times) the total spin-lowering operator $S^-$ to the highest-weight state. This operator can be interpreted as the creation of a magnon with zero momentum, corresponding to a rapidity at infinity, see Eq.~\eqref{eq:momentum}. Infinite rapidities decouple from the Bethe equations and the newly obtained state remains an eigenstate of the Hamiltonian. A generic state $|s,s^z, a\rangle$ can be therefore seen as a Bethe state with $N/2-s$ finite rapidities, supplemented by $s-s^z$ infinite rapidities. 

Let us define the operator $\hat{N}_\infty$, counting the number of infinite rapidities, {\it i.e.}, $ \hat{N}_\infty |s,s^z, a\rangle = (s-s^z)|s,s^z, a\rangle$. Note that $\hat{N}_\infty$ is a conserved quantity. We are interested in the expectation value of the number of infinite rapidities on the N\'eel state. For a generic zero-magnetization state $\ket{\Psi}$ we easily find
\begin{equation}
\langle \Psi | \hat{N}_\infty | \Psi\rangle =  \sum_{s=0}^{N/2} s \sum_{a=1}^{A_N(s)}  \left|\langle \Psi | s,0,a \rangle \right|^2 = \sum_{s=0}^{N/2} s \, C_s \epc
\end{equation}
where $C_s$ can be interpreted as a measure of how much overlap the state $|\Psi\rangle$ has with the total spin-$s$ sector.

To find this ``spin content'' of a generic state, define the function $f_N$ as the Fourier transform of the coefficient $C_s$,
\begin{equation}\label{eq:def_f}
	f_N(x) = \sum_{s=0}^{N/2} C_s e^{2s(s+1)x/N}\epp
\end{equation}
The inverse transformation exists and yields
\begin{equation}\label{eq:inv_FT}
	\frac{2}{i\pi N}\int\limits_{0}^{i\pi N/2}\mathrm{d}x\: f_N(x)e^{-2t(t+1)x/N} = \sum_{s=0}^{N/2} C_s \left(\frac{2}{i\pi N}\int\limits_{0}^{i\pi N/2}\mathrm{d}x\: e^{2[s(s+1)-t(t+1)]x/N}\right) = C_t\epc
\end{equation}
where we used that $[s(s+1)-t(t+1)]=0$ if and only if $s=t$ for non-negative integers $s$ and $t$. The coefficient $C_s$ is thus determined by the function $f_N$, which can be expressed by its Taylor series around $x=0$,
\begin{align} \label{eq:f_taylor}
	f_N(x) &= \sum_{n=0}^{\infty}\frac{1}{n!}f_N^{(n)}(0)x^n = \sum_{n=0}^{\infty}\frac{1}{n!}\sum_{s=0}^{N/2} C_s s^n(s+1)^n\left(\frac{2x}{N}\right)^n \notag \\
	&= \sum_{n=0}^{\infty}\frac{1}{n!}\left(\frac{2x}{N}\right)^n \langle\Psi | \left(S^+S^-\right)^n|\Psi\rangle \epp
\end{align}
For the last equality, we used Eq.~\eqref{eq:S_squared}, the zero-magnetization property and the following expression for the expectation value of the total-spin operator
\begin{equation}
\langle \Psi | \left(S^2\right)^n | \Psi\rangle = \sum_{s=0}^{N/2} s^n(s+1)^n \sum_{a=1}^{A_N(s)}\left|\langle \Psi | s,0,a \rangle \right|^2 = \sum_{s=0}^{N/2} s^n(s+1)^n C_s \epp
\end{equation}
It is convenient to bring the operators $S^+$ and $S^-$ of the product $\left(S^+S^-\right)^n$ in an appropriate order,
\begin{equation}\label{eq:exp_val_pow_S_squared}
 \bra{\Psi} \left(S^+S^-\right)^n \ket{\Psi} = \sum_{m=0}^n c_m^{(n)} \bra{\Psi} \left(S^+ \right)^m \left(S^- \right)^m \ket{\Psi} \epp
\end{equation}
As shown in \ref{sec:LS_triangle}, the coefficients $c_m^{(n)}$ are Legendre-Stirling numbers and given by
\begin{equation} \label{eq:result_legendre_sterling}
c_0^{(0)}=1 , \qquad c_{m}^{(n)} = \sum_{r=1}^m\frac{(-1)^{r+m}(2r+1)r^n(r+1)^n}{(m+r+1)!(m-r)!}
\end{equation}
for $n\geq 1$. Furthermore, the expectation values of the operator $(S^+S^-)^m$ on an arbitrary zero-magnetization state cannot be evaluated in general. However, let us focus on a special class of states that can be expressed in the local spin basis as a single product of local spin lowering operators acting on the fully-polarized state ({\it e.g.}~the N\'eel state), 
\begin{equation}\label{eq:initial_state}
 \ket{\Psi} = 	|\{n_j\}_{j=1}^{N/2}\rangle = \prod_{j=1}^{N/2}s_{n_j}^-\left|\uparrow\right\rangle^{\otimes N}\epp
\end{equation}
The integers $\{n_j\}_{j=1}^{N/2}$ with $1\leq n_1 <\ldots <  n_{N/2}\leq N$ label the positions of the downspins. One easily finds
\begin{equation}  \label{eq:result_expectation_splussmin}
	\bra{\Psi} \left(S^+\right)^m \left(S^-\right)^m \ket{\Psi} = \langle\{n_j\}_{j=1}^{N/2} | \left(S^+\right)^m \left(S^-\right)^m | \{n_j\}_{j=1}^{N/2} \rangle = (m!)^2\left(\!\!\!\begin{array}{c} N/2 \\ m \end{array}\!\!\!\right)\epp
\end{equation}
Plugging Eqs~\eqref{eq:result_legendre_sterling} and~\eqref{eq:result_expectation_splussmin} into Eq.~\eqref{eq:f_taylor}, we eventually obtain
\begin{align}
	f_N(x) &= c_0^{(0)} + \sum_{n=1}^{\infty} \sum_{m=1}^{n}\frac{(m!)^2}{n!} \left(\!\!\!\begin{array}{c} N/2 \\ m \end{array}\!\!\!\right) \sum_{r=1}^m\frac{(-1)^{r+m}(2r+1)r^n(r+1)^n}{(m+r+1)!(m-r)!}\left(\frac{2x}{N}\right)^n \notag\\
	&= 1 + \sum_{m=1}^{N/2} \sum_{r=1}^m (m!)^2\left(\!\!\!\begin{array}{c} N/2 \\ m \end{array}\!\!\!\right) \frac{(-1)^{r+m}(2r+1)}{(m+r+1)!(m-r)!}\sum_{n=1}^\infty\frac{1}{n!} \left(\frac{2r(r+1)x}{N}\right)^n \notag\\
	&= 1+ \sum_{m=1}^{N/2} \sum_{r=1}^m \left(\!\!\!\begin{array}{c} N/2 \\ m \end{array}\!\!\!\right) \frac{(-1)^{r+m}(m!)^2(2r+1)}{(m+r+1)!(m-r)!}\left(e^{2r(r+1)x/N}-1\right)\epp
\end{align}
We used that $c_m^{(n)}=0$ if $m=0$ or $m>n$, as can be seen from Eq.~\eqref{eq:result_legendre_sterling}. Using now the inverse Fourier transform~\eqref{eq:inv_FT} we can read off the coefficients $C_s$. They are given by
\begin{equation}\label{eq:main_result}
	C_s =  \sum_{m=s}^{N/2}\left(\!\!\!\begin{array}{c} N/2 \\ m \end{array}\!\!\!\right) \frac{(-1)^{s+m}(m!)^2(2s+1)}{(m+s+1)!(m-s)!} = \frac{(2s+1)\ (N/2)!^2}{(N/2-s)!(N/2+s+1)!} = \frac{A_N(s)}{\begin{pmatrix}
	N \\ N/2
	\end{pmatrix}}\epp
\end{equation}
The fact that $C_s$ is directly proportional to $A_N(s)$, the number of all zero-magnetization states in a fixed $s$-sector, is remarkable. It means that the average overlap squared is the same ($
=(N/2)!^2/N!$) for each sector. Therefore, one cannot argue that overlaps with higher $s$, {\it i.e.}, with more rapidities at infinity, $N_\infty = s$, decrease with increasing $s$. Only the number of zero-magnetization states $A_N(s)$ per $s$-sector decreases with increasing $s$ for sufficiently large $s$.

\subsection{Limit of large number of lattice sites}\label{sec:large_N}
The formula for $C_s$, which is a measure of how much spin $s$ is contained in a zero-magnetization state of the form~\eqref{eq:initial_state} and which is directly proportional to the number $A_N(s)$ of $(2s+1)$-multiplets for a given $N$, can be further analyzed in the limit of large lattice site $N$.

In the limit $N\to\infty$ we use Stirling's formula to manipulate Eq.~\eqref{eq:main_result}. After a straightforward calculation one obtains the scaling of the coefficient $C_s$ with large $N$,
\begin{equation}
C_s \sim \frac{2(2s+1)}{N}e^{-2s(s+1)/N} \epp
\end{equation}
This function has a maximum at $s_0 = (\sqrt{N}-1)/2\sim \sqrt{N}/2$ or, to be more precise, at the integer which lies as close as possible to this generally irrational number. Furthermore, the expectation value of the number of infinite rapidities can be computed analytically,
\begin{equation}
\langle \Psi | \hat{N}_\infty | \Psi\rangle =  \sum_{s=0}^{N/2} s \, C_s = \frac{1}{2}\left(\frac{2^N(N/2)!^2}{N!}-1\right)\epp
\end{equation}
Using Stirling's formula one finds that
\begin{equation}
\lim_{N \to \infty} \frac{\langle \Psi | \hat{N}_\infty | \Psi\rangle }{\sqrt{N}} = \sqrt{\frac{\pi}{8}} \epp
\end{equation}
In the thermodynamic limit, the number of infinite rapidities of the steady state is negligible compared to the total number of rapidities, {\it i.e}, $n_\infty = \lim_{N\to\infty} N_\infty / N =~0$. This serves as additional evidence for the correctness of the application of the quench action approach to the N\'eel-to-XXX quench.

\subsection{Legendre-Stirling numbers of the second kind}\label{sec:LS_triangle}
The coefficients $c_m^{(n)}$ appear in the reordering of operators $S^\pm$ in the product $(S^+S^-)^n$ to get terms like $(S^+)^m(S^-)^m$, see Eq.~\eqref{eq:exp_val_pow_S_squared}. Since we consider this inside expectation values $\langle \cdot \rangle$ of zero-magnetization states and since for these states
\begin{align}
	\left\langle S^+S^-\left(S^+\right)^m\left(S^-\right)^m\right\rangle &= \left\langle\left(S^+\right)^{m+1}\left(S^-\right)^{m+1} \right\rangle  +\left(2 + 4 + \ldots + 2m\right) \left\langle\left(S^+\right)^m\left(S^-\right)^m\right\rangle \notag \\
	&= \left\langle\left(S^+\right)^{m+1}\left(S^-\right)^{m+1} \right\rangle  +m(m+1) \left\langle\left(S^+\right)^m\left(S^-\right)^m \right\rangle \epc
\end{align}
we obtain
the relations ($c_m^{(n)}:=0$ for $m>n$ or $m<0$)
\begin{subequations}\label{eq:recursion_relations}
\begin{equation}
	c_0^{(0)} = 1\epc \qquad
	c_{m}^{(n+1)} = m(m+1)c_m^{(n)} + c_{m-1}^{(n)}\qquad\text{for}\quad 0\leq m \leq n+1\epc \quad n\geq 0\epp
\end{equation}
\end{subequations}
These recursion relations define the triangle of Legendre-Stirling numbers of second kind, which have an explicit representation for $n\geq 1$,
\begin{equation}\label{eq:c_m_n}
	c_{m}^{(n)} = \sum_{r=1}^m\frac{(-1)^{r+m}(2r+1)r^n(r+1)^n}{(m+r+1)!(m-r)!}\epp
\end{equation}

\section{Sumrule $N=12$} \label{app:overlaps_N_12}
Table~\ref{table:RV:sumruleN12} shows all Bethe states with nonzero overlap to the N\'eel state at $N=12$. The rapidities of the Bethe states were obtained by iteratively solving a parametrization for the Bethe equations for deviated strings~\cite{2007_Hagemans_JPA_40} and subsequently plugged into Eq.~\eqref{eq:overlap_XXX_onshell}.

Note that Bethe states with a single even-length string with quantum number zero, {\it i.e.}, with string center at zero, have identically zero overlaps with the N\'eel state. These states are not displayed in the table. For an even number of even-length strings at the origin, the string deviations keep the overlap finite. This is for example the case with the coinciding 4- and 2-string. The rapidities of this Bethe state were obtained in Ref.~\cite{2013_Hao_PRE_88} by homotopy continuation.

\begin{table}[h]
\scriptsize
\centering
Bethe states with nonzero N\'eel overlap ($N=12$)\\[1ex]
\begin{tabular}{rrrrr}
String content & $2I^+_n$ & E & $|\langle \{\lambda\}| \Psi_0 \rangle|^2$ & $\sum |\langle \{\lambda\}| \Psi_0 \rangle|^2$ \\[0.3em]
\toprule
6 inf & - & $0$ & $0.002164502165$ & $0.002164502165$ \\
\midrule
2 one, 4 inf &$1_1 $ & $-3.918985947229$ & $0.096183409244$ & $0.116883116883$ \\
 &$3_1 $ & $-3.309721467891$ & $0.011288497947$ \\
 &$5_1 $ & $-2.284629676547$ & $0.004542580506$ \\
 &$7_1 $ & $-1.169169973996$ & $0.002752622983$ \\
 &$9_1 $ & $-0.317492934338$ & $0.002116006203$ \\
\midrule
4 one, 2 inf &$1_1 3_1 $ & $-7.070529325964$ & $0.310133033838$ &$ 0.554809782804$ \\
  &$1_1 5_1 $ & $-5.847128730477$ & $0.129277023687$ \\
  &$ 1_1 7_1$ & $-4.570746557876$ & $0.085992436024$ \\
  &$ 3_1 5_1$ & $-5.153853093221$ & $0.015256395523$ \\
  &$3_1 7_1 $ & $-3.916336243695$ & $0.010091113504$ \\
  &$5_1 7_1 $ & $-2.817696043731$ & $0.004059780228$ \\
  \midrule
2 two, 2 inf &$1_2 $ & $-1.905667167442$ & $0.001207238321$ & $0.005468702625$\\
  &$3_2 $ & $-1.368837200825$ & $0.002340453815$ \\
  &$5_2 $ & $-0.681173793635$ & $0.001921010489$ \\
    \midrule
1 one, 1 three, 2 inf &$0_1 0_3 $ & $-2.668031843135$ & $0.034959609810$ & $0.034959609810$ \\
    \midrule
6 one &$1_1 3_1 5_1 $ & $-8.387390917445$ & $0.153412152966$ & $0.153412152966$ \\
  \midrule
2 two, 2 one &$1_1 1_2 $ & $-5.401838225870$ & $0.040162686361$ & $0.046134750850$ \\
  &$3_1 1_2 $ & $-4.613929948329$ & $0.004636541934$ \\
  &$5_1 1_2 $ & $-3.147465758841$ & $0.001335522556$ \\
    \midrule
1 three, 3 one &$0_1 2_1 0_3$ & $-6.340207488736$ & $0.052743525774$ & $0.078910020729$ \\
  &$0_1 4_1 0_3$ & $-5.203653009936$ & $0.015022005621$ \\
  &$0_1 6_1 0_3$ & $-3.788693957250$ & $0.011144489334$ \\
      \midrule
1 five, 1 one &$0_1 0_5$ & $-2.444293750583$ & $0.005887902992$ & $0.005887902992$ \\
      \midrule
2 three &$1_3$ & $-1.111855930538$ & $0.001342476001$ & $0.001342476001$ \\
      \midrule
1 two, 1 four &$0_2 0_4$ & $-1.560671012472$ & $0.000026982174$ & $0.000026982174$ \\
  \bottomrule
 \end{tabular}
\caption{All Bethe states for $N=12$ with nonzero overlap with the zero-momentum N\'eel state. The overlap squares add up to $1$ up to the precision in which the Bethe equations were solved. The $2I^+_n$ in the second column give the positive $n$-string quantum numbers of the parity-invariant Bethe states.}
\label{table:RV:sumruleN12}
\end{table}

\newpage
\section*{References}

\bibliographystyle{iopart-num}
\bibliography{XXZ_TBA_biblio_Long_paper_new}

\end{document}